\pdfoutput=1

\documentclass[sn-mathphys, Numbered, iicol]{sn-jnl}

\usepackage{graphicx}%
\usepackage{multirow}%
\usepackage{amsmath,amssymb,amsfonts}%
\usepackage{amsthm}%
\usepackage{mathrsfs}%
\usepackage[title]{appendix}%
\usepackage{xcolor}%
\usepackage{textcomp}%
\usepackage{manyfoot}%
\usepackage{booktabs}%
\usepackage{algorithm}%
\usepackage{algorithmicx}%
\usepackage{algpseudocode}%
\usepackage{listings}%
\usepackage{cleveref}
\usepackage{ulem}
\usepackage{xspace}
\usepackage{mhchem}
\usepackage{subcaption}
\usepackage{float}

\usepackage{tabularx}
\usepackage{fontawesome}
\usepackage{xr}
\usepackage{filecontents}

\DeclareUnicodeCharacter{0308}{\"}  
\DeclareUnicodeCharacter{030C}{\v}  
\DeclareUnicodeCharacter{0301}{\'{}}  

\makeatletter

\newcommand*{\addFileDependency}[1]{
  \typeout{(#1)}
  \@addtofilelist{#1}
  \IfFileExists{#1}{}{\typeout{No file #1.}}
}
\makeatother

\newcommand*{\myexternaldocument}[1]{
    \externaldocument{#1}
    \addFileDependency{#1.tex}
    \addFileDependency{#1.aux}
}

\myexternaldocument{supplement}

\newcolumntype{Y}{>{\centering\arraybackslash}X}

\begin{document}

\title[Article Title]{Advancing Surface Chemistry with Large-Scale Ab-Initio Quantum Many-Body Simulations}


\author*[1]{\fnm{Zigeng} \sur{Huang}}\email{huangzigeng@bytedance.com}
\author[1]{\fnm{Zhen} \sur{Guo}}\nomail
\author[1]{\fnm{Changsu} \sur{Cao}}\nomail
\author[2]{\fnm{Hung Q.} \sur{Pham}}\nomail
\author[1]{\fnm{Xuelan} \sur{Wen}}\nomail
\author*[3]{\fnm{George H.} \sur{Booth}}\email{george.booth@kcl.ac.uk}
\author*[4,5]{\fnm{Ji} \sur{Chen}}\email{ji.chen@pku.edu.cn}
\author*[1]{\fnm{Dingshun} \sur{Lv}}\email{lvdingshun@bytedance.com}

\affil[1]{ByteDance Research, Fangheng Fashion Center, No. 27, North 3rd Ring West Road, Haidian District, Beijing 100098, People’s Republic of China}
\affil[2]{ByteDance Research, San Jose, CA 95110, United States}
\affil[3]{Department of Physics, King’s College London, Strand, London WC2R 2LS, United Kingdom}
\affil[4]{School of Physics, Peking University, Beijing 100871, People’s Republic of China}
\affil[5]{Interdisciplinary Institute of Light-Element Quantum Materials, Frontiers Science Center for Nano-Optoelectronics, Peking University, Beijing 100871, People’s Republic of China}

\abstract{
Predictive simulation of surface chemistry is of paramount importance for progress in fields from catalysis to electrochemistry and clean energy generation. \textit{Ab-initio} quantum many-body methods should be offering deep insights into these systems at the electronic level, but are limited in their efficacy by their steep computational cost. In this work, we build upon state-of-the-art correlated wavefunctions to reliably converge to the `gold standard' accuracy in quantum chemistry for application to extended surface chemistry. Efficiently harnessing graphics processing unit acceleration along with systematically improvable multiscale resolution techniques, we achieve linear computational scaling up to 392 atoms in size. These large-scale simulations demonstrate the importance of converging to these extended system sizes, achieving a validating handshake between simulations with different boundary conditions for the interaction of water on a graphene surface. We provide a new benchmark for this water-graphene interaction that clarifies the preference for water orientations at the graphene interface. This is extended to the adsorption of carbonaceous molecules on chemically complex surfaces, including metal oxides and metal-organic frameworks, where we consistently achieve chemical accuracy compared to experimental references, and well inside the scatter of traditional density functional material modeling approaches. This pushes the state of the art for simulation of molecular adsorption on surfaces, and marks progress into a post-density functional era for more reliable and improvable approaches to first-principles modeling of surface problems at an unprecedented scale and accuracy using \textit{ab-initio} quantum many-body methods.
}

\maketitle
\small
\clearpage

\section{Main}\label{sec:Introduction}

\begin{figure*}[htbp]
\centering
\includegraphics[width=1.0\linewidth]{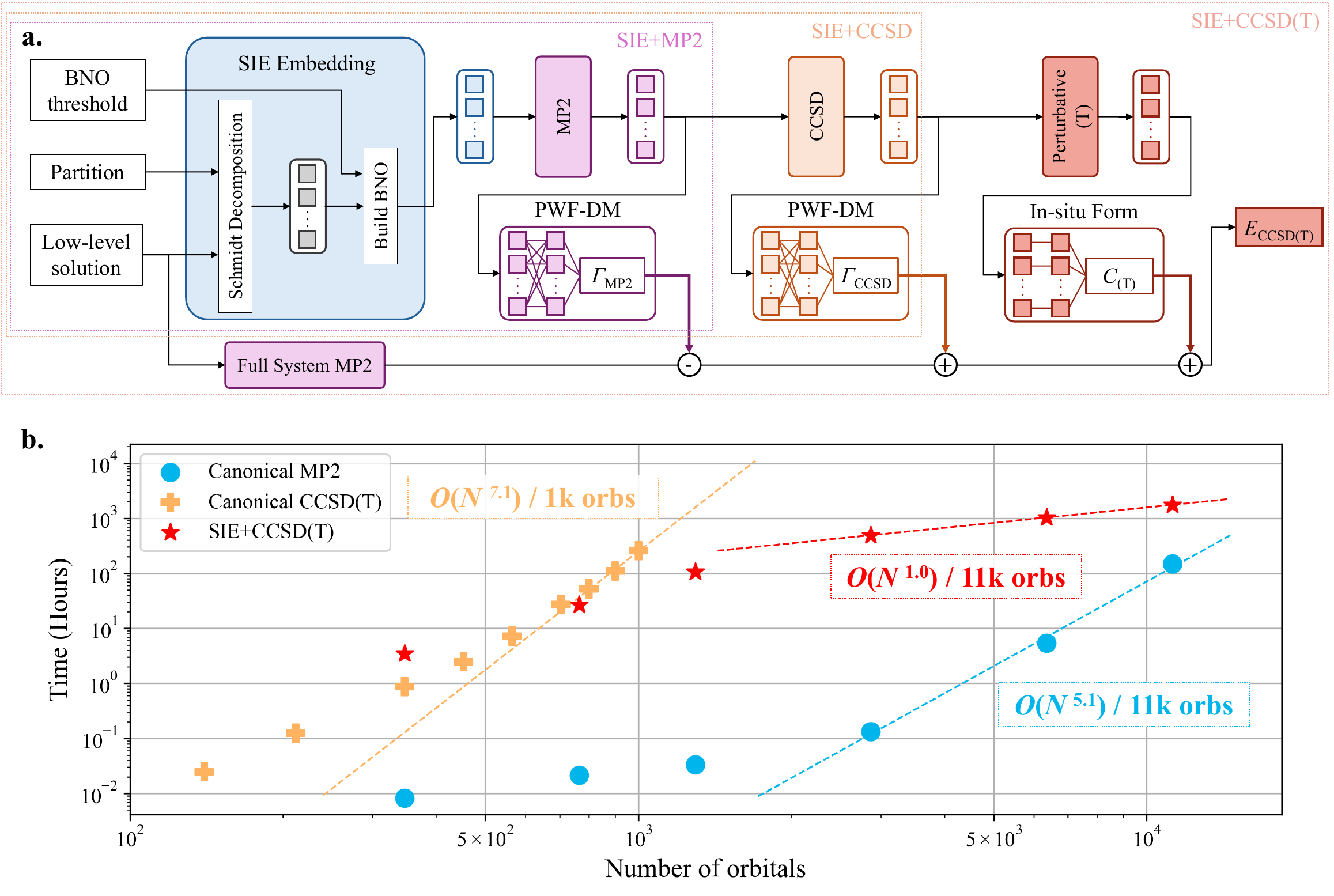}
\caption{\label{fig:workflow} \textbf{Systematically improvable quantum embedding~(SIE) framework and our GPU-accelerated quantum embedding for surfaces.} Panel \textbf{(a)} illustrates the general workflow for a controllable multi-scale resolution of correlations at different length scales, comprising SIE+MP2~(purple box), SIE+CCSD~(orange box), and SIE+CCSD(T)~(red box), with further details in the Methods Sec.~\ref{sec:Methods} and supplemental information~(SI) Sec.~\ref{sec:SI_method}. Panel \textbf{(b)} presents the computational time as a function of the total number of orbitals, 
measured on a single A100 GPU for systems comprising water monomer adsorbed on graphene under open boundary condition. The annotations beside the curves show the scaling obtained by fitting the last three points up to the maximum size of the test system. Further discussion on the observed linear scaling of SIE+CCSD(T) are provided in SI section~\ref{sec:SI_LS_SIE}.}
\end{figure*}

The chemical and physical properties of material surfaces play a central role in a wide range of scientific research areas and industrial applications~\cite{king2012chemical,schedin2007detection,farha2010novo}. However, predictive simulations of these systems without resorting to empiricism remains a formidable and long-standing challenge, requiring an accurate solution to the many-electron problem~\cite{schollwock2005density, riplinger2013efficient, riplinger2013natural, riplinger2016sparse, foulkes2001quantum, bartlett2007coupled}. The formal solution to this problem scales exponentially in computational effort with system size, and therefore to make progress necessitates a sufficiently accurate approximation to the electron correlations while also rendering the problem computationally efficient to scale to large systems at the bulk limit. Over the past few decades, density functional theory (DFT) has emerged as the standard bearer for this, favored for its wide applicability and low computational cost~\cite{HK_theorem,KS-DFT,kohn1999nobel}. However, DFT is not systematically improvable due to its reliance on semi-empirical exchange-correlation functionals, that are not universal and may not provide sufficient transferability or internal validation of its accuracy across different chemical environments. In contrast, correlated methods~\cite{pople1999nobel}, such as coupled cluster theories~\cite{bartlett2007coupled} or quantum Monte Carlo methods~\cite{foulkes2001quantum}, can achieve superior accuracy by explicitly describing electron correlation, and importantly their accuracy can be systematically improved~\cite{booth2013towards,austin2012quantum,li2022ab, ambrosetti2014hard,mcclain2017gaussian,liao2019comparative}. Among these, the coupled-cluster with single, double, and perturbative triple excitations~(CCSD(T))~\cite{raghavachari1989fifth}, is often regarded as the `gold standard' of electronic structure theory, and has been shown to accurately model the relevant long-range interactions on material surfaces~\cite{ye2024periodic, carbone2024co, ye2023ab}. However, in the worst case, the correlation present within the system could span across hundreds of atoms, and this poses a significant challenge for CCSD(T) due to its steep computational scaling with both system and basis set size, severely limiting its applicability for realistic surface chemistry.

This work considers advances that address the challenges of simulating realistic material surfaces with these quantum many-body methods, as well as using these developments to validate the accuracy and enable key insights into a chemically diverse range of outstanding surface absorption problems. This advance in simulation capabilities leverages the growing field of quantum embedding to introduce a controllable locality approximation that achieves a practical linear scaling in computational effort for system sizes considered~\cite{sun2016quantum}. Specifically, we extend the previously proposed `systematically improvable quantum embedding' (SIE) method~\cite{nusspickel2022systematic,nusspickel2023effective}, which itself builds on density matrix embedding theory and fragmentation methods of quantum chemistry~\cite{knizia2012density,wouters2016practical, ye2024periodic, nagy2019approaching, Zheng2017, Cui2022}. Our developments extend the approach to couple together layers of different resolutions of correlated effects at different length scales, up to the `gold standard' CCSD(T) level. We demonstrate exceptional accuracy with linear scaling, crucial for performing these simulations at the required scale. We also adapt the approach to utilize graphics processing units (GPUs) to eliminate key computational bottlenecks in our workflow, including the implementation of GPU-enhanced correlated solvers. With these, we can converge unprecedented simulations at the CCSD(T) level over solid-state systems of tens of thousands of orbitals, as summarized by the workflow of Fig.~\ref{fig:workflow}\textbf{a} and the observed computational scaling of Fig.~\ref{fig:workflow}\textbf{b}, with more details of these methodological advances described in Sec.~\ref{sec:Methods} and Supporting Information~(SI) Sec.~\ref{sec:SI_method}.

\section{Results}\label{sec:Results}

\subsection{Water molecule on graphene}
\label{sec:water@graphene}
\begin{figure*}[!ht]
\centering
\includegraphics[width=1.0\linewidth]{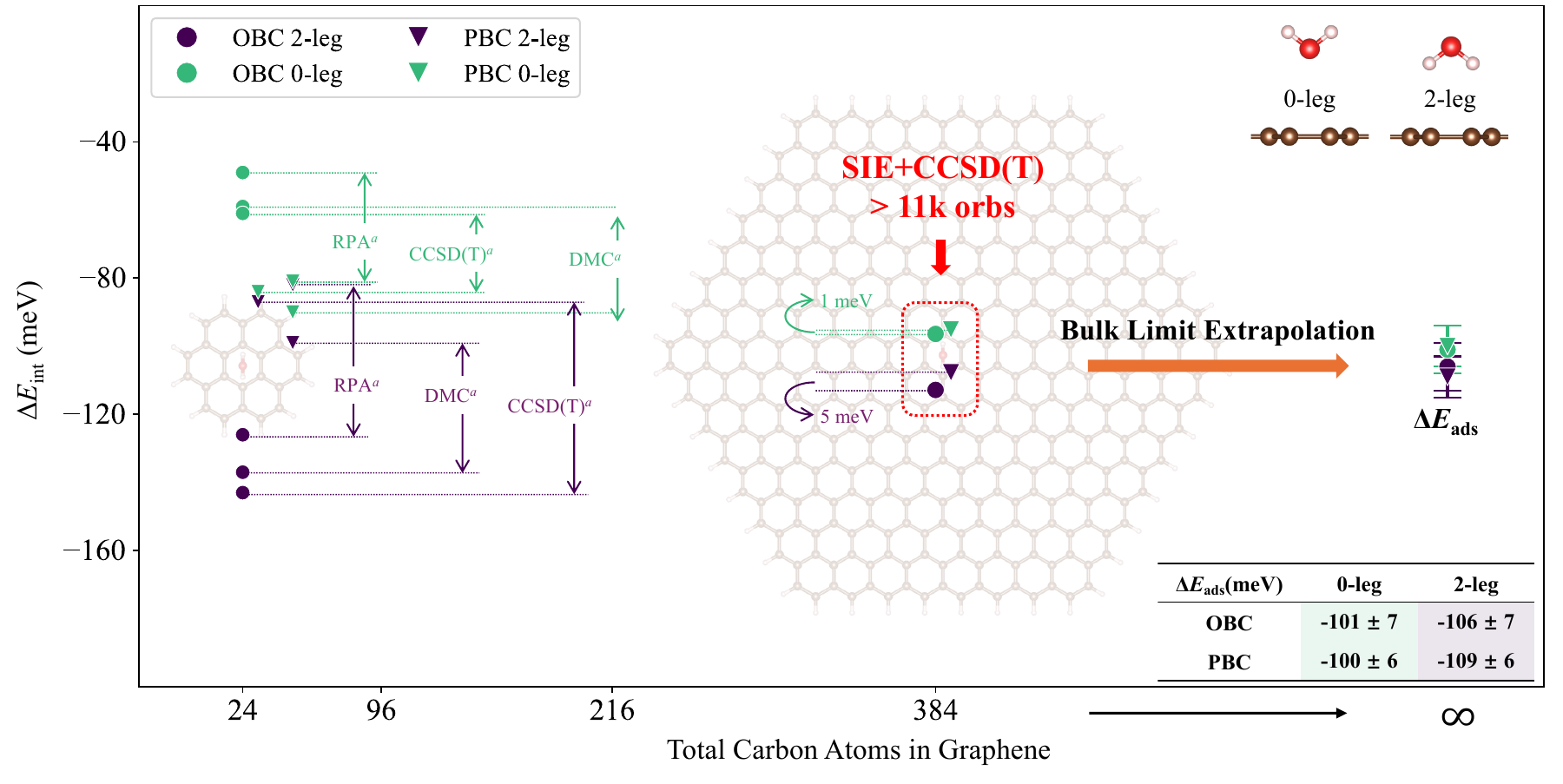}
\caption{\label{fig:WOG_data}\textbf{The adsorption energy of water on graphene at the bulk limit.} The computed interaction energies~($\Delta E_{\text{int}}$) of H$_2$O@Graphene are shown for both 0-leg (green) and 2-leg (purple) configurations (shown top right) as well as with open (circles) and periodic (triangles) boundary conditions. These are obtained for different system sizes with both the SIE+CCSD(T) workflow, and compared to other methods. The infinity symbol represents an extrapolation to the bulk limit. The final absorption energies, $\Delta E_{\text{ads}}$, shown in the inset table combine both an extrapolation of the computed interaction energies to the bulk limit at the SIE+CCSD(T) level, with a geometry relaxation correction at DFT level. The superscript \textit{a} represents results taken from Ref.~\cite{brandenburg2019physisorption}.
}
\end{figure*}

The adsorption of water on surfaces plays a central role in fields from desalination~\cite{seo2018anti} to clean energy~\cite{siria2017new, wang2020blue} and many more besides. The case of graphene represents a remarkably fundamental surface for this process, with its semi-metallic native state harboring a number of novel properties and promising applications including blue energy~\cite{wang2020blue} and quantum friction~\cite{kavokine2022fluctuation}. However, the dominating weak and long-range van der Waals interactions between water and graphene lead to a highly non-local interaction that poses significant technical challenges to achieve convergence with respect to the size of the graphene sheet~\cite{ma2011adsorption, rubes2009structure, jenness2010benchmark, voloshina2011physisorption, brandenburg2019physisorption, lau2021regional, ajala2019assessment}. A method capable of reliably computing the adsorption energies of water on graphene (H$_2$O@Graphene) therefore serves as a meaningful indicator of the feasibility of simulating realistic surface chemistry more broadly. 

This limiting finite-size error can be qualitatively estimated via the difference between adsorption energies calculated for structures of similar size under both open and periodic boundary conditions (OBC and PBC respectively), regarded as the OBC-PBC gap~\cite{brandenburg2019physisorption}. The OBC-PBC gap arises because the OBC and PBC models exhibit a different convergence of their finite-size errors, stemming from their contrasting physical origins. In OBC models, the error arises from the artificially truncated range of the interactions of the finite-sized substrate with the adsorbate, whereas under PBC conditions it is caused by the spurious periodic interactions between all the particles with their images in neighbouring cells. Our developments provide an opportunity to confidently eliminate the finite-size error via a handshake between the absorption energies computed under these different boundary conditions. 

\begin{figure*}[!ht]
\centering
\includegraphics[width=1.0\linewidth]{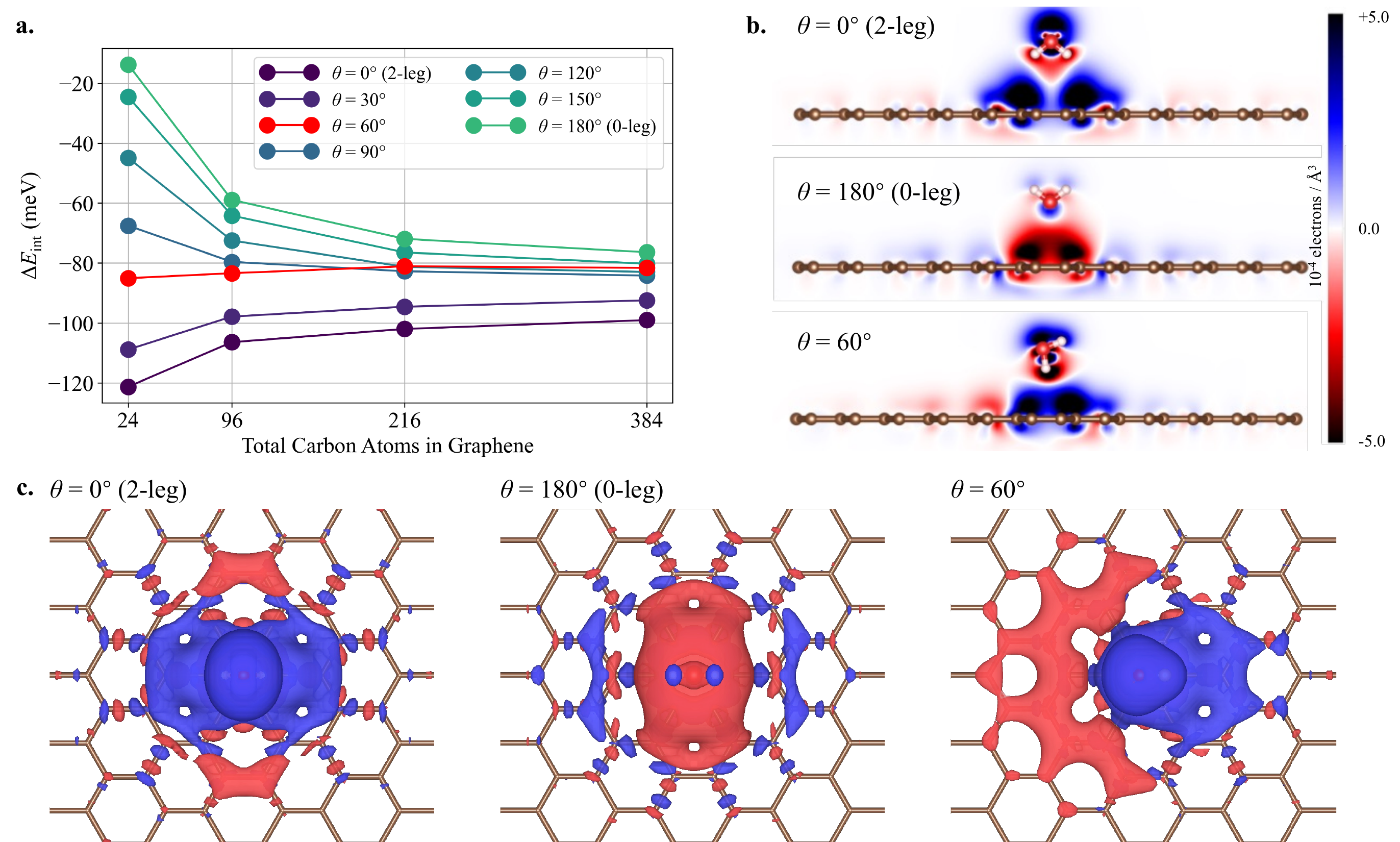}
\caption{\label{fig:water_rotate}\textbf{Adsorption of water on graphene at different orientations.} \textbf{(a)} Interaction energy of water adsorption on graphene at various orientation angles $\theta$. Calculations are performed at the SIE+CCSD level under OBC using PAH structures ranging from PAH(2) to PAH(8). \textbf{(b)} The cross-sectional views of the electron density rearrangement for PAH(6) with different water orientations. \textbf{(c)} Top views of the isosurface of electron density rearrangement, plotted with a cutoff value of $6.0 \times 10^{-5}$ electrons/Å$^3$.
}
\end{figure*}

We first consider two configurations of the water-graphene adsorption system, namely the 0-leg and the 2-leg configurations, as shown in the upper right of Fig.~\ref {fig:WOG_data}, with the substrate size systematically enlarged. For OBC models, we extend the substrate up to C$_{384}$H$_{48}$, which is a hexagonal-shaped polycyclic aromatic hydrocarbon structure with a formula C$_{6h^{2}}$H$_{6h}$~($h=8$), also referred to PAH(8), as shown in the background of Fig.~\ref{fig:WOG_data}. For PBC models, a $14 \times 14$ supercell of 392 carbon atoms is the closest in size to PAH(8). Both systems contain more than 11,000 orbitals, at which point the OBC-PBC gaps are reduced to 5 meV for the 2-leg configuration and 1 meV for the 0-leg configuration. The table inset in Fig.~\ref{fig:WOG_data} therefore presents the final predictions of adsorption energies at the CCSD(T) level, including the effects of geometry relaxation and bulk limit extrapolation. We find that the difference between the OBC and PBC models is merely 1 meV for the 0-leg configuration and 3 meV for the 2-leg configuration. These small OBC-PBC gaps suggest that our results are effectively free from finite-size errors, an important aspect that was absent in previous computational studies of correlated methods for this system, where significant OBC-PBC gaps persisted~\cite{brandenburg2019physisorption}. With bulk limit extrapolations, our calculations also demonstrate that the interaction range for 0- and 2-leg water adsorption extends over distances exceeding 18~Å, which requires 400 carbon atoms in the computational models. As shown in Fig. ~\ref{fig:water_rotate}\textbf{a}, the interaction energies converge quite slowly as a function of the graphene size, which explains the large OBC-PBC gaps observed in previous works where high-level \textit{ab-initio} calculations were employed on systems within only 50 carbon atoms~\cite{brandenburg2019physisorption}.

The 0- and 2-leg configurations are unique cases in which the dipole moment of the water molecule is perpendicular to the graphene surface. Water adsorption on graphene with varying orientations has not been extensively explored, but represents an important problem to address to understand the dynamics of water across graphene and subsequent implications in tribology applications. We can characterize the orientation of the water by an angle of rotation, $\theta$, with the 2-leg configuration defined at $\theta=0^\circ$ and the 0-leg configuration defined at $\theta=180^\circ$, with additional computational details in the SI Sec.~\ref{sec:SI_distance_opt_water_orientation}. SIE+CCSD calculations were conducted on PAH(\textit{h}) under OBC with $h=2,4,6,8$, and the results are shown in Fig.~\ref{fig:water_rotate}\textbf{a}. Again, we observe significant finite-size effects which change both the relative ordering and absolute scales of these absorption energies over the different orientations. In particular, we find that the long-range interaction depends significantly on the orientation of water and is maximized for the 0-leg and 2-leg configurations. Furthermore, long-range interactions stabilize the adsorption for $\theta > 60^\circ$, and destabilizes the adsorption for $\theta < 60^\circ$. This emphasizes the importance of being able to converge to the bulk limit for all orientations to correctly describe their relative stability. Notably, for the configuration at $\theta=60^\circ$ (red line), the interaction energy remains nearly constant as the graphene substrate size increases, indicating that finite-size errors are particularly small in this specific case.

For more insight into these observations, we analyze the electron density rearrangement~(see the definition in SI Sec.~\ref{sec:SI_EDR}) for the 2-leg, 0-leg, and $\theta=60^\circ$ configurations, with their cross-sectional side views shown in Fig.~\ref{fig:water_rotate}\textbf{b}, and isosurface top views in Fig.~\ref{fig:water_rotate}\textbf{c}, respectively. This adsorption-induced electron density rearrangement is considered as another descriptor for interaction range. For the 2-leg configuration, the dipole of the water points toward the substrate, pulling electrons from the graphene to the adsorption site and enhancing the water-graphene interaction, as shown in Fig.~\ref{fig:water_rotate} \textbf{b} and \textbf{c}. 
The electron density rearrangement around the absorbed site converges to the bulk limit faster than the free surface, thereby weakening the interaction energy as the system size increases to the converged limit for $\theta < 60^\circ$. 
For the 0-leg configuration, electrons in the substrate are pushed away from the adsorption site. Larger substrates allow for the repelled electrons to more effectively rearrange and relax, thereby lowering the energy of the absorbed system and strengthening the interaction towards bulk limits for $\theta > 60^\circ$. For the special case of $\theta=60^\circ$, a balance is reached between these two effects as electrons are pulled from one side and pushed to the other side, as shown in Fig.~\ref{fig:water_rotate}\textbf{c}. This balance largely cancels out the finite-size effects of the interaction energy. However, this does not mean that the effects of the interaction for $\theta=60^{\circ}$ are particularly short ranged. As can be seen from the electron density rearrangement in Fig.~\ref{fig:water_rotate} \textbf{b} and \textbf{c}, the extent of density changes in the $\theta=60^{\circ}$ system is not smaller than that for the 0-leg and 2-leg, and this cancellation for rapid convergence of the interaction energy with system size is likely to be only found for energetic properties. As a counter-example, we can also compute the adsorption-induced dipole moment (shown in SI Sec.~\ref{sec:SI_DP}). Unlike the interaction energy, it is found that the adsorption-induced dipole moment for the $\theta=60^{\circ}$ configuration exhibits pronounced changes as a function of substrate size. Based on this we can conclude that a full description of the interaction-driven effects between water and graphene are all long ranged.

The substantial differences in finite-size convergence for these rotations underline the importance of being able to rigorously converge the size of the graphene substrate before conclusions can be drawn. Considering that the extrapolated SIE+CCSD(T) absorption energies for the 0-leg and 2-leg configurations are already within their associated uncertainties of each other, it is expected that the adsorption energies among all configurations could be comparably close. This means that graphene does not exhibit a strong preference for any specific water orientation, which is unusual for water, as the polar and directional hydrogen-bonding nature of water molecules usually results in a pronounced preference for specific adsorption configurations on surfaces~\cite{al2017properties, gruber2018applying}. Such behavior suggests that water-graphene interaction is dominated by non-directional van der Waals forces, unlike the interaction between water and other surfaces~\cite{gruber2018applying, kubas2016surface, li2023understanding, alessio2018chemically, hodgson2009water}. This is supported by weak interaction analysis~\cite{lu2022independent, lu2012multiwfn, lu2024comprehensive} presented in SI Sec.~\ref{sec:SI_weak_interaction_analysis}. Recent studies of quantum friction phenomena ~\cite{secchi2016massive, thiemann2022water, kavokine2022fluctuation} and anomalous dielectric response~\cite{fumagalli2018anomalously, dufils2024origin} at water-graphitic interfaces indicate that the standing descriptions of water at these interfaces based on lower-level theories are insufficient. Our results further highlight the necessity to pursue a high-level theoretical description of the water-graphene interface to explain these counter-intuitive emergent phenomena.

\subsection{Carbonaceous molecules on various surfaces} \label{sec:benchmark}
\begin{figure*}[!ht]
\centering
\includegraphics[width=1.0\linewidth]{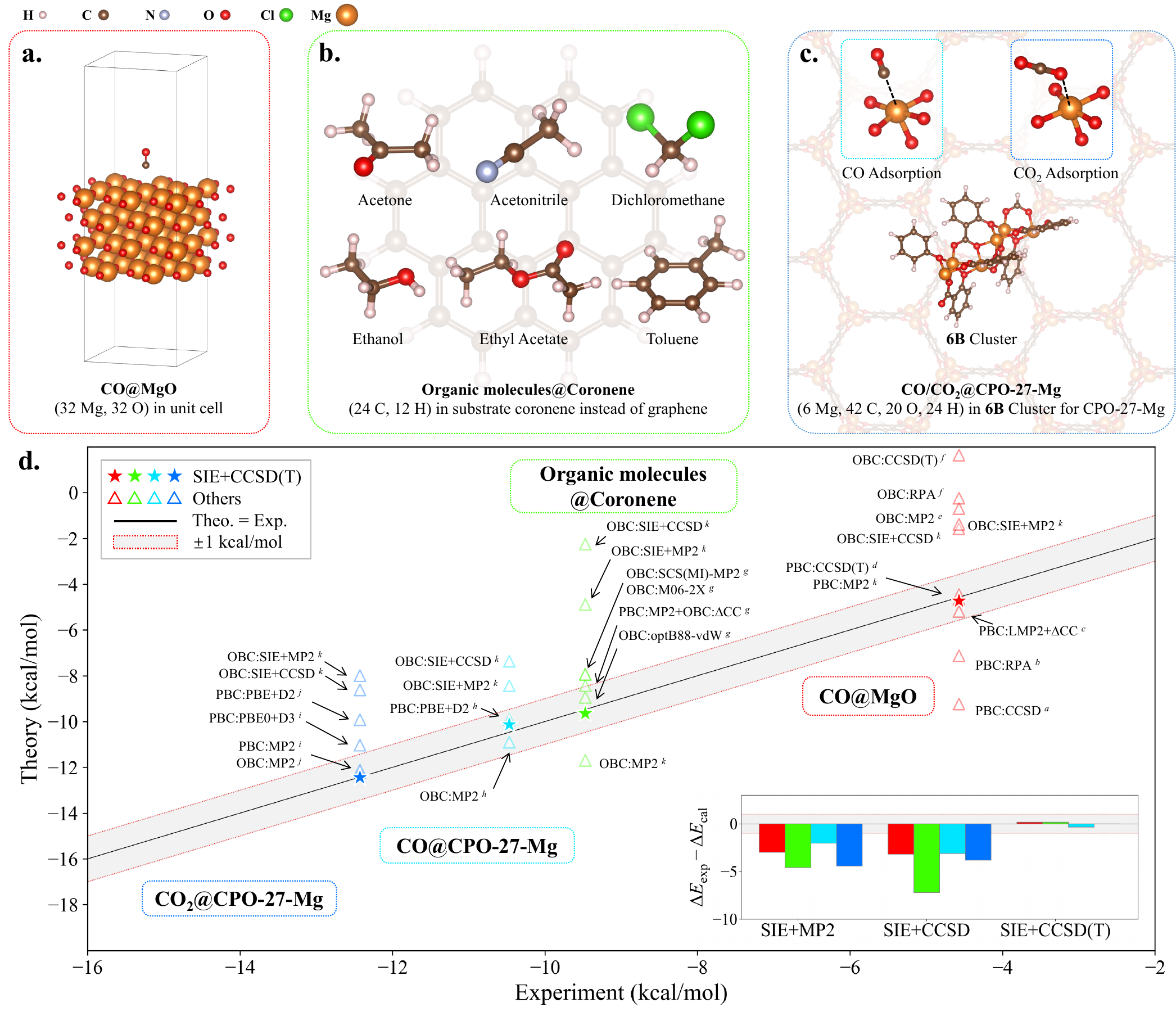}
\caption{\label{fig:benchmark} 
\textbf{Structures and adsorption energies for carbonaceous molecules on different systems.} \textbf{(a)} CO on a 4-layer MgO(001) surface with PBC. \textbf{(b)} 6 organic molecules used as the adsorbate on graphene which is replaced by Coronene under OBC. \textbf{(c)} CO/CO$_2$ adsorption configurations in CPO-27-Mg. The 6\textbf{B} cluster is cut from CPO-27-Mg, used to calculate interacting energies under OBC. Structures are rendered by using VESTA~\cite{momma2011vesta}. \textbf{(d)} Adsorption energies obtained through SIE+CCSD(T) (stars), while other methods shown with hollow triangles; An inset shows differences between SIE+MP2/CCSD/CCSD(T) and experimental data in kcal/mol. The references are labeled as superscript from (\textit{a-j}) corresponding to published data from references~\cite{mitra2022periodic,bajdich2015surface,alessio2018chemically,jacs_mgo_co,li2015lsqc,mazheika2016ni,jacs_graphene_small_organic_molecules,jacs_MOF_CO,jpcc_MOF_CO2_2023,jpcc_MOF_CO2_2017}, respectively, while \textit{k} refers to this work.}
\end{figure*}

Carbonaceous molecules are ubiquitous in nature and synthetic materials. The adsorption of these carbonaceous molecules plays a crucial role in various fields, such as pollution control, drug design and catalyst optimization~\cite{suNanocarbonsDevelopment2013, goudReviewCatalyst2020, longApplicationsCarbon2021, yanMultiscaleCO22023, salahshooriAdvancementsMolecular2024}. As a key quantity in determining surface chemistry, adsorption energies have been computationally probed for various carbonaceous molecules and surfaces to understand their behavior~\cite{jacs_graphene_small_organic_molecules,jacs_mgo_co,jacs_MOF_CH4,jacs_MOF_CO,jpcc_MOF_CO2_2017,jpcc_MOF_CO2_2023,boese2013accurate,alessio2018chemically, schimka2010accurate, kubas2016surface, hansen2010quantum, tosoni2010accurate, boese2016accurate, rehak2020including, sauer2019ab, Shi2024}. To demonstrate the wide applicability of our method, we consider three representative systems with chemically diverse surfaces selected: carbon monoxide on the MgO(001) surface, denoted as `CO@MgO', six organic molecules on coronene, a simplified model to mimic graphene, denoted as `Organic molecules@Coronene'~\cite{jacs_graphene_small_organic_molecules}, and carbon monoxide~\cite{jacs_MOF_CO}/carbon dioxide~\cite{jpcc_MOF_CO2_2017, jpcc_MOF_CO2_2023} on the metal-organic framework CPO-27-Mg, denoted as `CO/CO$_2$@CPO-27-Mg'. Fig.~\ref{fig:benchmark}\textbf{a-c} shows the adsorption configurations and the surfaces/clusters. The adsorption energies across these systems range from $4$ to $14$ kcal/mol. This range spans from typical weak physisorption to strong chemisorption, highlighting the diversity of these distinct surface types. 

To demonstrate the robustness and versatility of the SIE+CCSD(T) protocol, we compare our results to a variety of computational approaches across the systems. Detailed descriptions are provided in the SI Sec.~\ref{sec:SI_benchmark}, and we present only the main results here. The computed values, compared to their corresponding experimental data, are shown in Fig.~\ref{fig:benchmark}\textbf{d}, where the black line represents perfect agreement between theory and experiment, and the gray shade indicates deviation within a `chemical accuracy' of $\pm1$ \mbox{kcal/mol}. We note that the accuracy of the same method, for example MP2, can vary among different adsorption systems. Furthermore, even in the same system (such as CO@MgO), different implementations can result in significant differences, due to the convergence of technical parameters. Notably, the SIE+CCSD(T) method achieves a sub-chemical accuracy in agreement with experimental data across all studied systems. 

The inset in Fig.~\ref{fig:benchmark}\textbf{d} further illustrates the deviations of SIE+MP2, SIE+CCSD, and SIE+CCSD(T) relative to experimental values, highlighting the significant improvement in accuracy of SIE+CCSD(T) over SIE+MP2 and SIE+CCSD. Another important reason is that SIE+CCSD(T) maintains the size consistency inherent in Coupled Cluster theory, which is crucial for achieving consistent high accuracy across systems of different sizes. A contrasting example is MP2, which is widely known as lacking size consistency. Although it performs very well in systems involving the adsorption of small molecules, such as CO/CO$_2$@CPO-27-Mg and CO@MgO, in the case of Organic molecules@Coronene, where all six types of organic molecules are larger than the CO/CO$_2$ examples, MP2 performs poorly. As described in SI Sec.~\ref{sec:SI_OM@Coronene}, as the size of organic molecules increases, the differences between MP2 and experimental values significantly widen, whereas the differences between SIE+CCSD(T) and experimental values do not show as marked a change. This consistent high-accuracy performance bolsters our confidence that SIE+CCSD(T) possesses good versatility and transferability to be applied across complex surfaces and with a wide variety of adsorbates.

\section{Conclusions and outlook}\label{sec:Discussion}

High-accuracy simulations of realistic material surfaces using quantum many-body methods inevitably encounter two main challenges: (1) accurately capturing the essential electron correlations within materials, and (2) favorable computational scaling to reach the bulk limit, which can involve hundreds of atoms and tens of thousands of basis functions or more to account for long-range interactions. In this work, we demonstrate that our highly efficient GPU-accelerated SIE+CCSD(T) framework effectively addresses these challenges for typical adsorption systems. For water on graphene — a famously challenging system due to its pronounced long-range interactions — we scaled our calculations to converge to CCSD(T)-level accuracy through a systematic computational approach. This allowed us to eliminate finite-size errors which stymied previous studies. At the bulk limit, the adsorption energy converges to approximately 100 meV for both the 0-leg and 2-leg configurations, both lying inside the uncertainty of the calculations. The interaction energies for different orientations of water are also within this range, suggesting for the first time that graphene is insensitive to the orientation of water monomer adsorbed on its surface, contrary to previous findings. For structurally more complex systems of carbonaceous molecules on chemically diverse surfaces, our predictions for adsorption energies are within chemical accuracy of experimental values, demonstrating a good versatility and accuracy of SIE+CCSD(T), paving the way for use in tackling a broad array of material surface problems.

This large scale of quantum chemistry simulations for material surfaces has been enabled by our newly developed SIE+CCSD(T) framework, which leverages the linear scaling nature of fragmentation approaches alongside the computational power of modern GPUs. This efficient framework sets the stage for understanding the broader mechanisms of molecular surface chemistry. Integrating SIE with solvers that can assess excited states will allow for the accurate modeling of photochemistry at surfaces and non-radiative transitions in surface defects and quantum dots. Furthermore, combining this framework with reaction path searching tools will enable detailed mapping of surface reaction processes, and potentially offers predictive understanding of complex chemical reactions on surfaces.

\section{Methods}\label{sec:Methods}
\subsection{Framework of SIE+CCSD(T)}
\label{sec:framework}

Quantum embedding methods strike a balance between accuracy and computational simplicity by selecting a set of overlapping \textit{clusters} from the system for a high-level numerical method treatment~(such as FCI, CCSD(T), QMC, etc), integrated into a low-level (e.g. mean-field) representation of the surrounding \textit{environment}. These clusters are comprised of local atomic-like fragment spaces over each atomic species, along with a \textit{bath} space to span the physics of quantum fluctuations from the fragment into its environment as described by the low-level theory. However, the intrinsic mismatch between low-level and high-level solutions in reconstructing the total energy can lead to a lack of clear improvability in quantum embedding making it difficult to systematically improve. SIE provides a more reliable route towards convergence in energetic quantities by expanding the bath space with carefully designed bath natural orbitals~(BNOs) relying on a perturbative description of these fluctuations~\cite{nusspickel2022systematic}. 

The SIE framework, outlined in Fig.~\ref{fig:workflow}\textbf{a}, requires three key components: a low-level solution, precomputed using mean-field theory (such as the Hartree-Fock method employed in this work); a partitioning strategy designed to generate clusters for quantum embedding; and a BNO threshold to decide the number of BNOs at each level of theory in which the clusters are described. The SIE embedding procedure begins by forming a mean-field bath through the use of the Schmidt decomposition, as employed in density matrix embedding theory~(DMET)~\cite{knizia2012density, wouters2016practical}. The BNOs augment this bath space based on the entanglement between the impurity and its surrounding environment obtained through second-order Møller-Plesset perturbation theory~(MP2) over the full system orbitals. Only BNOs entanglement greater than the threshold are included in the cluster at each level of theory. Following this, the many-body problem over each cluster is solved at their respective level of theory, first undergoing MP2, then CCSD on top of the MP2 results, and then the final CCSD(T) calculation. We refer to this process as the SIE+MP2, SIE+CCSD, and SIE+CCSD(T) algorithms, as shown in Fig.~\ref{fig:workflow}\textbf{a}. It is important to note that although MP2 and CCSD can be used directly as solvers, CCSD(T) requires directly downfolding correlation into fragments. Therefore, CCSD(T) in SIE is actually a composite approach. A detailed description of perturbative (T) in SIE, along with the energy expression for SIE+CCSD(T), can be found in the SI Sec.~\ref{sec:SIE_(T)}. 

The correlation outside the clusters is estimated by the difference between the full sysetm MP2 and SIE+MP2~\cite{kurian2023toward}.
The full system MP2 is not the only candidate to catch outside correlation outside. Any method, which could estimates the correlation in a larger cluster space than SIE+MP2, can do the samething, such as a SIE+MP2 with a lower BNO threshold.

To compute the total energy and other observables, we use the partitioned density matrix approach~(PWF-DM) introduced in Ref.~\cite{nusspickel2023effective}. This improves the convergence of energies with respect to bath size by including the inter-cluster interactions between different clusters treated with high-level calculation. 

Additionally, we have implemented a GPU-accelerated version of this quantum embedding workflow to dramatically increase computational throughput. As shown in Fig.~\ref{fig:workflow}\textbf{b}, the observed complexity of the method has been fitted. Due to noticeable inflection points for those methods in larger systems, only the last three data points were selected for fitting. Interestingly, we observed a linear scaling for SIE+CCSD(T), indicating that with effective engineering optimizations, theoretical bottlenecks do not dominate practical computations for system sizes up to 11,423 orbitals in SIE+CCSD(T). For further discussion on the linear scaling of SIE+CCSD(T), we refer to SI Sec.~\ref{sec:SI_LS_SIE}.
Specific settings for SIE, including the selection of BNOs, partition schemes, choice of basis sets, as well as detailed error estimation and error correction, can be found in Sec.~\ref{sec:SI_H2O_Graphene}-~\ref{sec:SI_setting_summary}.

\subsection{Adsorption Energy}\label{sec:adsorption_energy}

The \textit{adsorption energy} is defined as
\begin{equation}
\Delta E_{\text{ads}} = E({\text{AB}}) - E({\text{A}}) - E({\text{B}}),
\end{equation}
where $\Delta E_{\text{ads}}$ represents the adsorption energy, $E({\text{AB}})$ is the total energy of the system A adsorbed on B, and $E({\text{A}})$ and $E({\text{B}})$ are the total energies of A and B in their isolated thermal equilibrium structures, respectively. In practice, the adsorption energy can be decomposed into two components:
\begin{equation}
\Delta E_{\text{ads}} = \Delta E_{\text{int}} + \Delta E_{\text{geom}},
\end{equation}
where $\Delta E_{\text{int}}$ denotes the energy arising from the interaction between A and B in their equilibrium geometries, and $\Delta E_{\text{geom}}$ is the energy resulting from \textit{geometry relaxation} of the absorbate relative to its isolated equilibrium structure~\cite{jacs_mgo_co}.
Generally, $\Delta E_{\text{geom}}$ is very small for surface adsorption and thus be estimated using DFT calculation as a correction to the adsorption energy. For more details, we refer to Sec.~\ref{sec:SI_geom_relax} of the SI. The \textit{interaction energy}, which constitutes the major part of $\Delta E_{\text{ads}}$, requires accurate computation. However, for the method using Gaussian type orbitals (GTO) as basis sets, the basis set superposition error~(BSSE) can arise in adsorption energy calculations~\cite{boys1970calculation}. A common approach to address the BSSE is counterpoise correction, applied to $\Delta E_{\text{int}}$ using equation
\begin{equation}
\label{eq:CP_correction}
\Delta E_{\text{int}} = E({\text{AB}}) - E({\text{A[B]}}) - E({\text{[A]B}}),
\end{equation}
where the $E({\text{A[B]}})$ refers to the energy of the AB geometry, calculated with only the basis functions of B included, while the electrons and nuclei of B are excluded, and similarly for [A] in $E({\text{[A]B}})$.

\backmatter

\bmhead{Acknowledgments}

We extend our gratitude to Dr. Hang Li and ByteDance Research for their invaluable support. We thank Dr. Hong-Zhou Ye for his insightful discussions on the project. We thank Dr. Kaido Sillar and Dr. Joachim Sauer for generously sharing the CPO-27-Mg structure. Additionally, we appreciate Dr. Weiluo Ren, Weizhong Fu and Yibo Wu for their suggestions for this article.
J.C. is supported by the National Key R\&D Program of China under Grant No. 2021YFA1400500 and the National Science Foundation of China under Grant No. 12334003. 

\bmhead{Contribution}

Z.H. and D.L. conceived the study; Z.H. implemented the main code with contributions from Z,G; X.W., J.C., D.L. and G.H.B. suggested the experiments; Z.H. performed simulations and analyzed the results; X.W., C.C, J.C., H.Q.P. and G.H.B. performed the chemical or physical analyses; Z.H., D.L., H.Q.P, J.C., X.W. and C.C. performed figure designing; Z.H. and D.L. organized the project; Z.H. C.C. H.Q.P, X.W., G.H.B., J.C. and D.L. wrote the paper.

\bibliography{ms}




\end{document}


\maketitle

\tableofcontents

\section{Methods}
\label{sec:SI_method}

\subsection{Systematically Improvable Quantum Embedding Workflow}
\label{sec:SI_SIE_framework}

The main workflow of systematically improvable quantum embedding~(SIE) follows ref.~\cite{nusspickel2022systematic}. SIE starts from the low-level processing. Hatree-Fock~(HF) method is used as the low-level solver throughout this work. With the HF result, the low-level one-body reduced density matrix~(1-RDM) and the localized atomic orbitals are obtained. The intrinsic atomic orbital~(IAO) localization method is used to localize the molecular orbitals obtained from HF. The fragment is formed by IAOs. By using low-level 1-RDM in IAO, the bath that corresponds to \textbf{x}-th fragment is constructed by Schmidt decomposition. Here, we perform the Schmidt decomposition by diagonalizing the environment part of the low-level 1-RDM that does not include the orbitals of fragment \textbf{x}.  To do this, the $\bm{\gamma}^{\text{HF}}$ under localized orbitals basis~(LO) should be reorganized by moving each row and column in the form of
\begin{equation}
    \bm{\gamma}^{\text{HF}} = \left[ 
    \begin{array}{cc}
    \bm{\gamma}^{\text{frag}} & \bm{\gamma}^{\text{inter}} \\
    \bm{\gamma}^{\text{inter}\dag} & \bm{\gamma}^{\text{env}}
    \end{array} 
    \right].
\end{equation}
The bath orbitals are obtained by diagonalizing $\bm{\gamma}^{\text{env}}$, like
\begin{equation}\label{eq:SI_diagonalization}
    \bm{\gamma}^{\text{env}} = \bm{C}^{\text{LO} \rightarrow \text{env}} \bm{\Lambda}^{\text{env}} \bm{C}^{\text{LO} \rightarrow \text{env}\dagger},
\end{equation}
where the diagonal eigenvalue matrix $\bm{\Lambda}^{\text{env}}$ represents the occupation numbers of the environment orbitals. 
For a closed-shell system, orbitals with occupation numbers between 0 and 2 that exhibit entanglement with fragment orbitals at the HF level are termed bath orbitals. Conversely, the fully occupied or unoccupied orbitals signify unentangled environment orbitals.
The matrix $\bm{C}^{\text{LO} \rightarrow \text{env}}$ represents the transformation coefficients from LO to these environment orbitals. By incorporating fragment orbitals, we derive a new set of orbitals, termed embedding orbitals~(EO), with the coefficient transformation from LO to EO represented as
\begin{equation}
    \bm{C}^{\text{LO} \rightarrow \text{EO}} = \left[ 
    \begin{array}{cc}
         \bm{I} & 0 \\
         0 & \bm{C}^{\text{LO} \rightarrow \text{env}}
    \end{array}
    \right].
\end{equation}
EO is constituted of fragment, bath, fully occupied and virtual unentangled environment orbitals. Therefore, $\bm{C}^{\text{LO} \rightarrow \text{EO}}$ is splitted into 4 parts in column dimension, likes
\begin{equation}
    \bm{C}^{\text{LO} \rightarrow \text{EO}} = \bm{C}^{\text{LO} \rightarrow \text{EO(frag)}} \oplus \bm{C}^{\text{LO} \rightarrow \text{EO(bath)}} \oplus \bm{C}^{\text{LO} \rightarrow \text{EO(occ)}} \oplus \bm{C}^{\text{LO} \rightarrow \text{EO(vir)}}, 
\end{equation}
where the 
\begin{equation}
    \bm{C}^{\text{LO} \rightarrow \text{EO(frag)}} = [\bm{I}, 0]^{T},
\end{equation}
\begin{equation}
    \bm{C}^{\text{LO} \rightarrow \text{EO(bath)}} = [0, \bm{C}^{\text{LO} \rightarrow \text{env(bath)}} ]^{T},
\end{equation}
\begin{equation}
    \bm{C}^{\text{LO} \rightarrow \text{EO(occ)}} = [0, \bm{C}^{\text{LO} \rightarrow \text{env(occ)}}]^{T},
\end{equation}
\begin{equation}
    \bm{C}^{\text{LO} \rightarrow \text{EO(vir)}} = [0, \bm{C}^{\text{LO} \rightarrow \text{env(vir)}}]^{T}.
\end{equation}

At the mean-field level, unentangled environment orbitals are indistinguishable from the impurity, as they are either fully occupied or fully empty. However, typically, the number of bath orbitals is commensurate with that of the fragment, which makes the impurity scale insufficient for capturing significant correlations that may contribute from these unentangled environment orbitals. To relieve this problem, going beyond mean-field approaches, like correlated methods, is necessary to capture the entanglement between the unentangled environment and the impurity, comprised of the fragment and bath orbitals. Within the SIE framework, this entanglement is assessed through MP2 within two separate subspaces: one formed by the occupied orbitals of the environment and the virtual orbitals of the impurity, and another formed by the virtual orbitals of the environment and the occupied orbitals of the impurity. As MP2 calculations are confined to these subspaces, and considering the number of orbitals in the impurity is small enough to be considered constant, the MP2 costs are confined to $O(N^3)$. We then construct Fock matrix $\bm{F}^{\text{imp}}$ by projecting the entire system's Fock matrix into the impurity space, following the equation 
\begin{equation}
    \bm{F}^{\text{imp}} = \bm{C}^{\text{LO} \rightarrow \text{EO(imp)}\dagger} \bm{F} \bm{C}^{\text{LO} \rightarrow \text{EO(imp)}},
\end{equation}
where the $\bm{F}$ is the full system Fock matrix in LO basis, and $\bm{C}^{\text{LO} \rightarrow \text{EO(imp)}}$ is the transformation matrix from LO to impurity, defined as
\begin{equation}
    \bm{C}^{\text{LO} \rightarrow \text{EO(imp)}} = [\bm{C}^{\text{LO} \rightarrow \text{EO(frag)}}, \bm{C}^{\text{LO} \rightarrow \text{EO(bath)}}].
\end{equation}
Then, by diagonalizing the $\bm{F}^{\text{imp}}$, the occupied and virtual orbitals in impurity could be described as a transformation matrix from EO to these orbitals, like
\begin{equation}
    \bm{F}^{\text{imp}} = \bm{C}^{\text{EO(imp)} \rightarrow \text{imp}} \bm{\varepsilon}^{\text{imp}} \bm{C}^{\text{EO(imp)} \rightarrow \text{imp}\dagger}.
\end{equation}
The $\bm{C}^{\text{EO(imp)} \rightarrow \text{imp}}$ denotes the transformation matrix. The $\bm{\varepsilon}^{\text{imp}}$ here denotes the molecular orbitals energies in impurity.
And the transformation matrix from LO to impurity molecular orbitals is obtained as 
\begin{equation}
    \bm{C}^{\text{LO} \rightarrow \text{imp}} = \bm{C}^{\text{LO} \rightarrow \text{EO(imp)}} \bm{C}^{\text{EO(imp)} \rightarrow \text{imp}}.
\end{equation}
The transformation matrix from LO to the subspace composed of unentangled environment occupied orbital~(eo) and impurity virtual orbitals~(iv) denotes $\bm{C}^{\text{LO} \rightarrow \text{sub(eo,iv)}}$ as
\begin{equation}
    \bm{C}^{\text{LO} \rightarrow \text{sub(eo,iv)}} = [\bm{C}^{\text{LO} \rightarrow \text{EO(occ)}}, \bm{C}^{\text{LO} \rightarrow \text{imp(vir)}} ],
\end{equation}
and the transformation matrix from LO to the subspace composed of impurity occupied orbital~(io) and unentangled environment virtual orbitals~(ev) denotes $\bm{C}^{\text{LO} \rightarrow \text{sub(io,ev)}}$ as
\begin{equation}
    \bm{C}^{\text{LO} \rightarrow \text{sub(io,ev)}} = [\bm{C}^{\text{LO} \rightarrow \text{imp(occ)}}, \bm{C}^{\text{LO} \rightarrow \text{EO(vir)}} ].
\end{equation}
Using these transformation matrices, the Hamiltonian can be truncated from the full system under the LO basis into the subspace, allowing for the performance of MP2 calculations within this subspace. The central double excitation $t_{ij}^{ab}$ in MP2 is denoted as
\begin{equation}
    t_{ij}^{ab} = -(ia|jb) / D_{ij}^{ab},
\end{equation}
\begin{equation}
    D_{ij}^{ab} = \epsilon_a + \epsilon_b - \epsilon_i - \epsilon_j
\end{equation}
where the $\epsilon_i$($\epsilon_j$) and $\epsilon_a$($\epsilon_b$) are the occupied orbital energy, the virtual orbital energy, respectively, and $ijkl$ ($abcd$) denote for the occupied(virtual) orbital. The 1-RDM could be obtained by
\begin{equation}
    \gamma_{ab} = 2 \sum_{ij}\sum_{c} t_{ij}^{ac}(2t_{ij}^{bc} - t_{ij}^{cb}),
\end{equation}
\begin{equation}
    \gamma_{ij} = 2\delta_{ij} - 2 \sum_{k}\sum_{ab} t_{ik}^{ab}(2t_{jk}^{ab} - t_{jk}^{ba}),
\end{equation}
where the $\gamma_{ab}$($\gamma_{ij}$) denotes the virtual(occupied) part of 1-RDM. The inter part between the virtual and occupied part in 1-RDM is 0 for MP2. Diagonizing the 1-RDM yields the bath natural orbitals~(BNOs) as eigenvectors, where the eigenvalues correspond to the population on the BNOs. The variation in population (comparing MP2 with HF) shows the level of entanglement between the BNOs and the impurity, allowing for the ranking of BNOs. Therefore, we focus only on the part of the subspace attributed to the environment. Specifically, in the subspace composed of the occupied orbital in the unentangled environment and virtual orbitals in impurity, all occupied orbitals come from the environment. Therefore, we only need to consider the $\gamma_{ij}$, similarly for the virtual orbitals from the unentangled environment. The diagonalization process is similar to the Eq.~\ref{eq:SI_diagonalization} and not repeated here.

The complete set of BNOs is derived from both the occupied and virtual orbitals within the unentangled environment.
The corresponding transformation matrix from LO to BNO could be obtained by
\begin{equation}
    \bm{C}^{\text{LO} \rightarrow \text{BNO}} = [\bm{C}^{\text{LO} \rightarrow \text{EO(occ)}}\bm{C}^{\text{EO(occ)} \rightarrow \text{BNO(occ)}}, \bm{C}^{\text{LO} \rightarrow \text{EO(vir)}}\bm{C}^{\text{EO(vir)} \rightarrow \text{BNO(vir)}}].
\end{equation}
The BNO threshold, $\eta$, is utilized to truncate the BNOs. Orbitals with population variation surpassing $\eta$ are deemed significant at the MP2 level due to their correlation with the impurity and are incorporated into the impurity to form the SIE cluster. Therefore, the transformation matrix $\bm{C}^{\text{LO} \rightarrow \text{clu}}$ from LO to the SIE cluster is defined by
\begin{equation}
    \bm{C}^{\text{LO} \rightarrow \text{clu}} = [\bm{C}^{\text{LO} \rightarrow \text{EO(imp)}}, \bm{C}^{\text{LO} \rightarrow \text{BNO}(\eta)}], 
\end{equation}
where the $\bm{C}^{\text{LO} \rightarrow \text{BNO}(\eta)}$ get the BNOs whose population variation is larger than $\eta$. After obtaining the cluster's orbitals, the cluster Hamiltonian can be truncated from the full system, allowing for direct computation using a high-level solver. Note that the addition of bath and BNO will normally introduce extra electrons, which must be considered in actual calculations. With this, the main part of the SIE workflow is complete.

The MP2-ranked BNOs are identified as a source for the systematic improvability in the SIE framework. Orbitals within the unentangled environment exhibit a significant correlation with the impurity at the MP2 level.
Therefore, incorporating more of these orbitals into the cluster allows for a better representation of the impurity-environment entanglement. This improved representation leads to more accurate final results~\cite {nusspickel2022systematic, nusspickel2023effective}.

The origin of BNOs is not strictly limited to the unentangled environment. Ref.~\cite{nusspickel2022systematic} mentioned the direct usage of the full system's molecular orbitals to construct BNOs. In this context, before BNO construction, the MP2-derived 1-RDM is truncated to the unentangled environment space to ensure there is no overlap with the impurity. The outcomes sourced from these two origins of BNOs do not substantially diverge because the environment is considerably large compared to the impurity.

\subsection{Partition wave function density matrix scheme}
\label{sec:SI_PWF_RDM}

The partition wave function density matrix~(PWF-DM) is constructed from all SIE clusters' CCSD or MP2 solutions. Here we use the restricted CCSD high-level solver as an example to get the full system RDM. To achieve RDM at the CCSD(T) level, further refinement is necessary. This can be achieved by incorporating methodologies such as the ex-situ form of perturbative (T), which will be discussed in future work. When utilizing MP2 as the high-level solver, it is important to note that MP2 has no single excitation amplitude. Thus, in calculating PWF-DM for MP2, terms related to single excitation amplitude can be simply omitted. 

In this paper, we follow the original recommendations for calculating RDMs using PWF-DM. For one-body RDM(1-RDM), we construct a global 1-RDM for the entire system.
This approach introduces contractions between different cluster solutions, effectively incorporating cluster interactions and thereby enhancing accuracy. 
Due to computational limitations, we employ an in-cluster method for calculating the two-body RDM, meaning that the calculation is performed within each SIE cluster without inter-cluster computations.
Although the in-cluster 2-RDM lacks explicit cluster interactions, a certain error cancellation mechanism inherent to this approach ensures that it outperforms a global 2-RDM. For further details, please refer to the original article~\cite {nusspickel2023effective}.

\subsubsection{Global 1-RDM}
\label{sec:SI_global_1RDM}

The calculation of the PWF-DM necessitates the solution of the CCSD for each cluster. 
Specifically, this includes $t_{i}^{a}$ and $t_{ij}^{ab}$, which are obtained from solving the CCSD equations. Additionally, it requires $\lambda_{a}^{i}$ and $\lambda_{ab}^{ij}$, which are derived from solving the CCSD $\Lambda$-equation
Here, we adhere to the convention that $ijklm$($abcde$) denotes the occupied(virtual) orbital index within the SIE cluster, $f$ denotes the fragment orbital index in the LO basis, \textbf{x} and \textbf{y} denote the indices of a different cluster. The $IJKLM$($ABCDE$) denotes the occupied(virtual) orbitals index of the full system. 
For the 1-RDM, the original paper~\cite {nusspickel2023effective} suggests a global construction. This approach ensures the inclusion of cluster-cluster interactions.
For the normal step in building the occupied part in CCSD 1-RDM~\cite{gauss1991coupled}, it would proceed as follows
\begin{equation}
    \gamma_{IJ} = 2\delta_{IJ} - P_{IJ} \left( \sum_{A} t_{I}^{A} \lambda^{J}_{A} + \sum_{KAB} \theta_{IK}^{AB} \lambda^{JK}_{AB} \right),
\end{equation}
where $\theta$ defines as
\begin{equation}
    \theta_{IJ}^{AB} = 2t_{IJ}^{AB} - t_{IJ}^{BA}.
\end{equation}
$P_{IJ}$ denotes the summation operation with indices $I$ and $J$ permutation like
\begin{equation}
    P_{IJ}\left( X \right) = X_{IJ} + X_{JI}.
\end{equation}
The $t_{I}^{A}$ and $\lambda^{J}_{A}$ are the single excitation amplitudes for full system, which are formed by using the amplitudes in each cluster, like 
\begin{equation}
\label{sec:SI_PWF_DM_t1}
    t_{I}^{A} = \sum_{\textbf{x}}\sum_{f_\textbf{x} a_\textbf{x}} t_{f_\textbf{x}}^{a_\textbf{x}} C_{f_\textbf{x} I} C_{a_\textbf{x} A},
\end{equation}
\begin{equation}
\label{sec:SI_PWF_DM_l1}
    \lambda^{I}_{A} = \sum_{\textbf{x}}\sum_{f_\textbf{x} a_\textbf{x}} \lambda^{f\textbf{x}}_{a_\textbf{x}} C_{f_\textbf{x} I} C_{a_\textbf{x} A},
\end{equation}
where the $C_{f_\textbf{x} I}$ $(C_{a_\textbf{x} A})$ is the transformation matrix from the full system occupied~(virtual) orbitals to fragment~(virtual) orbitals of the cluster \textbf{x}. Those transformation matrices could be obtained by contracting the inter basis. For example, the transformation matrix from full system occupied orbitals to the cluster \textbf{x} fragment orbitals could be obtained like
\begin{equation}
    \bm{C}^{\text{EO(frag)} \rightarrow \text{MO(occ)}} = \bm{C}^{\text{LO} \rightarrow \text{EO(frag)}\dagger}_{\textbf{x}}\bm{C}^{\text{LO} \rightarrow \text{MO(occ)}},
\end{equation}
where the $\bm{C}^{\text{LO} \rightarrow \text{EO(frag)}}_{\textbf{x}}$ is the transformation matrix from localized orbitals to the cluster \textbf{x} fragment orbitals and the $\bm{C}^{\text{LO} \rightarrow \text{MO(occ)}}$ is the transformation matrix from localized orbitals to the full system occupied orbitals. The $t_{f_\textbf{x}}^{a_\textbf{x}}$ and $\lambda^{f\textbf{x}}_{a_\textbf{x}}$ are obtained from truncating the single excitation amplitudes of SIE cluster, like
\begin{equation}
    t_{f_\textbf{x}}^{a_\textbf{x}} = \sum_{i_\textbf{x}}t_{i_\textbf{x}}^{a_\textbf{x}} C_{i_\textbf{x} f_\textbf{x}},
\end{equation}
\begin{equation}
    \lambda^{f_\textbf{x}}_{a_\textbf{x}} = \sum_{i_\textbf{x}}\lambda^{i_\textbf{x}}_{a_\textbf{x}} C_{i_\textbf{x} f_\textbf{x}},
\end{equation}
where $C_{i_\textbf{x} f_\textbf{x}}$ is the transformation matrix in cluster \textbf{x} from the occupied orbitals to the fragment orbitals. Therefore, the first part in occupied-occupied 1-RDM denotes
\begin{equation}
\label{eq:SI_1st_occ_DM}
    t_{I}^{A} \lambda^{J}_{A} = \left( \sum_{\textbf{x}}\sum_{f_\textbf{x} i_\textbf{x} a_\textbf{x}} t_{i_\textbf{x}}^{a_\textbf{x}} C_{i_\textbf{x} f_\textbf{x}} C_{f_\textbf{x} I} C_{a_\textbf{x} A} \right) \left( \sum_{\textbf{y}}\sum_{f_\textbf{y} i_\textbf{y} a_\textbf{y}} \lambda^{i\textbf{y}}_{a_\textbf{y}} C_{i_\textbf{y} f_\textbf{y}} C_{f_\textbf{y} I} C_{a_\textbf{y} A} \right).
\end{equation}

The memory complexity scales as $O(N^2)$ for single excitation amplitudes, posing no significant consumption on memory resources, which allows for the computation of complete $t_{I}^{A}$ and $\lambda^{J}_{A}$ before proceeding with the calculations of global 1-RDM. However, for double excitation amplitudes, the size escalates to $O(N^4)$ for the full system, presenting a potential storage challenge. Hence, when computing the second part of the occupied-occupied 1-RDM, $\sum_{KAB} \theta_{IK}^{AB} \lambda^{JK}_{AB}$, simplifications are necessary to avoid excessive memory consumption.

The same process in Eq.~\ref{sec:SI_PWF_DM_t1} and Eq.~\ref{sec:SI_PWF_DM_l1} could be performed on $\theta_{IJ}^{AB}$ and $\lambda^{IJ}_{AB}$, like
\begin{equation}
    \theta_{IJ}^{AB} = \sum_{\textbf{x}}\sum_{f_\textbf{x} i_\textbf{x} j_\textbf{x} a_\textbf{x} b_\textbf{x}} \theta_{i_\textbf{x} j_\textbf{x}}^{a_\textbf{x} b_\textbf{x}} C_{i_\textbf{x} f_\textbf{x}} C_{I f_\textbf{x}} C_{J j_\textbf{x}} C_{A a_\textbf{x}} C_{B b_\textbf{x}},
\end{equation}
\begin{equation}
    \lambda^{IJ}_{AB} = \sum_{\textbf{x}}\sum_{f_\textbf{x} i_\textbf{x} j_\textbf{x} a_\textbf{x} b_\textbf{x}} \lambda^{i_\textbf{x} j_\textbf{x}}_{a_\textbf{x} b_\textbf{x}} C_{i_\textbf{x} f_\textbf{x}} C_{I f_\textbf{x}} C_{J j_\textbf{x}} C_{A a_\textbf{x}} C_{B b_\textbf{x}}.
\end{equation}
Therefore, the $\sum_{KAB} \theta_{IK}^{AB} \lambda^{JK}_{AB}$ could be rewritten as
\begin{equation}
\label{eq:SI_1RDM_2p}
\begin{split}
    \sum_{KAB} \theta_{IK}^{AB} \lambda^{JK}_{AB} = &\sum_{KAB} \left( \sum_{\textbf{x}}\sum_{f_\textbf{x} i_\textbf{x} j_\textbf{x} a_\textbf{x} b_\textbf{x}} \theta_{i_\textbf{x} j_\textbf{x}}^{a_\textbf{x} b_\textbf{x}} C_{i_\textbf{x} f_\textbf{x}} C_{I f_\textbf{x}} C_{K j_\textbf{x}} C_{A a_\textbf{x}} C_{B b_\textbf{x}} \right) \\
    & \left( \sum_{\textbf{y}}\sum_{f_\textbf{y} i_\textbf{y} j_\textbf{y} a_\textbf{y} b_\textbf{y}} \lambda^{i_\textbf{y} j_\textbf{y}}_{a_\textbf{y} b_\textbf{y}} C_{i_\textbf{y} f_\textbf{y}} C_{J f_\textbf{y}} C_{K j_\textbf{y}} C_{A a_\textbf{y}} C_{B b_\textbf{y}} \right) \\
    =&\sum_{\textbf{xy}} \sum_{f_\textbf{x}f_\textbf{y}}\sum_{j_\textbf{x}j_\textbf{y}}\sum_{a_\textbf{x}a_\textbf{y}}\sum_{b_\textbf{x}b_\textbf{y}} \theta_{f_\textbf{x} j_\textbf{x}}^{a_\textbf{x} b_\textbf{x}} \lambda^{f_\textbf{y} j_\textbf{y}}_{a_\textbf{y} b_\textbf{y}} C_{j_\textbf{x} j_\textbf{y}} C_{a_\textbf{x} a_\textbf{y}} C_{b_\textbf{x} b_\textbf{y}} C_{I f_\textbf{x}} C_{J f_\textbf{y}},
\end{split}
\end{equation}
where the truncated double excitation amplitudes $\theta_{f_\textbf{x} j_\textbf{x}}^{a_\textbf{x} b_\textbf{x}}$ and $\lambda^{f_\textbf{y} j_\textbf{y}}_{a_\textbf{y} b_\textbf{y}}$ are defined as
\begin{equation}
    \label{eq:SI_truncated_t2}
    \theta_{f_\textbf{x} j_\textbf{x}}^{a_\textbf{x} b_\textbf{x}} = \sum_{i_\textbf{x}} \theta_{i_\textbf{x} j_\textbf{x}}^{a_\textbf{x} b_\textbf{x}} C_{i_\textbf{x} f_\textbf{x}},
\end{equation}
\begin{equation}
    \label{eq:SI_truncated_l2}
    \lambda^{f_\textbf{y} j_\textbf{y}}_{a_\textbf{y} b_\textbf{y}} = \sum_{i_\textbf{y}} \lambda^{i_\textbf{y} j_\textbf{y}}_{a_\textbf{y} b_\textbf{y}} C_{i_\textbf{y} f_\textbf{y}}.
\end{equation}
And the corresponding transformation matrices are formulated as
\begin{equation}
    C_{j_\textbf{x} j_\textbf{y}} = \sum_{K} C_{K j_\textbf{x}} C_{K j_\textbf{y}},
\end{equation}
\begin{equation}
    C_{a_\textbf{x} a_\textbf{y}} = \sum_{A} C_{A a_\textbf{x}} C_{A a_\textbf{y}},
\end{equation}
\begin{equation}
    C_{b_\textbf{x} b_\textbf{y}} = \sum_{B} C_{B b_\textbf{x}} C_{B b_\textbf{y}}.
\end{equation}
Note the double excitation amplitudes should have the symmetry $\theta_{IJ}^{AB} = \theta_{JI}^{BA}$ and $\lambda^{IJ}_{AB} = \lambda^{JI}_{BA}$ for full system, thus, we can rewrite the summation $\sum_{KAB} \theta_{IK}^{AB} \lambda^{JK}_{AB}$ as
\begin{equation}
\label{eq:SI_exchange_principle}
    \sum_{KAB} \theta_{IK}^{AB} \lambda^{JK}_{AB} = \frac{1}{4} \left( \sum_{KAB} \theta_{IK}^{AB} \lambda^{JK}_{AB} + 
    \sum_{KAB} \theta_{KI}^{AB} \lambda^{KJ}_{AB} +
    \sum_{KAB} \theta_{IK}^{AB} \lambda^{JK}_{BA} +
    \sum_{KAB} \theta_{KI}^{AB} \lambda^{KJ}_{BA}
    \right).
\end{equation}
Although those 4 summations are equal for the full system, 
it would be slightly different when using amplitudes constituted from the SIE cluster amplitudes, which are truncated within the fragment,
because the symmetry has been broken when truncating the occupied index, shown in Eq.~\ref{eq:SI_truncated_t2} and Eq.~\ref{eq:SI_truncated_l2}. To address this problem, the Eq.~\ref{eq:SI_1RDM_2p} will be modified based on the equation above, like
\begin{equation}
\begin{split}
    \sum_{KAB} \theta_{IK}^{AB} \lambda^{JK}_{AB} = &
    \frac{1}{4} \sum_{\textbf{xy}} \sum_{f_\textbf{x}f_\textbf{y}}\sum_{j_\textbf{x}j_\textbf{y}}\sum_{a_\textbf{x}a_\textbf{y}}\sum_{b_\textbf{x}b_\textbf{y}} \theta_{f_\textbf{x} j_\textbf{x}}^{a_\textbf{x} b_\textbf{x}} \lambda^{f_\textbf{y} j_\textbf{y}}_{a_\textbf{y} b_\textbf{y}} C_{j_\textbf{x} j_\textbf{y}} C_{a_\textbf{x} a_\textbf{y}} C_{b_\textbf{x} b_\textbf{y}} C_{I f_\textbf{x}} C_{J f_\textbf{y}} \\
    +& \frac{1}{4} \sum_{\textbf{xy}} \sum_{f_\textbf{x}f_\textbf{y}}\sum_{j_\textbf{x}j_\textbf{y}}\sum_{a_\textbf{x}a_\textbf{y}}\sum_{b_\textbf{x}b_\textbf{y}} \theta_{f_\textbf{x} j_\textbf{x}}^{a_\textbf{x} b_\textbf{x}} \lambda^{f_\textbf{y} j_\textbf{y}}_{a_\textbf{y} b_\textbf{y}} C_{f_\textbf{x} f_\textbf{y}} C_{a_\textbf{x} a_\textbf{y}} C_{b_\textbf{x} b_\textbf{y}} C_{I j_\textbf{x}} C_{J j_\textbf{y}} \\
    +& \frac{1}{4} \sum_{\textbf{xy}} \sum_{f_\textbf{x}f_\textbf{y}}\sum_{j_\textbf{x}j_\textbf{y}}\sum_{a_\textbf{x}a_\textbf{y}}\sum_{b_\textbf{x}b_\textbf{y}} \theta_{f_\textbf{x} j_\textbf{x}}^{a_\textbf{x} b_\textbf{x}} \lambda^{f_\textbf{y} j_\textbf{y}}_{a_\textbf{y} b_\textbf{y}} C_{f_\textbf{x} j_\textbf{y}} C_{a_\textbf{x} b_\textbf{y}} C_{b_\textbf{x} a_\textbf{y}} C_{I j_\textbf{x}} C_{J f_\textbf{y}} \\
    +& \frac{1}{4} \sum_{\textbf{xy}} \sum_{f_\textbf{x}f_\textbf{y}}\sum_{j_\textbf{x}j_\textbf{y}}\sum_{a_\textbf{x}a_\textbf{y}}\sum_{b_\textbf{x}b_\textbf{y}} \theta_{f_\textbf{x} j_\textbf{x}}^{a_\textbf{x} b_\textbf{x}} \lambda^{f_\textbf{y} j_\textbf{y}}_{a_\textbf{y} b_\textbf{y}} C_{j_\textbf{x} f_\textbf{y}} C_{a_\textbf{x} b_\textbf{y}} C_{b_\textbf{x} a_\textbf{y}} C_{I f_\textbf{x}} C_{J j_\textbf{y}}.
\end{split}
\end{equation}
With such rearrangement, the memory usage is reduced from $O(N^4)$ to $O(n^3)$, where $n$ represents the size of the SIE clusters, typically no larger than a few hundred orbitals, and normally the number of fragment orbitals is exceedingly small and generally constant. This eliminates the concern of memory pressure. Further simplifications in computation can be achieved by utilizing certain symmetries, such as the need for summation over cluster \textbf{x} and \textbf{y}, where \textbf{x} and \textbf{y} are exchange-symmetrical, thereby allowing calculations for only half of the summation. Additionally, since fragments are selected in real space, spatial symmetries between different fragments may be leveraged to further reduce the computation cost. Moreover, even though the computation of the RDM seems complex, the actual theoretical complexity is at most $O(n^4)$. 

Finally, we give the entire global 1-RDM equation without proving. The occupied-occupied part denotes
\begin{equation}\label{eq:SI_1RDM_1}
\begin{split}
    \gamma_{IJ} = 2\delta_{IJ} - P_{IJ} &\left( \sum_{A} t_{I}^{A} \lambda^{J}_{A} + \sum_{KAB} \theta_{IK}^{AB} \lambda^{JK}_{AB} \right)\\
    = 2\delta_{IJ} - P_{IJ} &\left( \sum_{A} \left( \sum_{\textbf{x}}\sum_{f_\textbf{x} i_\textbf{x} a_\textbf{x}} t_{i_\textbf{x}}^{a_\textbf{x}} C_{i_\textbf{x} f_\textbf{x}} C_{f_\textbf{x} I} C_{a_\textbf{x} A} \right) \left( \sum_{\textbf{y}}\sum_{f_\textbf{y} i_\textbf{y} a_\textbf{y}} \lambda^{i\textbf{y}}_{a_\textbf{y}} C_{i_\textbf{y} f_\textbf{y}} C_{f_\textbf{y} I} C_{a_\textbf{y} A} \right) \right. \\
    &+ \frac{1}{4} \sum_{\textbf{xy}} \sum_{f_\textbf{x}f_\textbf{y}}\sum_{j_\textbf{x}j_\textbf{y}}\sum_{a_\textbf{x}a_\textbf{y}}\sum_{b_\textbf{x}b_\textbf{y}} \theta_{f_\textbf{x} j_\textbf{x}}^{a_\textbf{x} b_\textbf{x}} \lambda^{f_\textbf{y} j_\textbf{y}}_{a_\textbf{y} b_\textbf{y}} C_{j_\textbf{x} j_\textbf{y}} C_{a_\textbf{x} a_\textbf{y}} C_{b_\textbf{x} b_\textbf{y}} C_{I f_\textbf{x}} C_{J f_\textbf{y}} \\
    &+ \frac{1}{4} \sum_{\textbf{xy}} \sum_{f_\textbf{x}f_\textbf{y}}\sum_{j_\textbf{x}j_\textbf{y}}\sum_{a_\textbf{x}a_\textbf{y}}\sum_{b_\textbf{x}b_\textbf{y}} \theta_{f_\textbf{x} j_\textbf{x}}^{a_\textbf{x} b_\textbf{x}} \lambda^{f_\textbf{y} j_\textbf{y}}_{a_\textbf{y} b_\textbf{y}} C_{f_\textbf{x} f_\textbf{y}} C_{a_\textbf{x} a_\textbf{y}} C_{b_\textbf{x} b_\textbf{y}} C_{I j_\textbf{x}} C_{J j_\textbf{y}} \\
    &+ \frac{1}{4} \sum_{\textbf{xy}} \sum_{f_\textbf{x}f_\textbf{y}}\sum_{j_\textbf{x}j_\textbf{y}}\sum_{a_\textbf{x}a_\textbf{y}}\sum_{b_\textbf{x}b_\textbf{y}} \theta_{f_\textbf{x} j_\textbf{x}}^{a_\textbf{x} b_\textbf{x}} \lambda^{f_\textbf{y} j_\textbf{y}}_{a_\textbf{y} b_\textbf{y}} C_{f_\textbf{x} j_\textbf{y}} C_{a_\textbf{x} b_\textbf{y}} C_{b_\textbf{x} a_\textbf{y}} C_{I j_\textbf{x}} C_{J f_\textbf{y}} \\
    & \left. + \frac{1}{4} \sum_{\textbf{xy}} \sum_{f_\textbf{x}f_\textbf{y}}\sum_{j_\textbf{x}j_\textbf{y}}\sum_{a_\textbf{x}a_\textbf{y}}\sum_{b_\textbf{x}b_\textbf{y}} \theta_{f_\textbf{x} j_\textbf{x}}^{a_\textbf{x} b_\textbf{x}} \lambda^{f_\textbf{y} j_\textbf{y}}_{a_\textbf{y} b_\textbf{y}} C_{j_\textbf{x} f_\textbf{y}} C_{a_\textbf{x} b_\textbf{y}} C_{b_\textbf{x} a_\textbf{y}} C_{I f_\textbf{x}} C_{J j_\textbf{y}} \right),
\end{split}
\end{equation}
and the virtual-virtual part denotes
\begin{equation}\label{eq:SI_1RDM_2}
\begin{split}
    \gamma_{AB} = P_{AB} &\left( \sum_{I} t_{I}^{A} \lambda^{I}_{B} + \sum_{IJC} \theta_{IJ}^{AC} \lambda^{IJ}_{BC} \right)\\
    = P_{AB} &\left( \sum_{I} \left( \sum_{\textbf{x}}\sum_{f_\textbf{x} i_\textbf{x} a_\textbf{x}} t_{i_\textbf{x}}^{a_\textbf{x}} C_{i_\textbf{x} f_\textbf{x}} C_{f_\textbf{x} I} C_{a_\textbf{x} A} \right) \left( \sum_{\textbf{y}}\sum_{f_\textbf{y} i_\textbf{y} a_\textbf{y}} \lambda^{i\textbf{y}}_{a_\textbf{y}} C_{i_\textbf{y} f_\textbf{y}} C_{f_\textbf{y} I} C_{a_\textbf{y} A} \right) \right. \\
    &+ \frac{1}{4} \sum_{\textbf{xy}} \sum_{f_\textbf{x}f_\textbf{y}}\sum_{j_\textbf{x}j_\textbf{y}}\sum_{a_\textbf{x}a_\textbf{y}}\sum_{b_\textbf{x}b_\textbf{y}} \theta_{f_\textbf{x} j_\textbf{x}}^{a_\textbf{x} b_\textbf{x}} \lambda^{f_\textbf{y} j_\textbf{y}}_{a_\textbf{y} b_\textbf{y}} C_{f_\textbf{x} f_\textbf{y}} C_{j_\textbf{x} j_\textbf{y}} C_{b_\textbf{x} b_\textbf{y}} C_{A a_\textbf{x}} C_{B a_\textbf{y}} \\
    &+ \frac{1}{4} \sum_{\textbf{xy}} \sum_{f_\textbf{x}f_\textbf{y}}\sum_{j_\textbf{x}j_\textbf{y}}\sum_{a_\textbf{x}a_\textbf{y}}\sum_{b_\textbf{x}b_\textbf{y}} \theta_{f_\textbf{x} j_\textbf{x}}^{a_\textbf{x} b_\textbf{x}} \lambda^{f_\textbf{y} j_\textbf{y}}_{a_\textbf{y} b_\textbf{y}} C_{f_\textbf{x} f_\textbf{y}} C_{j_\textbf{x} j_\textbf{y}} C_{a_\textbf{x} a_\textbf{y}} C_{A b_\textbf{x}} C_{B b_\textbf{y}} \\
    &+ \frac{1}{4} \sum_{\textbf{xy}} \sum_{f_\textbf{x}f_\textbf{y}}\sum_{j_\textbf{x}j_\textbf{y}}\sum_{a_\textbf{x}a_\textbf{y}}\sum_{b_\textbf{x}b_\textbf{y}} \theta_{f_\textbf{x} j_\textbf{x}}^{a_\textbf{x} b_\textbf{x}} \lambda^{f_\textbf{y} j_\textbf{y}}_{a_\textbf{y} b_\textbf{y}} C_{f_\textbf{x} j_\textbf{y}} C_{j_\textbf{x} f_\textbf{y}} C_{b_\textbf{x} a_\textbf{y}} C_{A a_\textbf{x}} C_{B b_\textbf{y}} \\
    &\left. + \frac{1}{4} \sum_{\textbf{xy}} \sum_{f_\textbf{x}f_\textbf{y}}\sum_{j_\textbf{x}j_\textbf{y}}\sum_{a_\textbf{x}a_\textbf{y}}\sum_{b_\textbf{x}b_\textbf{y}} \theta_{f_\textbf{x} j_\textbf{x}}^{a_\textbf{x} b_\textbf{x}} \lambda^{f_\textbf{y} j_\textbf{y}}_{a_\textbf{y} b_\textbf{y}} C_{f_\textbf{x} j_\textbf{y}} C_{j_\textbf{x} f_\textbf{y}} C_{a_\textbf{x} b_\textbf{y}} C_{A b_\textbf{x}} C_{B a_\textbf{y}} \right) ,\\
\end{split}
\end{equation}
and the occupied-virtual part denotes
\begin{equation}\label{eq:SI_1RDM_3}
\begin{split}
    \gamma_{IA} = & t_{I}^{A} + \lambda^{I}_{A} - \sum_{CK} t_{I}^{C} \lambda^{K}_{C} t_{K}^{A} - \sum_{J} d_{IJ} t_{J}^{A} - \sum_{B} d_{AB} t_{I}^{B} + \sum_{KC} \theta_{IK}^{AC} \lambda^{K}_{C} \\
    = & t_{I}^{A} + \lambda^{I}_{A} - \sum_{CK} t_{I}^{C} \lambda^{K}_{C} t_{K}^{A} - \sum_{J} d_{IJ} t_{J}^{A} - \sum_{B} d_{AB} t_{I}^{B} \\
    + & \frac{1}{2} \sum_{KC} \sum_{\textbf{x}} \left( \sum_{k_\textbf{x} c_\textbf{x}} \theta_{f_\textbf{x} k_\textbf{x}}^{a_\textbf{x} c_\textbf{x}} \lambda_{C}^{K} C_{K k_\textbf{x}} C_{C c_\textbf{x}} C_{I f_\textbf{x}} C_{A a_\textbf{x}} + \sum_{k_\textbf{x} c_\textbf{x}} \theta_{f_\textbf{x} k_\textbf{x}}^{a_\textbf{x} c_\textbf{x}} \lambda_{C}^{K} C_{K f_\textbf{x}} C_{C a_\textbf{x}} C_{I k_\textbf{x}} C_{A c_\textbf{x}} \right),
\end{split}
\end{equation}
where the $t_{I}^{A}$ and $\lambda_{I}^{A}$ are defined in Eq.~\ref{sec:SI_PWF_DM_t1} and Eq.~\ref{sec:SI_PWF_DM_l1}, which should be obtained first before building the global 1-RDM, $d_{IJ} = \sum_{KAB} \theta_{IK}^{AB} \lambda^{JK}_{AB}$ and $d_{AB} = \sum_{IJC} \theta_{IJ}^{AC} \lambda^{IJ}_{BC}$ are the intermediate variables when building the occupied-occupied part and virtual-virtual part in 1-RDM. By following a similar principle shown in Eq.~\ref{eq:SI_exchange_principle}, $\sum_{KC} \theta_{IK}^{AC} \lambda^{K}_{C}$ could also be split into $\left( \sum_{KC} \theta_{IK}^{AC} \lambda^{K}_{C} + \sum_{KC} \theta_{KI}^{CA} \lambda^{K}_{C} \right)/2$ then the occupied-virtual part could be rewritten as the last line in the above equation.

\subsubsection{In-cluster 2-RDM}
\label{sec:SI_in_cluster_2RDM}
The key point for building the in-cluster 2-RDM is to use the fragment contracted $\lambda^{i'_\textbf{x}}_{a_\textbf{x}}$ and $\lambda^{i'_\textbf{x} j_\textbf{x}}_{a_\textbf{x} b_\textbf{x}}$ instead of using the standard amplitudes, $\lambda^{i_\textbf{x}}_{a_\textbf{x}}$ and $\lambda^{i_\textbf{x} j_\textbf{x}}_{a_\textbf{x} b_\textbf{x}}$, to build the 2-RDM within the cluster \textbf{x}, where the $\lambda^{i'_\textbf{x}}_{a_\textbf{x}}$ and $\lambda^{i'_\textbf{x} j_\textbf{x}}_{a_\textbf{x} b_\textbf{x}}$ denote
\begin{equation}
    \lambda^{i'_\textbf{x}}_{a_\textbf{x}} = \sum_{i_\textbf{x}}  \lambda^{i_\textbf{x}}_{a_\textbf{x}} C^{f_\textbf{x}}_{i_\textbf{x} i'_\textbf{x}},
\end{equation}
\begin{equation}
    \lambda^{i'_\textbf{x} j_\textbf{x}}_{a_\textbf{x} b_\textbf{x}} = \sum_{ix}  \lambda^{i_\textbf{x} j_\textbf{x}}_{a_\textbf{x} b_\textbf{x}} C^{f_\textbf{x}}_{i_\textbf{x} i'_\textbf{x}},
\end{equation}
where the $C^{f_\textbf{x}}_{i_\textbf{x} i'_\textbf{x}}$ is constructed by contracting the fragment orbitals with the same transformation matrix as 
\begin{equation}
    C^{f_\textbf{x}}_{i_\textbf{x} i'_\textbf{x}} = \sum_{f_\textbf{x}} C_{i_\textbf{x} f_\textbf{x}} C_{i'_\textbf{x} f_\textbf{x}},
\end{equation}
where the $C_{i_\textbf{x} f_\textbf{x}}$ and $C_{i'_\textbf{x} f_\textbf{x}}$ are the same transformation matrix projecting occupied orbitals of the \textbf {x} cluster to its fragment orbitals. However, the projected $\lambda^{i'_\textbf{x} j_\textbf{x}}_{a_\textbf{x} b_\textbf{x}}$ would lose the symmetry which means $\lambda^{i'_\textbf{x} j_\textbf{x}}_{a_\textbf{x} b_\textbf{x}} \neq \lambda^{j_\textbf{x} i'_\textbf{x}}_{b_\textbf{x} a_\textbf{x}}$. To keep this symmetry, one approach is
\begin{equation}
    \tilde{\lambda}^{i_\textbf{x} j_\textbf{x}}_{a_\textbf{x} b_\textbf{x}} = \frac{1}{2} \left( \lambda^{i'_\textbf{x} j_\textbf{x}}_{a_\textbf{x} b_\textbf{x}} + \lambda^{j_\textbf{x} i'_\textbf{x}}_{b_\textbf{x} a_\textbf{x}} \right),
\end{equation}
and use $\tilde{\lambda}^{i_\textbf{x} j_\textbf{x}}_{a_\textbf{x} b_\textbf{x}}$ instead.
The subsequent processes for constructing the 2-RDM do not differ from the normal CCSD processing and will not be further elaborated here. It is highly recommended to investigate the CCSD 2-RDM construction for more details see the article~\cite{gauss1991coupled} and the PySCF code~\cite{sun2020recent, sun2018pyscf}.

\subsubsection{Energy Calculation in PWF-DM}
Theoretically, once the 1-RDM and 2-RDM are obtained, all two-body or single-body observables can be computed. Since the focus of the paper is on the energy calculation, 
here we demonstrate how to use the PWF-DM 1-RDM and 2-RDM for energy computation.
The Hamiltonian under Born-Oppenheimer approximation of the entire system can be represented as
\begin{equation}
    \hat{H}_e = E_{\text{nuc}} + \sum_{PQ} d_{PQ} \hat{a}^\dagger_P \hat{a}_Q + \frac{1}{2} \sum_{PQRS} (PQ|RS) \hat{a}^\dagger_P \hat{a}^\dagger_Q \hat{a}_S \hat{a}_R
\end{equation}
where the $PQRS$ denote the full molecular orbitals, $E_{\text{nuc}}$ is the nuclear repulsion energy and $ d_{PQ}$ represents single electron integration coefficients in the molecular basis. Therefore, the total energy for PWF-DM could be obtained as
\begin{equation}
    E = E_{\text{nuc}} + \sum_{PQ} d_{PQ} \gamma_{PQ} + \frac{1}{2} \sum_{\textbf{x}}\sum_{p_\textbf{x}q_\textbf{x}r_\textbf{x}s_\textbf{x}}(p_\textbf{x}q_\textbf{x}|r_\textbf{x}s_\textbf{x}) \Gamma_{p_\textbf{x}q_\textbf{x}r_\textbf{x}s_\textbf{x}}, 
\end{equation}
where the $\gamma_{PQ}$ denotes the global 1-RDM comes from section~\ref{sec:SI_global_1RDM}, and the $\Gamma_{p_\textbf{x}q_\textbf{x}r_\textbf{x}s_\textbf{x}}$ denotes the in-cluster 2-RDM for \textbf{x} SIE cluster coming from section~\ref{sec:SI_in_cluster_2RDM}, the indices $p_\textbf{x}q_\textbf{x}r_\textbf{x}s_\textbf{x}$ denote the molecular orbitals indices within cluster \textbf{x} and the $(p_\textbf{x}q_\textbf{x}|r_\textbf{x}s_\textbf{x})$ denotes the double electron integration for cluster \textbf{x}, which is projected from full system double electron integration following the equation
\begin{equation}
    (p_\textbf{x}q_\textbf{x}|r_\textbf{x}s_\textbf{x}) = \sum_{PQRS} (PQ|RS) C_{P p_\textbf{x}} C_{Q q_\textbf{x}} C_{R r_\textbf{x}} C_{S s_\textbf{x}}.
\end{equation}

\subsection{The In-situ Form of Perturbative (T) in SIE}\label{sec:SIE_(T)}
As illustrated in Fig.~\ref{fig:workflow}\textbf{a}, the SIE+CCSD(T) includes the SIE+MP2/CCSD, perturbative (T), the final treatment of results through the partition wavefunction density matrix (PWF-DM) sections, and the downfolding error correction. Both the SIE embedding part and PWF-DM have been discussed in detail in prior papers~\cite{nusspickel2022systematic,nusspickel2023effective}, with comprehensive formula expressions provided in SI sections~\ref {sec:SI_SIE_framework} and ~\ref {sec:SI_PWF_RDM}. The downfolding error correction is detailed in SI section~\ref {sec:SI_BTEC}; thus, these topics will not be reiterated here. In this section, we will focus our discussion on the in-cluster form perturbative (T) within SIE.

Our perturbative (T) calculations are performed within each SIE cluster without any amplitude contraction across clusters. Specifically, the closed-shell (T) correlation energy $E^{(T)}_{\textbf{x}}$ for the \textbf{x}-th SIE cluster can be described by the following formula:
\begin{equation}
    \label{eq:E_corr_CCSD_T}
    E^{(T)}_{\textbf{x}} = 2 \sum_{(ia \geq jb \geq kc )\in \textbf{x}} \sum_{i' \in \textbf{x}} A R\left[ Z^{abc}_{ijk} \right]  \tilde{W} ^{abc}_{i'jk}C^f_{ii'},
\end{equation}
where $ijkl$ and $i'$ denote the occupied orbital indexes within cluster \textbf{x}, while $abcd$ represents the virtual orbitals. The term $A$ is a conditional constant, expressed as:
\begin{equation}
A =
\begin{cases} 
    \frac{1}{6} & \text{if } a = b \text{ and } b = c \\
    \frac{1}{2} & \text{if } a = b \text{ or } b = c  \\
    1 & \text{else} \\
\end{cases}.
\end{equation}
The $Z^{abc}_{ijk}$ denotes
\begin{equation}
    Z^{abc}_{ijk} =  W^{abc}_{ijk} + V^{abc}_{ijk} ,
\end{equation}
where $W^{abc}_{ijk}$, $V^{abc}_{ijk}$ and $D^{abc}_{ijk}$ are defined as 
\begin{equation}
    \label{eq:W_in_CCSD_T}
    W^{abc}_{ijk} = \mathcal{P}^{abc}_{ijk} \left[ \sum_l (ia,bl) t^{cl}_{kj} - \sum_d (ia,jd) t^{bc}_{dk} \right] / D^{abc}_{ijk}, 
\end{equation}
\begin{equation}
    \label{eq:V_in_CCSD_T}
    V^{abc}_{ijk} = \frac{1}{2} \mathcal{P}^{abc}_{ijk} \left[ (ia,jb) t^c_k \right] / D^{abc}_{ijk},
\end{equation}
$D^{abc}_{ijk}$ and $\tilde{W} ^{abc}_{ijk}$ denotes
\begin{equation}
    D^{abc}_{ijk} = \epsilon_i + \epsilon_j + \epsilon_k - \epsilon_a - \epsilon_b - \epsilon_c, 
\end{equation}
\begin{equation}
    \tilde{W} ^{abc}_{ijk} = W^{abc}_{ijk} D^{abc}_{ijk}
\end{equation}

where $t_{ij}^{ab}$ and $t_{i}^{a}$ are the single and double excitation amplitude from CCSD, and $(pq,rs)$ is the electron repulsion integral where $pqrs$ denotes occupied or virtual orbital indexes here, and $\epsilon_i$ denotes the occupied orbital energy and $\epsilon_a$ denotes the virtual orbital energy. $\mathcal{P}^{abc}_{ijk}$ is the permutation operation, 
\begin{equation}
    \mathcal{P}^{abc}_{ijk}[X] = X^{abc}_{ijk} + X^{bca}_{jki} + X^{cab}_{kij}
    + X^{acb}_{ikj} + X^{cba}_{kji} + X^{bac}_{jik}.
\end{equation}
And $R$ in eq.~\ref{eq:E_corr_CCSD_T} is the operation defined as
\begin{equation}
    \label{eq:R_opt_in_CCSD(T)}
    R\left[ X \right]^{abc}_{ijk} = 4X^{abc}_{ijk} + X^{abc}_{klj} + X^{abc}_{jki}
    - 2X^{abc}_{kji} - 2X^{abc}_{ikj} - 2X^{abc}_{jik}.
\end{equation}
The $C^f_{ii'}$ is the occupied-occupied orbital coefficient truncated by fragment orbitals in \textbf{x} cluster, which reads
\begin{equation}
    C^f_{ii'} = \sum_{f \in \textbf{x}} C_{if}C_{i'f}, 
\end{equation}
where $f$ denotes the fragment orbitals index~(describe in SI section~\ref{sec:SI_SIE_framework} for detail) in the SIE cluster \textbf{x}. The total correlation energy contributed by perturbative (T) correction could be directly sum over all (T) correlation energy in each SIE cluster,
\begin{equation}
    \label{eq:C_f}
    E^{(T)} = \sum_{\textbf{x}} E^{(T)}_{\textbf{x}}.
\end{equation}

Even though the formula involves a sixth-order tensor like $W^{abc}_{ijk}$ and $V^{abc}_{ijk}$, thanks to the divisibility of the contraction in this forms, the actual programming implementation maintains an efficient memory usage, which does not exceed the memory and storage requirements of CCSD.

\subsection{Energy for SIE+CCSD(T)}\label{sec:energy_formula}

The energy formula of SIE+CCSD(T) in this work can be summarized as
\begin{equation}
    E_{\text{SIE+CCSD(T)}} = E^{\text{HF}} + E^{\text{corr}}_{\text{SIE+CCSD(T)}},
\end{equation}
\begin{equation}
    E^{\text{corr}}_{\text{SIE+CCSD(T)}} =  E_{\text{SIE+CCSD}}^{\text{PWF-DM}} + E_{\text{(T)}} + E^{\text{BTEC}},
\end{equation}
\begin{equation}
    E^{\text{BTEC}} =  E_{\text{MP2}} - E_{\text{SIE+MP2}}^{\text{PWF-DM}},
\end{equation}
where $E_{\text{HF}}$ and is the full system HF energy, and $E^{\text{corr}}_{\text{SIE+CCSD(T)}}$ is the correlation energy from SIE+CCSD(T), $E_{\text{SIE+CCSD}}^{\text{PWF-DM}}$ is the correlation energy from SIE+CCSD with PWF-DM form described in SI section~\ref{sec:SI_PWF_RDM}, $E_{\text{(T)}}$ is the perturbative (T) correlation energy coming from Eq.~\ref{eq:E_corr_CCSD_T}. $E^{\text{BTEC}}$ denotes the bath truncation error correction by adding a difference between full system MP2 energy $E_{\text{MP2}}$ and SIE+MP2 energy $E_{\text{SIE+MP2}}^{\text{PWF-DM}}$ back to total correlation energy, where $E_{\text{SIE+MP2}}^{\text{PWF-DM}}$ is obtained using the same setting with SIE+CCSD but using MP2 as the high-level solver. See more details about this correction in SI section~\ref{sec:SI_BTEC}.

\subsection{Complete Basis Set Extrapolation\label{sec:SI_CBS_extrapolation}}
We use Dunning’s correlation-consistent series of Gaussian basis sets throughout this work. Therefore, we follow the two-point extrapolation scheme proposed in ref.~\cite{halkier1999basis}. The extrapolated HF energy is written as
\begin{equation}
    E^{\rm{HF}}_{\infty} = E^{\rm{HF}}_{n} - \frac{E^{\rm{HF}}_{n}-E^{\rm{HF}}_{n+1}}{1-e^{-B}} ,
\end{equation}
where $B$ is a constant number 1.637 and $n$ is the $\zeta$ cardinality for basis set. The extrapolated correlation energy follow the formula~\cite{halkier1998basis2}
\begin{equation}
    E^{\rm{corr}}_{\infty} = \frac{n^3 E^{\rm{corr}}_n - m^3 E^{\rm{corr}}_m}{n^3 - m^3} ,
\end{equation}
where $n$ and $m$ is the $\zeta$ cardinality for basis set.

\subsection{The choice of BNO Threshold}
A smaller BNO threshold generally leads to more accurate outcomes at the expense of increased memory usage. When using the canonical CCSD(T)as a solver, the cluster is limited to 800 orbitals in size due to the out-of-memory (OOM) issue. Therefore, in this work, the guiding principle to set the BNO threshold is as small as possible to make the largest clusters not exceed 800 orbitals.
$10^{-8}$ is found to be a suitable number.
Note in most adsorption cases, with the same BNO threshold, the cluster size constructed from the fragment in the substrate is usually smaller than that in adsorbate. Therefore, in practice, different thresholds are often set for fragments in the substrate and adsorbate to prevent the OOM issue in CCSD(T) calculations.

\subsection{Linear scaling for SIE+CCSD(T)}
\label{sec:SI_LS_SIE}

According to the main workflow of SIE+CCSD(T), the complexity of SIE+CCSD(T) is influenced mainly by three aspects. First, the full system MP2 is used to estimate correlations outside the clusters, and thus, it is typical to perform a canonical MP2 calculation on the full system, which has a computational complexity of $O(N^5)$ where $N$ represents the full system size. However, if SIE+MP2 with the smaller BNO threshold or other local MP2 methods are employed to estimate correlations outside the clusters, the complexity of this component will be reduced. The second aspect is the BNO building, which is the theoretical complexity in the original SIE, marked by $O(N^3)$ for a cluster~\cite{nusspickel2022systematic}, the BNO building cost for all $m$ clusters takes $O(mN^3)$. Clearly, the BNO building cost is not comparable to the full system MP2 cost, thus can be omitted in the total scaling of SIE+CCSD(T). The final part comes from the time consumption of CCSD(T); performing a CCSD(T) calculation within a SIE cluster of size $n$ is characterized by a complexity of $O(n^7)$. If there are $m$ clusters in the SIE calculation, this part of the cost is $O(mn^7)$. Therefore, a general scaling for SIE+CCSD(T) using full system MP2 can be formulated as
\begin{equation}
    O = \overset{\text{Full System MP2}}{O(N^5)} + \overset{\text{Cluster CCSD(T)}}{O(mn^7)}.
\end{equation}
A remarkable feature of SIE is that the cluster size converges as the system size increases with a fixed BNO threshold. Therefore, in calculations where the system size systematically expands, such as in adsorption energy computations where the surface is enlarged to converge the interacting energy, the actual CCSD(T) time consumption becomes a constant number as $n$ converges. At this point, the complexity shifts from $O(mn^7)$ to a linear scaling $O(\text{C}m)$, where C represents the constant CCSD(T) time consumption for a size-converged cluster. Thus, the SIE+CCSD(T) scaling becomes
\begin{equation}
    O = \overset{\text{Full System MP2}}{O(N^5)} + \overset{\text{Cluster CCSD(T)}}{O(\text{C}m)}.
\end{equation}

However, in practical SIE+CCSD(T) calculations, there can be a competition between the components of the scaling. To illustrate this issue, consider the SIE+CCSD(T) calculation for H$_2$O@PAH(6) using the ccECP-cc-pVTZ basis set, which involves 6357 orbitals. The full system canonical MP2 is used to correct the truncation bath error. In Table~\ref{tab:SI_time_consumption_example} we list the computational time utilized for the full system MP2, BNO building in a cluster, and the CCSD(T) calculation in a cluster.

\renewcommand{\arraystretch}{1.1}
\begin{table}[ht]
    \centering
    \caption{
    The time consumption estimated on A100 of full system canonical MP2, one cluster BNO building and one cluster CCSD(T) calculation for H$_{2}$O@PAH(6) with ccECP-cc-pVTZ basis set.
    }
    \label{tab:SI_time_consumption_example}
    \begin{tabular}{cccc}
        \toprule\midrule
        & Full system MP2 & Cluster BNO building & Cluster CCSD(T) \\
        \midrule
        Theoretical scaling & $O(N^5)$ & $O(N^3)$ for a cluster & $O(n^7)$ for a cluster \\
        Size & 6357 & 6357 & 640 \\
        Time consumption & 6 hours & 3 minutes & 60 hours \\
        \midrule\bottomrule
    \end{tabular}
\end{table}

As observed, although the full system MP2 theoretically presents higher complexity than the other two parts. its inherently non-iterative, straightforward nature of the MP2 enables its efficient implementation on GPU and ensures that its computational demand remains remarkably low, even for the system with thousands of orbitals. However, even though the cluster size is only about one-tenth of the entire system size, the seventh power complexity of CCSD(T) results in the time consumed for a single cluster significantly exceeding the time required for the full system MP2 computation, especially considering that it merely includes the time required for only one cluster. The cumulative time cost for CCSD(T) across $m$ clusters ultimately renders the costs for full system MP2 and BNO building negligible. This dominance of CCSD(T) in the SIE+CCSD(T) setup ensures that the framework exhibits a linear complexity of $O(\text{C}m)$ in medium-sized systems.

\begin{figure}[ht]
    \centering
    \includegraphics[width=1.0\linewidth]{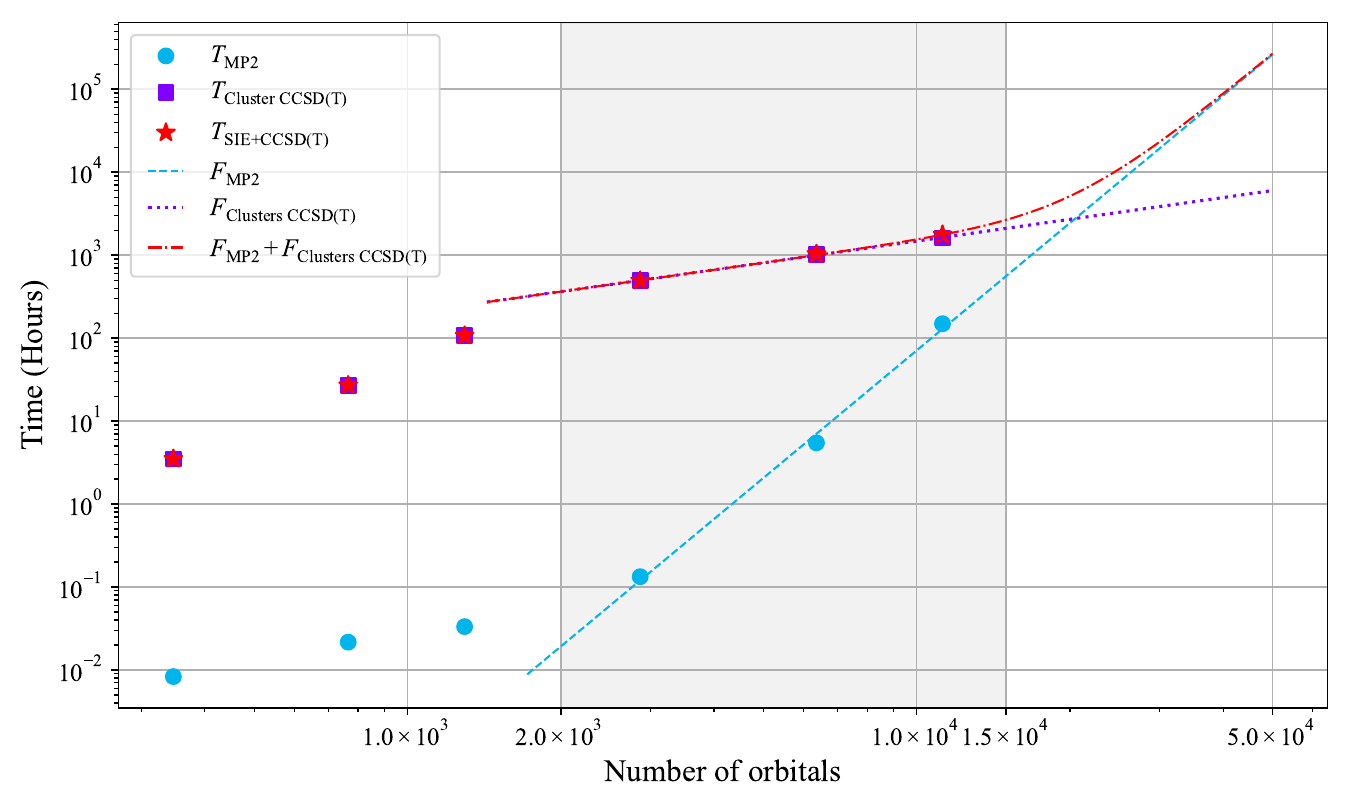}
    \caption{\label{fig:SI_linear_scaling}
    The time consumptions for the components in the SIE+CCSD(T) calculations for the H$_2$O@PAH systems. Here, the $T$ denotes the real time cost for the full system MP2 or clusters CCSD(T) in SIE+CCSD(T) or the total SIE+CCSD(T) time consumption. $F$ represents the fitted time cost based on the last 3 real time cost points.
    }
\end{figure}

It should be noted that the establishment of linear scaling requires certain conditions. Firstly, for a fixed reasonable BNO threshold, the cluster size must approach convergence so that the time consumed by a single cluster can trend towards a constant. Secondly, when employing full system MP2, it is essential that the MP2 computation does not become the primary consuming component. These conditions imply that the system sizes satisfying linear-scaling SIE+CCSD(T) should fall in an interval. We illustrate this conclusion using the computational cost of H$_2$O@PAH as the system size increases, as illustrated in Figure~\ref{fig:SI_linear_scaling}. The MP2 time consumption, $T_{\text{MP2}}$, and the total time consumption of SIE+CCSD(T), $T_{\text{SIE+CCSD(T)}}$, are both depicted in main text Fig.~\ref{fig:workflow}\textbf{b}. Apart from the time consumed by CCSD(T) in clusters and the full system MP2, other consumption can be disregarded in SIE+CCSD(T). Thus, the difference between $T_{\text{SIE+CCSD(T)}}$ and $T_{\text{MP2}}$ is considered to represent the consumption of clusters CCSD(T), i.e., $T_{\text{Clusters CCSD(T)}}$. According to the Figure~\ref{fig:SI_linear_scaling}, it is clear that the SIE+CCSD(T) scaling has three stages with the increasing system size. In the frist stage, the system is too small, with system size below 2k orbitals. Although the clusters CCSD(T) scaling $O(mn^7)$ dominates the SIE+CCSD(T), due to the non-convergence of the cluster size, the complexity may slightly exceed linearity. In the second stage, the system size is moderate, as shown by the gray area in the Figure~\ref{fig:SI_linear_scaling} with system sizes ranging from 2k to 15k. In this stage, the consumption of clusters CCSD(T) still remains predominant, but since the cluster size has already converged, the scaling of clusters CCSD(T) reduces to $O(\text{C}m)$, allowing SIE+CCSD(T) to achieve linear scaling. In the third stage, the system size becomes very large. Since the systems calculated using SIE+CCSD(T) still remain within the linear scaling region, we use the fitted consumption of cluster CCSD(T), $F_{\text{Clusters CCSD(T)}}$, combined with the fitted MP2 consumption, $F_{\text{MP2}}$, as the estimation of total SIE+CCSD(T) cost, $F_{\text{Clusters CCSD(T)}} + F_{\text{MP2}}$. Clearly, when the system size exceeds 15k orbitals, MP2 dominates the scaling of SIE+CCSD(T).

The linear complexity of SIE+CCSD(T) indeed appears in a specific range, and this range is related to factors such as the system type, partition strategy, and BNO threshold. However, it is observed that, for the H$_2$O@Graphene system, this region extends from 2k to 15k orbitals, which is sufficiently broad. This suggests that SIE+CCSD(T) can maintain high efficiency in most problems with a moderate system size.

\section{Water molecule on graphene}
\label{sec:SI_H2O_Graphene}

\subsection{Structures}
\label{sec:SI_WOG_structure}

\begin{figure}[ht]
    \centering
    \includegraphics[width=1.0\linewidth]{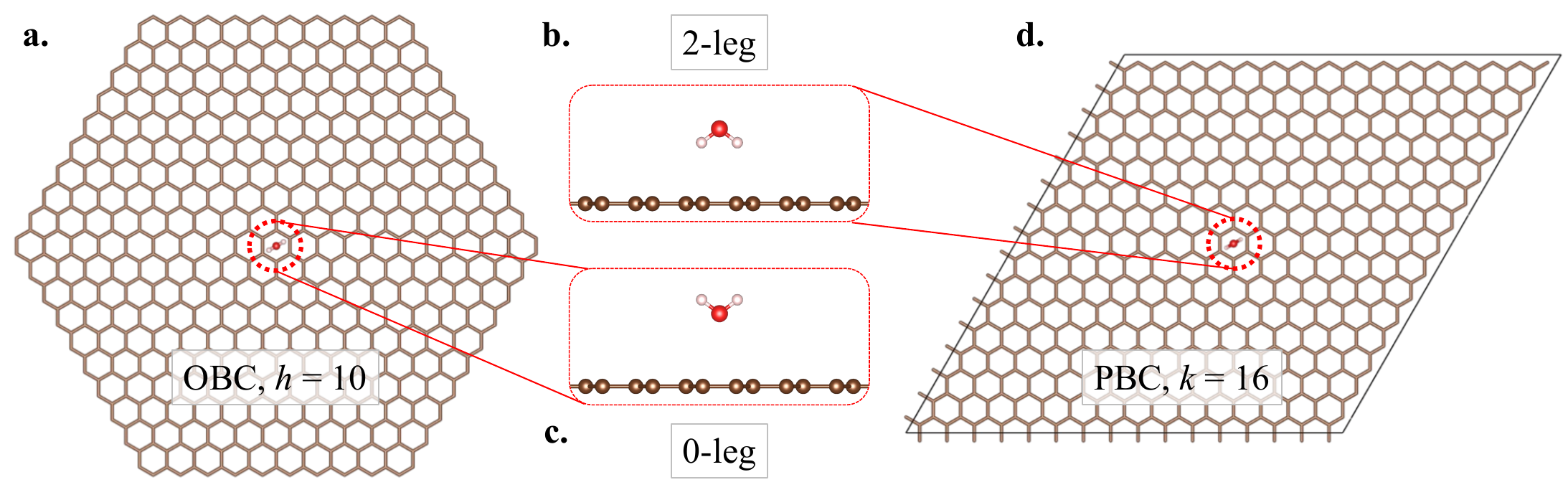}
    \caption{\label{fig:SI_sturactures}
    (\textbf{a}) OBC structure 0-leg@PAH(10). (\textbf{b}), (\textbf{c}) is the zoom-in picture for 2-leg configuration in picture (\textbf{d}) and 0-leg configuration in picture (\textbf{a}), respectively. (\textbf{d}) PBC structure 2-leg@Graphene with supercell  with $16 \times 16 \times 1$ unit cell. 
    }
\end{figure}

Various efforts in studying the calculation of adsorption energy for the system water monomer on graphene~\cite{ma2011adsorption, rubes2009structure, jenness2010benchmark, voloshina2011physisorption, brandenburg2019physisorption, lau2021regional, ajala2019assessment}, of which two frequently mentioned configurations, 0-leg and 2-leg, are the most likely candidates for the ground-state adsorption configurations. The 0-leg configuration, where two hydrogen atoms from the water monomer are oriented away from the graphene, is shown in Figure~\ref{fig:SI_sturactures}\textbf{b} and the 2-leg configuration, where the hydrogen atoms are oriented towards the graphene, is shown in Figure~\ref{fig:SI_sturactures}\textbf{c}. 

In this study, graphene under Open Boundary Conditions~(OBC) is modeled using a series of polycyclic aromatic hydrocarbons~(PAHs) as substitutes. PAHs can undergo systematic expansion by adding carbon atoms around the periphery. The number of rings denoted by \textit{h}, marks the size of the PAH, PAH(\textit{h}), and the molecular formula of PAHs can be expressed as C$_{6h^2}$H$_{6h}$. Under Periodic Boundary Conditions~(PBC), structures were computed at the $\Gamma$ point, and the corresponding overall expansion is achieved by extending the supercell at the $\Gamma$ point with $k \times k \times 1$ unit cell, containing $2k^2$ carbon atoms. Figure~\ref{fig:SI_sturactures} displays the largest systems calculated under the ccECP-cc-pVDZ basis set within this paper, with Figure~\ref{fig:SI_sturactures}\textbf{a} showing PAH(10) calculated under OBC, and Figure~\ref{fig:SI_sturactures}\textbf{b} showing a supercell with $16 \times 16 \times 1$ unit cell under PBC.

Considering that the adsorption energy between water and graphene is very weak, the deformations effect is relatively small. Therefore, in the SIE+CCSD(T) calculations, the configurations employed are those of water and graphene in their independent thermal equilibrium states. Here, we only optimize the distance between the water and graphene and use the equilibrium distance configurations to do the interacting energy calculations. The influence of geometry relaxation will be estimated by DFT. For the detailed information in graphene or PAHs, the bond length is 1.42~Å, and all C-C-C bond angles are 120$^{\circ}$. For the peripheral C-H bonds in PAHs, the bond length is 1.089~Å. The O-H bond length within water molecules is 0.957~Å, and the H-O-H bond angle is 104.5$^{\circ}$. The geometry relaxation resulting from adsorption is obtained by using DFT and subsequently corrected in the final adsorption energy.

\subsection{Graphene-Water Distance Optimization}
\label{sec:SI_distance_opt}

\begin{figure}[ht]
\centering
\includegraphics[width=1.0\linewidth]{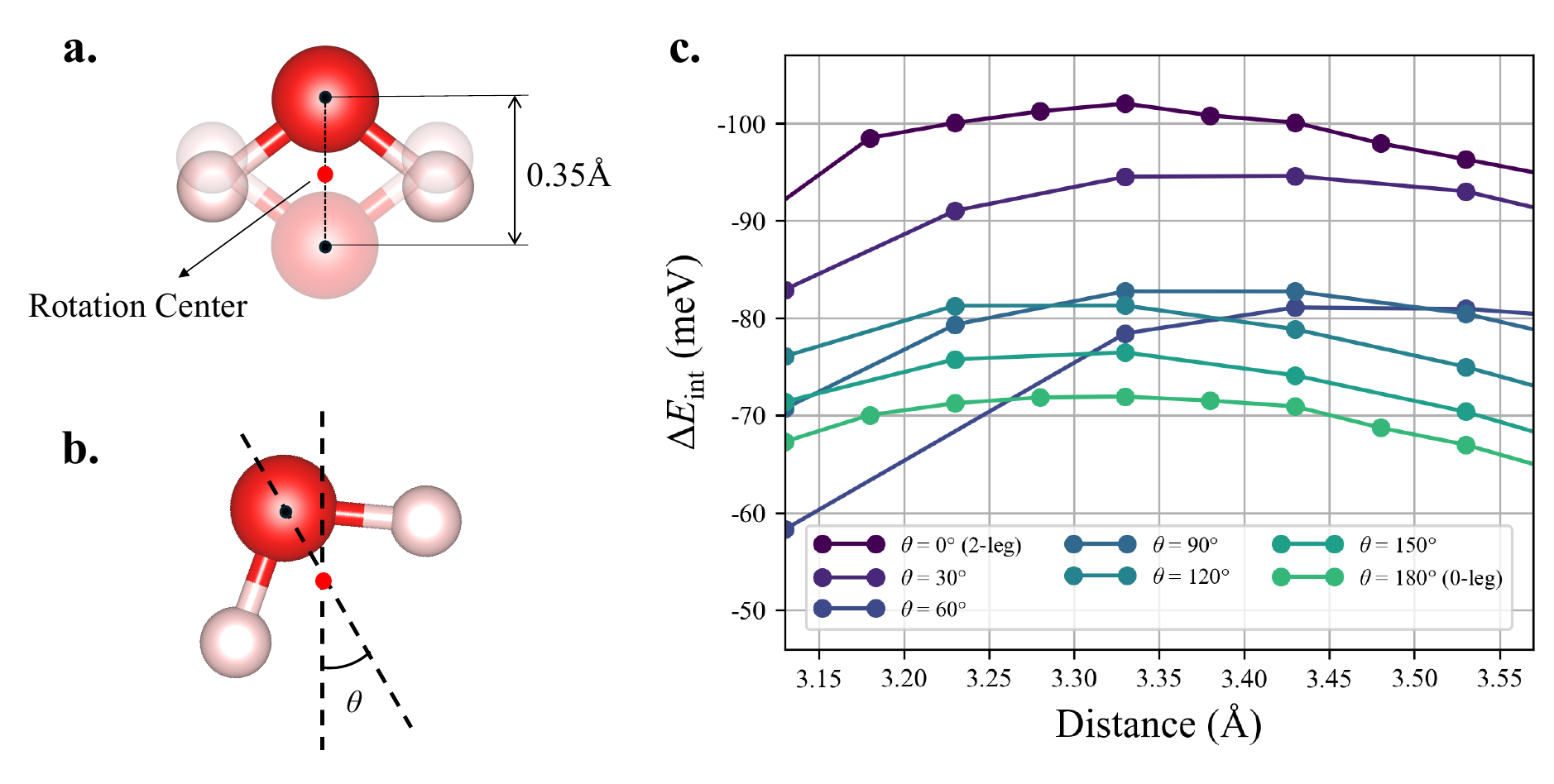}
\caption{\label{fig:SI_distance_opt}
(a) Side views of water monomer in 2-leg and 0-leg configurations. After optimization, the distance between the centers of the oxygen atoms in 2-leg and 0-leg configurations is 0.35 Å. A point located 0.175 Å from the oxygen atom in the direction from the negative polarization end towards the positive end is used as the rotation center. Water monomer rotates around the axis which is perpendicular to and through the rotation center. (b) Examples of the water molecule rotating around the rotation center, where $\theta$ refers to the rotational angle relative to the 2-leg configuration. (c) For different $\theta$ configurations, changes in the interacting energy relative to the distance from the rotation center of the monomer to the graphene plane.
}
\end{figure}
In previous studies, there has been a significant difference in the reported equilibrium distances from the oxygen atom in water monomer to the graphene surface in 2-leg and 0-leg configurations~\cite{ajala2019assessment,brandenburg2019physisorption,voloshina2011physisorption}. In table~\ref{tab:SI_WOG_CO_distance}, we summarize this equilibrium distances from various works, as well as those obtained after our optimization with SIE+CCSD.

\renewcommand{\arraystretch}{1.1}
\begin{table}[ht]
    \centering
    \caption{
    The equilibrium distances between the oxygen atom in water monomer and the graphene surface in 2-leg and 0-leg configurations.
    }
    \label{tab:SI_WOG_CO_distance}
    \begin{tabular}{ccc}
        \toprule\midrule
        Reference & 0-leg~(Å) & 2-leg~(Å) \\
        \midrule
        \cite{voloshina2011physisorption} & 3.06 & 3.20 \\
        \cite{ajala2019assessment} & 3.075 & 3.155 \\
        \cite{brandenburg2019physisorption} & 3.10 & 3.37 \\
        SIE+CCSD & 3.155 & 3.505 \\
        \midrule\bottomrule
    \end{tabular}
\end{table}

During the optimization process, the SIE+CCSD method was utilized, employing the structure of PAH(6), and the partition scheme as described in section~\ref{sec:SI_Partition}. This approach aimed to obtain relatively accurate interacting energies using the smallest possible basis sets. Therefore a mixed basis set strategy was employed: for the water monomer, the aug-cc-pVDZ basis set was used; for carbon atoms, the cc-pVDZ was selected, and for edge hydrogen atoms, the STO-3G basis set was applied. The BNO threshold for the water monomer fragment was set to 10$^{-6.5}$, while for the remaining fragments, it was set to 10$^{-8.5}$.

During optimization, the search for the equilibrium distance of the water-graphene in the 2-leg and 0-leg configurations started from 3.155~Å, with a step size of 0.05~Å. Given the sufficiently small step size, the difference in interacting energy between consecutive steps at the equilibrium distance range was very small, within 1~meV. Therefore, we did not perform any further numerical fitting and directly selected the point with the minimum interacting energy as the equilibrium distance for subsequent adsorption energy calculations.

\subsection{Graphene-Water Distance Optimization under different water orientation}
\label{sec:SI_distance_opt_water_orientation}

\renewcommand{\arraystretch}{1.1}
\begin{table}[ht]
    \centering
    \caption{
    $d_{\rm{C-G}}$ denotes the equilibrium distance from the rotation center to the graphene surface. And $d_{\rm{O-G}}$ denotes the equilibrium distance from O atom in water monomer to the graphene surface.
    }
    \label{tab:SI_WOG_center_graphene_distance}
    \begin{tabular}{cccccccc}
        \toprule\midrule
        $\theta$ & $0^\circ$~(2-leg) & $30^\circ$ & $60^\circ$ & $90^\circ$ & $120^\circ$ & $150^\circ$ & $180^\circ$~(0-leg) \\
        \midrule
        $d_{\rm{C-G}}$~(Å) & 3.33 & 3.43  & 3.43 & 3.33 & 3.33 & 3.33 & 3.33 \\
        \midrule
        $d_{\rm{O-G}}$~(Å) & 3.505 & 3.582 & 3.518 & 3.330 & 3.243 & 3.178 & 3.155 \\
        \midrule\bottomrule
    \end{tabular}
\end{table}

For different orientation water on graphene, to ensure consistency in the electronic response induced by the positive and negative polar ends of the rotating water molecule on the graphene surface, we designated the midpoint location of the oxygen atoms in the 0-leg and 2-leg configurations under equilibrium distance as the rotation center, as depicted in Figure~\ref{fig:SI_distance_opt}\textbf{a}. The water monomer rotates around an axis passing through the rotation center and perpendicular to the plane of the water monomer. The rotation initiates from the 2-leg configuration and we define the rotation configurations by the angle $\theta$ relative to the 2-leg configuration, as illustrated in Figure~\ref{fig:SI_distance_opt}\textbf{b}. Thus, the 2-leg configuration corresponds to $\theta = 0^{\circ}$, and the 0-leg configuration to $\theta = 180^{\circ}$, with additional rotations at theta = $30^{\circ}$, $60^{\circ}$, $90^{\circ}$, $120^{\circ}$, and $150^{\circ}$ also considered. 

Note these configurations are all optimized to find the equilibrium distance from the rotation center, instead of the O atom in water monomer, to the graphene surface, as shown in the Figure~\ref{fig:SI_distance_opt}\textbf{c}. The distance optimization setting is performed as section~\ref{sec:SI_distance_opt} discussed, but with step size of 0.1~Å. All equilibrium distances between the rotation center and the graphene surface are summarised in Table~\ref{tab:SI_WOG_center_graphene_distance}. It is important to note that, although the equilibrium distance between the rotation center and the graphene surface is very similar, due to the rotation angle $\theta$, there is still a slight difference in the final equilibrium distance between the oxygen atom in the water monomer and the graphene surface which also be calculated and shown in Table~\ref{tab:SI_WOG_center_graphene_distance}.

\subsection{Partition Strategy}
\label{sec:SI_Partition}

\begin{figure}[!ht]
\centering
\includegraphics[width=0.5\linewidth]{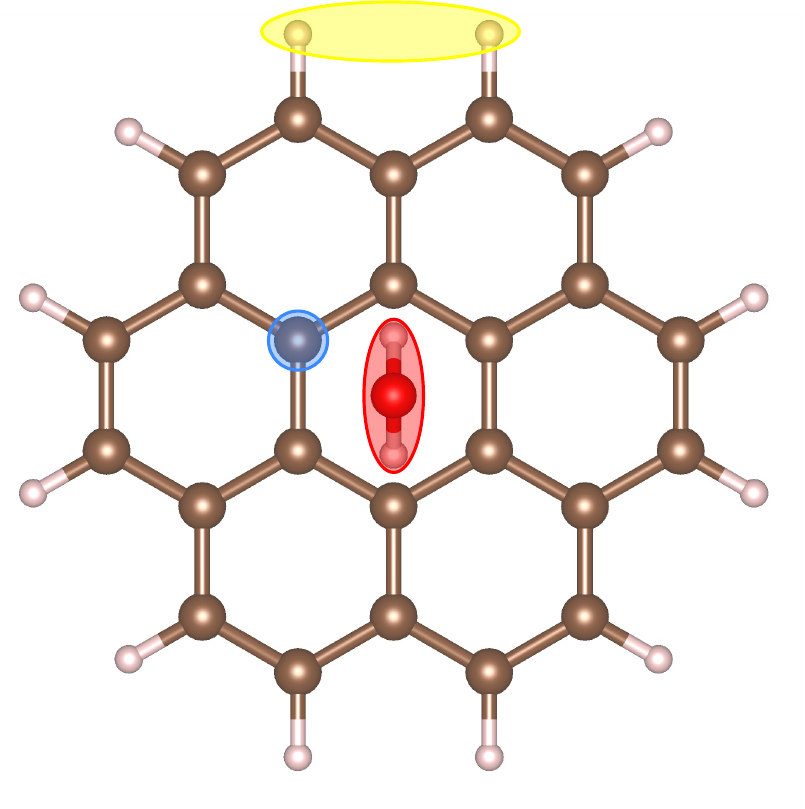}
\caption{\label{fig:SI_partition}
Each differently colored box represents a distinct kind of fragment: the red box denotes the water monomer fragment, the blue box denotes the carbon fragment in graphene, and the yellow box denotes the edge hydrogen atoms fragment.
}
\end{figure}

The water monomer was considered as an independent fragment, and each carbon atom on the graphene was also treated as a separate fragment. Specially, for OBC, the PAH essentially is a regular hexagon, where all the edge hydrogen atoms on each edge are considered as a single fragment. In Figure~\ref{fig:SI_partition}, we show the partition in H$_2$O@PAH(2) as an example.

\subsection{Interacting Energy Calculation with SIE+CCSD(T)}

In the calculations of H$_2$O@Graphene, the BNO threshold for the water monomer fragment is set to $10^{-6.5}$, which is the smallest BNO threshold we could take and results in the adsorbate cluster containing 700 orbitals at most.  While for other fragments, it is set to $10^{-8.0}$. The fragmentation used the strategy mentioned in section~\ref{sec:SI_Partition}, and the distance between the water monomer and graphene plane is fixed with the values mentioned in section~\ref{sec:SI_distance_opt}. Basis set used ccECP-cc-pV(D,T)Z~(ccECP-(D,T)Z). It is worth noting that, all calculation settings, including the basis set, the distance between water monomer and graphene, the threshold setting and others, are the same for the 0-leg and 2-leg configurations in OBC and PBC. Furthermore, it should be noted that the use of the ccECP basis set is not for the sake of reducing the computational cost. Typically, using the cc-pV\textit{x}Z basis sets for PBC calculations in SIE+CCSD(T) leads to CCSD diverge, whereas the ccECP-cc-pV\textit{x}Z does not encounter this issue. All data are listed in Table.~\ref{tab:SI_OBC_2-leg_int}-\ref{tab:SI_PBC_0-leg_int}, in a unit of meV.
\renewcommand{\arraystretch}{1.1}
\begin{table}[!ht]
    \centering
    \caption{
    Interacting energy (in meV) for 2-leg with OBC.
    }
    \label{tab:SI_OBC_2-leg_int}
    \begin{tabular}{lccc}
        \toprule\midrule
        Structure & ccECP-DZ & ccECP-TZ & CBS \\
        \midrule
        2-leg@PAH(2) & -78 & -121 & -138\\
        2-leg@PAH(4) & -68 & -111 & -128\\
        2-leg@PAH(6) & -62 & -105 & -122\\
        2-leg@PAH(8) & -60 & -104 & -121\\
        2-leg@PAH(10) & -58 & / & /\\
        \midrule\bottomrule
    \end{tabular}
\end{table}

\begin{table}[!ht]
    \centering
    \caption{
    Interacting energy (in meV) for 0-leg with OBC.
    }
    \label{tab:SI_OBC_0-leg_int}
    \begin{tabular}{lccc}
        \toprule\midrule
        Structure & ccECP-DZ & ccECP-TZ & CBS \\
        \midrule
        0-leg@PAH(2) & 12 & -26 & -40\\
        0-leg@PAH(4) & -29 & -72 & -90\\
        0-leg@PAH(6) & -46 & -84 & -100\\
        0-leg@PAH(8) & -47 & -87 & -104\\
        0-leg@PAH(10) & -48 & / & /\\
        \midrule\bottomrule
    \end{tabular}
\end{table}

\renewcommand{\arraystretch}{1.1}
\begin{table}[!ht]
    \centering
    \caption{
    Interacting energy (in meV) for 2-leg with PBC. 
    }
    \label{tab:SI_PBC_2-leg_int}
    \begin{tabular}{lccc}
        \toprule\midrule
        \textit{k}-point & ccECP-DZ & ccECP-TZ & CBS\\
        \midrule
        2-leg@$4\times4\times1$ & -30 & -65 & -79\\
        2-leg@$8\times8\times1$ & -51 & -95 & -112\\
        2-leg@$10\times10\times1$ & -57 & -100 & -116\\
        2-leg@$14\times14\times1$ & -57 & -99 & -116\\
        2-leg@$16\times16\times1$ & -58 & / & /\\
        \midrule\bottomrule
    \end{tabular}
\end{table}

\begin{table}[!ht]
    \centering
    \caption{
    Interacting energy (in meV) for 0-leg with PBC.
    }
    \label{tab:SI_PBC_0-leg_int}
    \begin{tabular}{lccc}
        \toprule\midrule
        \textit{k}-point & ccECP-DZ & ccECP-TZ & CBS \\
        \midrule
        0-leg@$4\times4\times1$ & -21 & -51 & -64\\
        0-leg@$8\times8\times1$ & -48 & -83 & -99\\
        0-leg@$10\times10\times1$ & -50 & -89 & -106\\
        0-leg@$14\times14\times1$ & -52 & -88 & -104\\
        0-leg@$16\times16\times1$ & -55 & / & /\\
        \midrule\bottomrule
    \end{tabular}
\end{table}

\subsection{Bath truncation error correction}
\label{sec:SI_BTEC}

\begin{figure}[ht]
\centering
\includegraphics[width=1.0\linewidth]{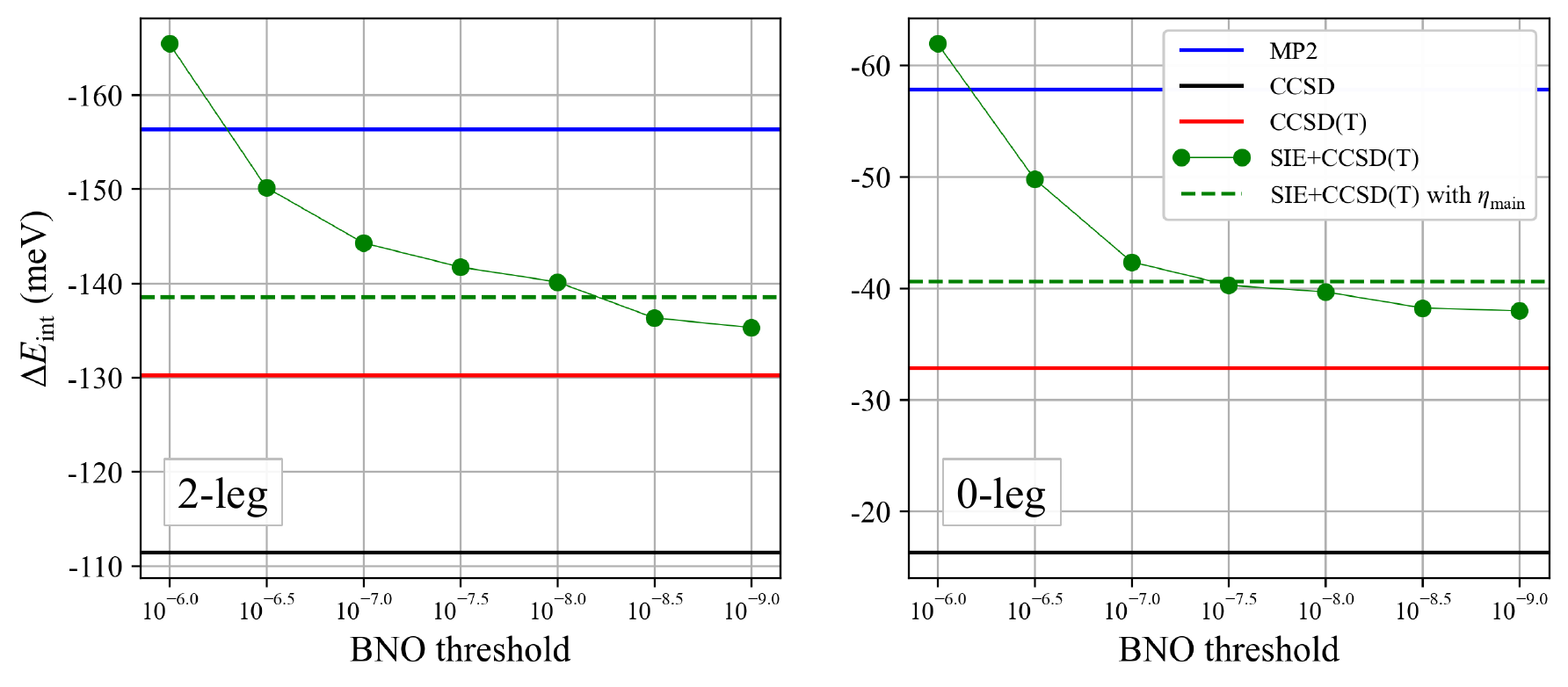}
\caption{\label{fig:SI_BNO_threshold}
In PAH(2), as reducing the BNO threshold, the variation of interacting energy calculated by SIE+CCSD(T) (green solid line) relative to canonical CCSD(T) (red line), canonical CCSD (black line), and canonical MP2 (blue line) are shown. The green dashed line SIE+CCSD(T) with $\eta_{\rm{main}}$ corresponds to the results calculated using the BNO threshold settings discussed in the main text, which is $10^{-6.5}$ for water monomer fragment and $10^{-8.0}$ for carbon fragments on graphene.
}
\end{figure}

For any method involving partitioning, there is an inherent error because the subspace size is generally much smaller than the entire system, leading to a poor handling of correlations outside the subspace. Such error stems from the Hamiltonian being truncated to within the subspace. To estimate this error within the SIE framework, a relatively inexpensive correlated method is chosen to measure the difference between the result obtained directly using this method for the full system and the result derived using it as a high-level solver for the SIE. MP2 is considered as a suitable choice. Thus, in the main text's method section~\ref{sec:Methods}, bath truncation error correction is detailed as
\begin{equation}
    \label{eq:MP2_BTEC}
    \Delta E^{\text{BTEC}}_{\text{MP2}} = E_{\text{MP2}} - E^{\text{PWF-DM}}_{\text{SIE+MP2}},
\end{equation}
where $\Delta E^{\text{BTEC}}_{\text{MP2}}$ denotes the bath truncation error correction at MP2-level, $E_{\text{MP2}}$ denotes the full system MP2 correlation energy, and $E^{\text{PWF-DM}}_{\text{SIE+MP2}}$ denotes the correlation energy calculated by SIE+MP2 with PWF-DM. It is noteworthy that making this correction does not necessarily require a full system MP2 calculation. 
All methods which as long as can capture more correlation outside the cluster space are acceptable to do this correction.
For instance, employing SIE+MP2 results with a smaller BNO threshold is acceptable, which can save some computational resources. In this article, thanks to the efficient engineering implementation of MP2, $E^{\text{BTEC}}_{\text{MP2}}$ is directly obtained from a full system MP2 calculation. In addition, $E^{\text{PWF-DM}}_{\text{SIE+MP2}}$ should utilize the same SIE settings as those used for obtaining $E^{\text{PWF-DM}}_{\text{SIE+CCSD(T)}}$.

The MP2-level bath truncation error correction is considered part of the SIE+CCSD(T) workflow, utilized in all computations within this paper. Unless specifically stated otherwise, it is assumed that this correction has been applied in the results of the SIE+CCSD(T). However, strictly speaking, the bath truncation error of SIE+CCSD(T) should be derived from the difference between full system CCSD(T) and SIE+CCSD(T) results. The correction obtained through MP2 still slightly deviates from the ground truth. Like the results demonstrated in Figure~\ref{fig:SI_BNO_threshold}, for the H$_2$O@PAH(2) system, although the SIE+CCSD(T) results already include the $\Delta E^{\text{BTEC}}_{\text{MP2}}$, a small gap still remains compared to canonical CCSD(T) outcomes.

Fortunately, for a series of systematically expanded systems, such as the H$_2$O@Graphene system, the CCSD(T)-level bath truncation error correction can be calculated on the smallest system and then applied this correction as a constant to larger systems, thereby enhancing the accuracy of SIE calculation. Specifically, for the OBC H$_2$O@PAH($h$) system, the bath truncation error correction can be estimated as
\begin{equation}
    \Delta E^{\text{BTEC}}(\text{OBC},h) = \Delta E^{\text{BTEC}}_{\text{MP2}}(\text{OBC},h) + \Delta E^{\text{BTEC}}_{\text{CCSD(T)}}(\text{OBC},2),
\end{equation}
where $\Delta E^{\text{BTEC}}_{\text{MP2}}(\text{OBC},h)$ denotes the MP2-level bath truncation error correction for the OBC system H$_2$O@PAH($h$) defined in Eq.~\ref{eq:MP2_BTEC}. And $E^{\text{BTEC}}_{\text{CCSD(T)}}(\text{OBC},2)$ denotes improved part of the bath truncation error correction at CCSD(T)-level for H$_2$O@PAH($2$), which could be defined as
\begin{equation}
    \Delta E^{\text{BTEC}}_{\text{CCSD(T)}}(\text{OBC},2) = E_{\text{CCSD(T)}}(\text{OBC},2,\text{CBS}) - E_{\text{SIE+CCSD(T)}}(\text{OBC},2,\text{CBS}),
\end{equation}
where $E_{\text{CCSD(T)}}(\text{OBC},2,\text{CBS})$ denotes the canonical CCSD(T) calculated correlation energy under CBS for H$_2$O@PAH(2), and $E_{\text{SIE+CCSD(T)}}(\text{OBC},2,\text{CBS})$ denotes the similar correlation energy calculated by SIE+CCSD(T) with MP2-level bath truncation error correction. Since the derivation of interacting energy is based on the direct addition and subtraction of total energies, the bath truncation error correction for interacting energy at the CCSD(T) level can be straightforwardly expressed as
\begin{equation}
    \Delta E^{\text{BTEC}}_{\text{CCSD(T)}}(\text{OBC},2) = \Delta E^{\text{CCSD(T)}}_{\text{int}}(\text{OBC},2,\text{CBS}) - \Delta E^{\text{SIE+CCSD(T)}}_{\text{int}}(\text{OBC},2,\text{CBS}),
\end{equation}
where the $\Delta E_{\text{int}}$ denotes the interacting energy. The final $\Delta E^{\text{BTEC}}_{\text{int,CCSD(T)}}(\text{OBC},2)$ values for 2-leg and 0-leg configurations are summarised in Table~\ref{tab:SI_OBC_BTEC}.

\renewcommand{\arraystretch}{1.1}
\begin{table}[ht]
    \centering
    \caption{
    OBC CCSD(T)-level bath truncation error correction (in meV) for interacting energy estimation on PAH(2).
    }
    \label{tab:SI_OBC_BTEC}
    \begin{tabular}{lcc}
        \toprule\midrule
        Configuration & 2-leg & 0-leg \\
        \midrule
        $\Delta E_{\text{int}}^{\text{CCSD(T)}}$ & -130 & -33 \\
        $\Delta E^{\text{BTEC}}_{\text{CCSD(T)}}$ & -138 & -41 \\
        \midrule
        $\Delta E^{\text{BTEC}}_{\text{CCSD(T)}}$ & 8 & 8 \\
        \midrule\bottomrule
    \end{tabular}
\end{table}

For PBC, the smallest structure has $4 \times 4 \times 1$ graphene supercell, and under ccECP-TZ basis set the system has nearly 1000 orbitals which is beyond the upper limit we could afford to do the canonical CCSD(T). 
This issue is addressed by implementing a SIE+CCSD(T) calculation with a smaller threshold, 
\begin{equation}
\begin{split}
    \Delta E^{\text{BTEC}}_{\text{CCSD(T)}}(\text{PBC},4) = & \Delta E^{\text{CCSD(T)}}_{\text{int}}(\text{PBC},4) - \Delta E^{\text{SIE+CCSD(T)}}_{\text{int}}(\text{PBC},4,\eta) \\
    = & [\Delta E^{\text{CCSD(T)}}_{\text{int}}(\text{PBC},4) - \Delta E^{\text{SIE+CCSD(T)}}_{\text{int}}(\text{PBC},4,\eta_{s})] \\
    + & [\Delta E^{\text{SIE+CCSD(T)}}_{\text{int}}(\text{PBC},4,\eta_{s}) - \Delta E^{\text{SIE+CCSD(T)}}_{\text{int}}(\text{PBC},4,\eta)]
\end{split}
\end{equation}
where BNO threshold $\eta_{s}$ is smaller than $\eta$. The $\eta_{s}$ is selected to a value such that the SIE+CCSD(T) calculation can be performed under CBS~(the largest basis is TZ here) to ensure that the difference between SIE+CCSD(T) with $\eta_{s}$ and $\eta$ can be estimated in CBS. While for the difference between canonical CCSD(T) and SIE+CCSD(T) with $\eta_{s}$, the difference is estimated under DZ basis set which is affordable. Therefore, the formulation could be rewritten as
\begin{equation}
\begin{split}
    \Delta E^{\text{BTEC}}_{\text{CCSD(T)}}(\text{PBC},4)
    = & [\Delta E^{\text{CCSD(T)}}_{\text{int}}(\text{PBC},4,\text{DZ}) - \Delta E^{\text{SIE+CCSD(T)}}_{\text{int}}(\text{PBC},4,\eta_{s},\text{DZ})] \\
    + & [\Delta E^{\text{SIE+CCSD(T)}}_{\text{int}}(\text{PBC},4,\eta_{s},\text{CBS}) - \Delta E^{\text{SIE+CCSD(T)}}_{\text{int}}(\text{PBC},4,\eta,\text{CBS})].
\end{split}
\end{equation}
Here, $\eta_{s}$ set as $10^{-9.0}$ for all fragments. Then PBC CCSD(T)-level bath truncation error corrections for interacting energy estimated in H$_2$O@Graphene with $4 \times 4 \times 1$ are listed in Table.\ref{tab:SI_PBC_FCSE}.

\renewcommand{\arraystretch}{1.1}
\begin{table}[ht]
    \centering
    \caption{
    PBC CCSD(T)-level bath truncation error correction (in meV) for interacting energy estimation on graphene with $4 \times 4 \times 1$ supercell. $\eta_{s}=10^{-9.0}$. Unit uses meV.
    }
    \label{tab:SI_PBC_FCSE}
    \begin{tabular}{lcc}
        \toprule\midrule
        Configuration & 2-leg & 0-leg \\
        \midrule
        $\Delta E_{\rm{int}}^{\rm{CCSD(T)}}(\text{DZ})$ & -25 & -16 \\
        $ \Delta E_{\rm{int}}^{\rm{SIE+CCSD(T)}}(\eta_s,\text{DZ})$ & -28 & -19 \\
        $\Delta E_{\rm{int}}^{\rm{SIE+CCSD(T)}}(\eta_{s}, \rm{CBS})$ & -74 & -58 \\
        $\Delta E_{\rm{int}}^{\rm{SIE+CCSD(T)}}(\eta, \rm{CBS})$ & -79 & -64 \\
        \midrule
        $\Delta E^{\text{BTEC}}_{\text{CCSD(T)}}$ & 8 & 8 \\
        \midrule\bottomrule
    \end{tabular}
\end{table}

\subsection{Bulk Limit Extrapolation}
\label{sec:SI_bulk_limit_exp}

\begin{figure}[ht]
\centering
\includegraphics[width=1.0\linewidth]{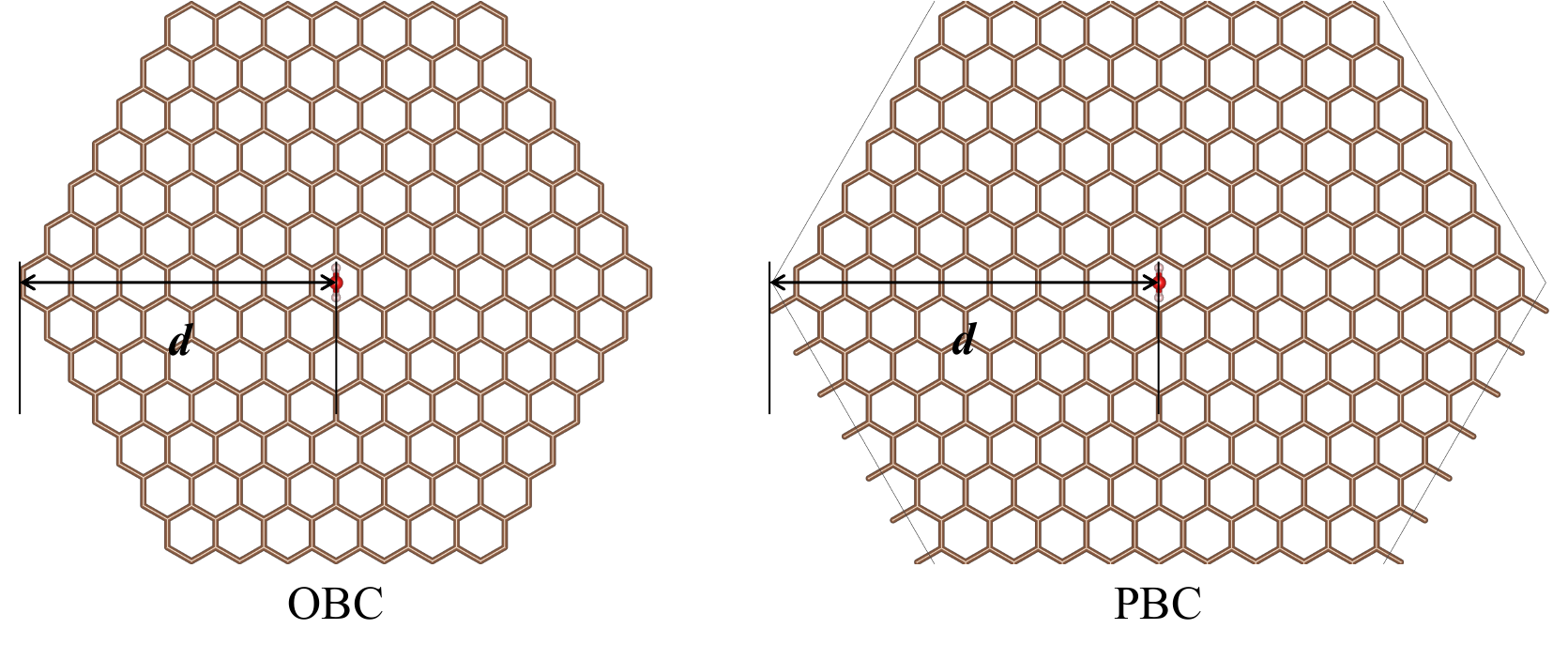}
\caption{\label{fig:SI_radius_def}
The definitions for radius in OBC and PBC. The edge H atoms are hidden in OBC structure.
}
\end{figure}

The relationship between the interacting energy and the system size is fitted with the modified nonlinear equation~\cite{brandenburg2019physisorption},
\begin{equation}
    \Delta E_{\rm{int}} = A + B / d^{\gamma},
\end{equation}
where \textit{A}, \textit{B}, and $\gamma$ are the parameters to be fitted, and \textit{d} represent the radius of the graphene substrate. The definition of substrate radius for OBC PAH and PBC graphene are shown in Figure.~\ref{fig:SI_radius_def}. Note, here we use four points for the extrapolation.

With ccECP-TZ basis set, we extended our calculations for OBC up to PAH(8) and for PBC supercell with $14\times14\times1$ unit cell. With ccECP-DZ basis set, we are able to reach PAH(10) and supercell with $16\times16\times1$ unit cell, thereby allowing us to verify the reliability of our extrapolations on the ccECP-DZ. All the ccECP-DZ basis set results have been listed in Table~\ref{tab:SI_bulk_limit_exp}.The difference between 4-point and 5-point extrapolated interacting energy on ccECP-DZ basis set would be regarded as the uncertainty estimated by bulk limit extrapolation, which is also listed in the Table.~\ref{tab:SI_bulk_limit_exp}.  

\begin{figure}[ht]
\centering
\includegraphics[width=0.6\linewidth]{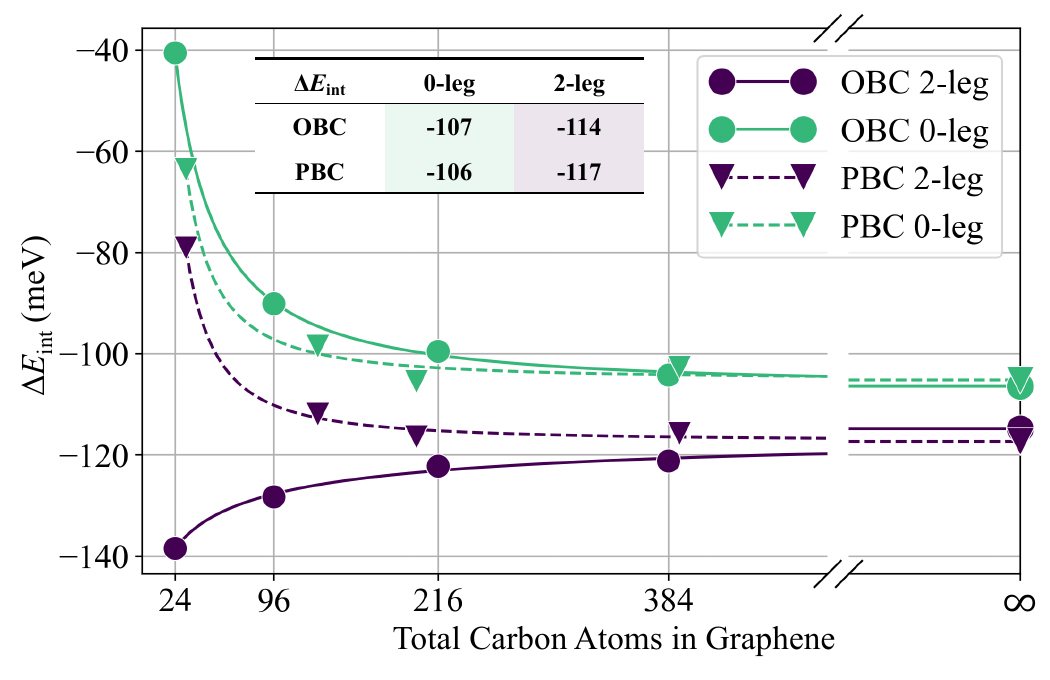}
\caption{\label{fig:SI_bulk_limit_extrapolation}
The interacting energy $\Delta E_{\text{int}}$ calculated by SIE+CCSD(T) under ccECP-(D,T)Z extrapolated to CBS. The infinite notation stands for the extrapolated bulk limit. The insert table shows the values of extrapolated CBS interacting energy at bulk limit.
}
\end{figure}

Figure~\ref{fig:SI_bulk_limit_extrapolation} shows the bulk limit extrapolation of CBS interacting energy of 0-leg and 2-leg under OBC and PBC, and the insert table of Figure~\ref{fig:SI_bulk_limit_extrapolation} shows the values of the interacting energy at bulk limit. Interestingly, we observe that in PBC, for both the 2-leg and 0-leg configurations, the fitted $\gamma$ closely approaches 3, meaning the dipole-dipole interaction between water monomers in different cells is the primary factor influencing the magnitude of finite size error.

\renewcommand{\arraystretch}{1.1}
\begin{table}[ht]
    \centering
    \caption{
    PAH(\textit{h}) scales as C$_{6h^2}$H$_{6h}$ and \textit{k} stands for the graphene supercell with $k\times k\times1$ unit cell. The list for \textit{h} and \textit{k} is referred as the structures which take part into the bulk limit extrapolation. $\Delta \Delta E_{\rm{BL}}$ denotes the uncertainty estimated from bulk limit extrapolation. All energies are in meV.
    }
    \label{tab:SI_bulk_limit_exp}
    \begin{tabular}{ccccccc}
        \toprule\midrule
        & \multicolumn{3}{c}{OBC} & \multicolumn{3}{c}{PBC} \\
        & $h$ in PAH & 2-leg & 0-leg & $k$ in supercell & 2-leg & 0-leg \\
        \midrule
        \multirow{2}{*}{ccECP-DZ} & [2,4,6,8] & -49 & -57 & [4,8,10,14] & -62 & -53 \\
        & [2,4,6,8,10] & -51 & -53 & [4,8,10,14,16] & -61 & -56 \\
        \midrule
        $\Delta \Delta E_{\rm{BL}}$ & / & $\pm 3$& $\pm 4$ & / & 0 & $\pm 3$ \\
        \midrule\bottomrule
    \end{tabular}
\end{table}

\subsection{DFT Study and Geometry Relaxation}
\label{sec:SI_geom_relax}
The graphene structure and water monomer used in interacting energy calculation are obtained from their respective thermal equilibrium geometries.
Although the interaction between graphene and water monomer is relatively weak, the graphene and water monomer may still undergo slight deformations due to adsorption. Therefore, the change in adsorption energy due to this geometry relaxation should be reasonably estimated and included in the overall calculation. Generally, a complete geometry optimization should be performed. However, such optimization is too costly for the current framework of SIE due to the gradient calculation. Therefore, geometry relaxation is estimated at the DFT level. 

In an early work~\cite{ajala2019assessment}, multiple DFT functionals are used to calculate the interacting energy of H$_2$O@Graphene. B3LYP+D3, BLYP+D3, PBE0+D3, PBE+D3, revPBE0+D3, revPBE+D3, B97M-V, $\omega$B97M-V are selected by experience from the work~\cite{ajala2019assessment} to address this task. D3 here is the dispersion correction for functionals~\cite{kristyan1994can, hobza1995density}. Inspired by this work, we performed all these functionals under cc-pVQZ basis on H$_2$O@PAH(4), which is selected from a series of OBC structures used for SIE+CCSD(T) to calculate interacting energy. The calculations in this section are performed by GPU4PySCF~\cite{wu2024python,lehtola2018recent,ekstrom2010arbitrary} and the results are listed in Table~\ref{tab:SI_DFT_benchmark_org}. As shown in Table~\ref{tab:SI_DFT_benchmark_org}, the 0-leg configuration of almost all functionals are comparable to SIE+CCSD(T), while for the 2-leg configuration, only double-hybrid functional B97M-V and $\omega$B97M-V achieved more reasonable results comparable to those of SIE+CCSD(T). Hence, B97M-V and $\omega$B97M-V are employed for the relaxation of the overall structure.  

\renewcommand{\arraystretch}{1.1}
\begin{table}[ht]
    \centering
    \caption{
     The adsorption energies (in meV) for original structures $\Delta E_{\text{org}}$.
    }
    \label{tab:SI_DFT_benchmark_org}
    \begin{tabular}{ccc}
        \toprule\midrule
        Method & $\Delta E_{\text{org}}^{\text{2-leg}}$ & $\Delta E_{\text{org}}^{\text{0-leg}}$ \\
        \midrule
        SIE+CCSD(T) & -120 & -82   \\
        B3LYP+D3 & -150 & -82  \\
        BLYP+D3 & -151 & -79   \\
        PBE0+D3 & -144 & -79 \\
        PBE+D3 & -146 & -81  \\
        revPBE0+D3 & -153 & -73  \\
        revPBE+D3 & -160 & -76  \\
        B97M-V & -133 & -96  \\
        $\omega$B97M-V & -131 & -98 \\
        \midrule\bottomrule
    \end{tabular}
\end{table}

We take certain constraints during the structure relaxation process. Firstly, since the effects of adsorption should decrease at the edges, thereby the distances between C-C and C-H on the edges of PAH(4) are fixed during relaxation. Secondly, according to our calculations, all selected functionals greatly underestimated the ground-state distance between graphene and water monomer. To mitigate this issue, the distance between the oxygen atom in the water and the nearest six carbon atoms in PAH(4) was also fixed. The final adsorption energies for optimized structures are also shown in Table~\ref{tab:SI_DFT_benchmark_opt}. And the adsorption energy differences between the original structure and the optimized structure for functionals are shown in Table~\ref{tab:SI_DFT_benchmark_difference}. The mean value of this difference is used as the geometry relaxation correction, $\Delta E_{\text{geom}}$ to ultimately obtain the adsorption energy, while the root mean square deviation from this average value $\Delta \Delta E_{\text{geom}}$ is considered as the uncertainty of geometry relaxation.

\renewcommand{\arraystretch}{1.1}
\begin{table}[ht]
    \centering
    \caption{
     The adsorption energies (in meV) for optimized structures $\Delta E_{\text{opt}}$.
    }
    \label{tab:SI_DFT_benchmark_opt}
    \begin{tabular}{ccc}
        \toprule\midrule
          & $\Delta E_{\text{opt}}^{\text{2-leg}}$ & $\Delta E_{\text{opt}}^{\text{0-leg}}$ \\
        \midrule
        B97M-V & -139 & -103 \\
        $\omega$B97M-V & -125 & -94 \\
        \midrule\bottomrule
    \end{tabular}
\end{table}

\renewcommand{\arraystretch}{1.1}
\begin{table}[ht]
    \centering
    \caption{
    The adsorption energy differences (in meV) between different functionals. 
    }
    \label{tab:SI_DFT_benchmark_difference}
    \begin{tabular}{ccc}
        \toprule\midrule
        & $\Delta E_{\text{opt}}^{\text{2-leg}} - \Delta E_{\text{org}}^{\text{2-leg}}$ & $\Delta E_{\text{opt}}^{\text{0-leg}} - \Delta E_{\text{org}}^{\text{0-leg}}$\\
        \midrule
        B97M-V & -6 & -8 \\
        $\omega$B97M-V & 7 & 5 \\
        \midrule
        $\Delta E_{\text{geom}}$ & $0 $ & $-2 $\\
        $\Delta \Delta E_{\text{geom}}$ & $\pm 6$ & $\pm 6$\\
        \midrule\bottomrule
    \end{tabular}
\end{table}

\subsection{H$_2$O@Graphene adsorption energy}

\begin{table}[!ht]
    \centering
    \caption{
    The final adsorption energy (in meV) with corrections.
    }
    \label{tab:SI_summary}
    \begin{tabular}{ccccc}
        \toprule\midrule
        & \multicolumn{2}{c}{OBC} & \multicolumn{2}{c}{PBC} \\
        & 2-leg & 0-leg & 2-leg & 0-leg \\
        \midrule
        $\Delta E^{\rm{SIE+CCSD(T)}}_{\rm{int}}$  & -114 & -107 & -117 & -106 \\
        \midrule
        $\Delta E^{\rm{BTEC}}_{\rm{CCSD(T)}}$ & 8 & 8 & 8 & 8 \\
        $\Delta E_{\rm{geom}}$ & $0$ & $-2$ & $0$ & $-2$ \\
        \midrule
        $\Delta \Delta E_{\rm{BL}}$ & $\pm 3$ & $\pm 4$ & $\pm 0$ & $\pm 3$ \\
        $\Delta \Delta E_{\rm{geom}}$ & $\pm 6$ & $\pm 6$ & $\pm 6$ & $\pm 6$ \\
        \midrule
        $\Delta E^{\rm{SIE+CCSD(T)}}_{\rm{ads}}$ & $-106 \pm 7$ & $-101 \pm 7$ & $-109 \pm 6$ & $-100 \pm 6$ \\
        \midrule\bottomrule
    \end{tabular}
\end{table}

As previously mentioned, in the calculations for H$_2$O@Graphene, there are two corrections applied to mitigate the error arising from downfolding Hamiltonian (as discussed in the section~\ref{sec:SI_BTEC}), $\Delta E^{\rm{BTEC}}_{\rm{CCSD(T)}}$ and the geometry relaxation (as discussed in the section~\ref{sec:SI_geom_relax}), $\Delta E_{\rm{geom}}$. To correct the bath truncation error, the gap between canonical CCSD(T)and SIE+CCSD(T) is estimated on smallest structure for OBC PAH(2) and for PBC graphene supercell with $4\times4\times1$ \textit{k}-point. The changes from geometry relaxation are fully studied by DFT using B97M-V and $\omega$B97M-V functionals. Those two corrections are summed up to the SIE+CCSD(T) calculated interacting energy to form the final adsorption energy.

The uncertainty originates from two sources. One is from the bulk limit extrapolation, $\Delta \Delta E_{\rm{BL}}$, which comes from using more data to do the bulk limit extrapolation on a smaller basis set, ccECP-DZ~(as discussed in the section~\ref{sec:SI_bulk_limit_exp}). The other source of uncertainty comes from estimating geometry relaxation effect under different functionals, $\Delta \Delta E_{\text{geom}}$~(as discussed in the section~\ref{sec:SI_geom_relax}). Those two uncertainties are taken in root-sum-square value to form the total uncertainty for adsorption energy.
All components have been summarised in the Table~\ref{tab:SI_summary}, with unit meV.

\subsection{Electron density rearrangement}
\label{sec:SI_EDR}
The electron density rearrangement for a adsorption system is defined as
\begin{equation}
\label{eq:SI_EDR_eq}
    \Delta D_{e} = D_{e}(\text{AB}) - D_{e}(\text{A[B]}) - D_{e}(\text{[A]B}),
\end{equation}
where $\Delta D_{e}$ denotes the electron density rearrangement and $D_{e}$ denotes the electron density distribution. $D_{e}(\text{AB})$ denotes the electron density distribution for the full adsorbate-substrate system, $D_{e}(\text{A[B]})$ denotes the electron density distribution for only adsorbate with substrate being settled as ghost atom and $D_{e}(\text{A[B]})$ denotes the electron density distribution for only substrate with adsorbate being settled as ghost atom. 

The advantage of this definition is that the electron density rearrangement can be considered as only induced by adsorption interactions, excluding other disturbances. In fact, the electron density distribution can be derived using the 1-RDM. Thanks to the specially designed partition wavefunction RDM approach~(see more details in~\ref{sec:SI_PWF_RDM}), the global 1-RDM can be obtained via SIE+CCSD, indicating that such electron density rearrangements are characterized with CCSD level accuracy. This undoubtedly provides a more accurate image for understanding adsorption interactions.

\subsection{Adsorption-Induced Dipole Moment}
\label{sec:SI_DP}

In the main text, for the configuration $\theta=60^{\circ}$, the interacting energy shows negligible variation with increasing PAH size, indicating an fake `short-range interaction' between water and graphene. This phenomenon is caused by error cancellation. In the main text Fig.~\ref{fig:water_rotate}\textbf{b}, the range of electron density rearrangement for configurations with 2-leg, 0-leg, and $\theta=60^\circ$ shows no significant difference, indicating that the interaction of the $\theta=60^\circ$ configuration is also long-range. The change in the system's dipole moment caused by adsorption can also provide additional evidence. Here, the change in dipole moment, also named adsorption-induced dipole moment in this paper, $\Delta \mu$, can be defined as
\begin{equation}
    \Delta \mu = \mu(\text{AB}) - \mu(\text{A[B]}) - \mu(\text{[A]B}),
\end{equation}
where $\mu(\text{AB})$ denotes the dipole moment of the combined AB system, while $\mu(\text{A[B]})$ denotes the dipole moment in the system comprising A with ghost B, and similarly for $\mu(\text{[A]B})$. Here, A and B represent the water monomer and the PAH respectively. This definition is very similar to those used for interacting energy~(Eq.~\ref{eq:CP_correction}) and electron density rearrangement~(Eq. \ref{eq:SI_EDR_eq}). Therefore, $\Delta \mu$ can be considered as entirely introduced due to adsorption.

\begin{figure}[ht]
\centering
\includegraphics[width=1.0\linewidth]{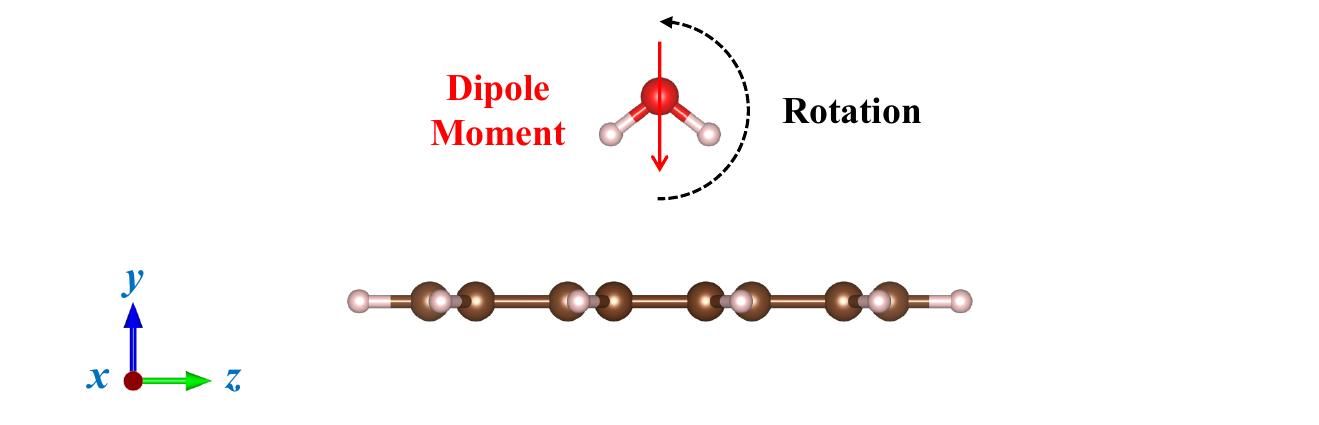}
\caption{\label{fig:SI_dp_structure}
The structure and coordinate axis of 2-leg@PAH(2). The dashed black arrow in the diagram indicates the direction of water rotation. The solid red arrow marks the direction of the dipole moment of the water monomer. Note that the $x$ axis is perpendicular to the plane of the paper and points outside.
}
\end{figure}

\begin{figure}[ht]
\centering
\includegraphics[width=1.0\linewidth]{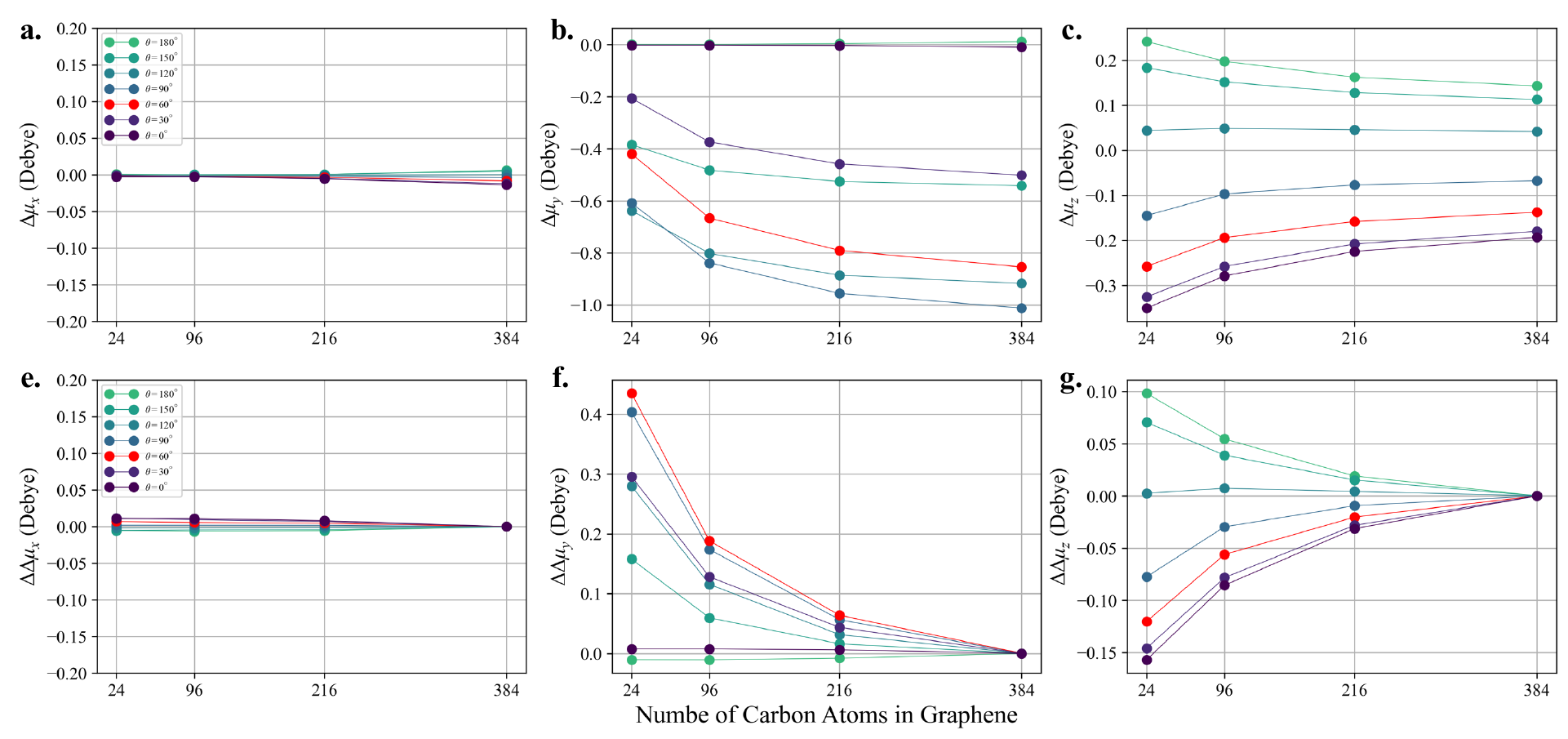}
\caption{\label{fig:SI_dp_moment}
The $\Delta \mu$ and $\Delta \Delta \mu$ components along the $x$, $y$, and $z$ axes changes with the size of PAH increasing from 2 to 8.
}
\end{figure}

The $\Delta \mu$ for H$_2$O@PAH($h$) with $h$ ranging from 2 to 8, were obtained from the global 1-RDM calculated using SIE+CCSD(T) and processed via PySCF. The $\Delta \mu$ components along the $x$, $y$, and $z$ axes are shown in Figure~\ref{fig:SI_dp_moment}. The definition of the $x$, $y$, and $z$ axes, as illustrated, has the water monomer rotating within the $yz$ plane. To more distinctly observe the variation in $\Delta \mu$ with changes in the size of the PAH, we aligned all the curves at zero to the $\Delta \mu$ of H$_2$O@PAH(8), resulting in the definition of $\Delta \Delta \mu$ is
\begin{equation}
    \Delta \Delta \mu (h) = \Delta \mu (h) - \Delta \mu (8).
\end{equation}

It can be observed that $\Delta \mu_{x}$ remains almost unchanged with the increasing size of the PAH, which is reasonable since the component of the water monomer dipole moment in the x-axis direction is nearly zero. Additionally, graphene does not spontaneously generate dipole moment along the x-axis, hence $\Delta \mu_{x}$ is almost zero. However, once the rotation angle of the water deviates from $0^\circ$~(2-leg) or $180^\circ$~(0-leg), a component of the dipole moment is generated along the y-axis, and a corresponding opposite dipole moment is induced on the graphene surface along the y-axis to partially counteract the component of the water's dipole moment. From Figure~\ref{fig:SI_dp_moment}\textbf{b} (or \textbf{f}), it is evident that for the configuration at $\theta=60^{\circ}$, $\Delta \mu_{y}$ (or $\Delta \Delta \mu_{y}$) changes dramatically with the size of the PAH, highlighting the long-range interaction between water and graphene. A similar conclusion can also be drawn from the changes observed in $\Delta \mu_{z}$ (or $\Delta \Delta \mu_{z}$) in Figure~\ref{fig:SI_dp_moment}\textbf{c} (or \textbf{g}).

\subsection{Weak Interaction Analysis}
\label{sec:SI_weak_interaction_analysis}

\begin{figure}[!ht]
\centering
\includegraphics[width=0.60\linewidth]{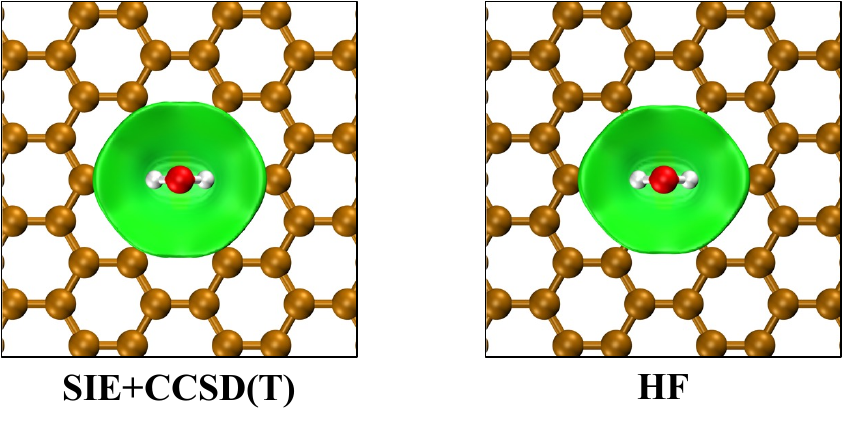}
\caption{\label{fig:SI_IGMH_comparison}
The comparison IGMH analysis between SIE+CCSD(T) and HF results. The figures are the top view of 2-leg@PAH(6).
}
\end{figure}

\begin{figure}[!ht]
\centering
\includegraphics[width=1.0\linewidth]{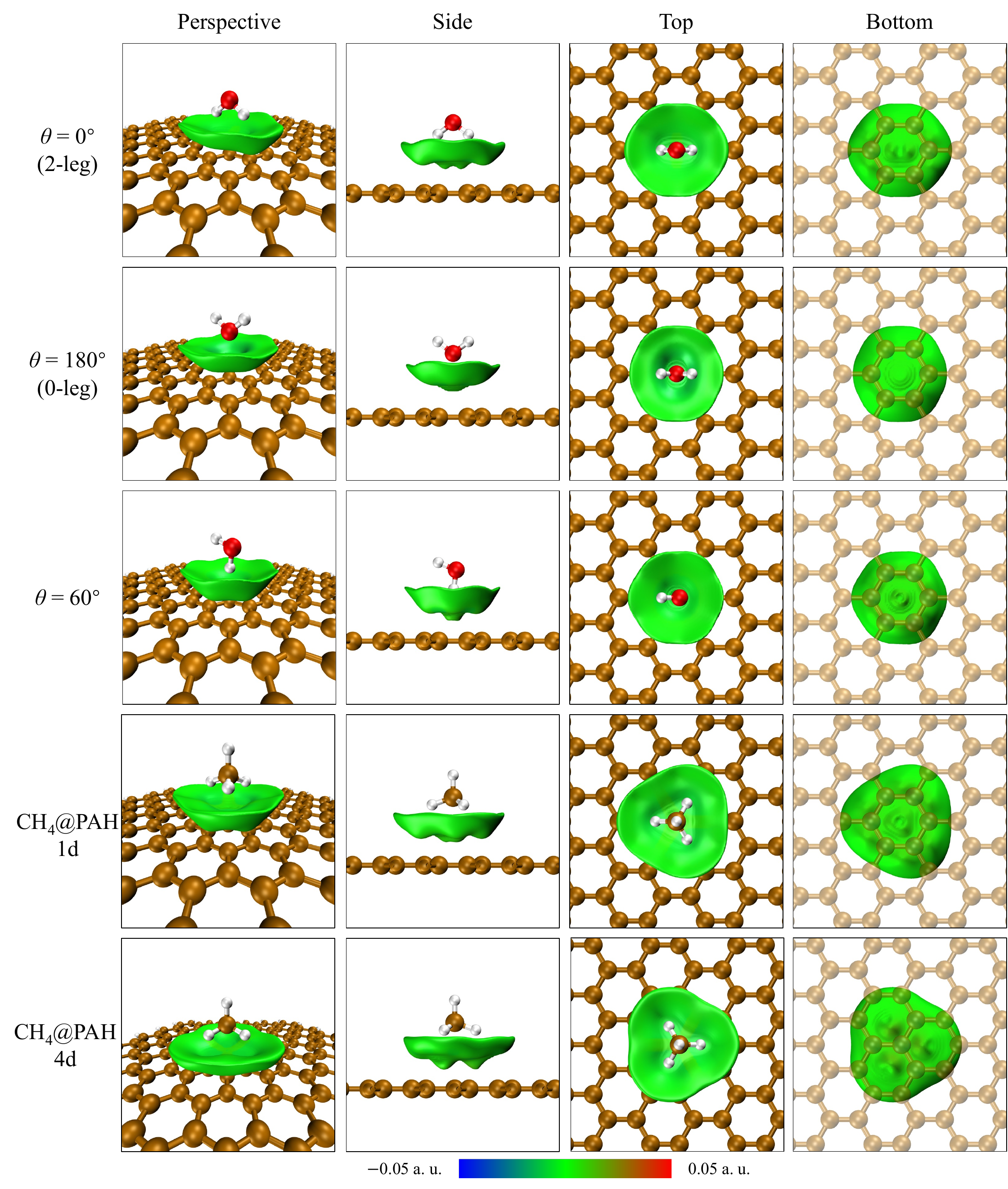}
\caption{\label{fig:SI_IGMH}
The IGMH weak interaction analysis for 2-leg, 0-leg, $\theta = 60^{\circ}$ configuration of H$_2$O@PAH(6)~(first three lines) and 1d/4d configuration of CH$_4$@PAH(6)~(last two lines). The isosurcfaces cutoff value for H$_2$O@PAH(6) and  CH$_4$@PAH(6) configurations is set as $1 \times 10^{-4}$ a.~u. to get a appropriate demonstration for isosurfaces. The color bar here represents the range of interaction strengths projected onto the isosurface, where the same settings are used for all configurations. The structures are rendered by VMD software~\cite{HUMP96}.
}
\end{figure}

As demonstrated in the main text, the water-graphene interaction is insensitive to the orientation of the adsorbed water monomer. Here, we provide additional rationalization using the weak interaction analysis of Independent Gradient Model based on Hirshfeld partitioning (IGMH)~\cite{lu2022independent}, a visualization method to illustrate weak interactions. We utilize Multiwfn software~\cite{lu2012multiwfn, lu2024comprehensive} to perform IGMH analysis to investigate the interaction regions and types between the water monomer and graphene. We have tested and found that the qualitative features of IGMH analyses are similar to the SIE+CCSD(T) and HF wavefunctions, as compared in Figure~\ref{fig:SI_IGMH_comparison}. 
Therefore, we provided more detailed comparison of IGMH analyses between different systems with HF wavefunctions computed with cc-pVDZ basis (Figure~\ref{fig:SI_IGMH}).

In the IGMH method, the interaction strength is characterized using a quantity related to electron density, $\delta g^{\text{inter}}$, as detailed in the referenced article~\cite{lu2022independent}. The final IGMH analysis results are shown in the Figure~\ref{fig:SI_IGMH}. The isosurfaces in the figures represent surfaces of equal interaction strength between adsorbate and substrate, while the colors on these surfaces represent the projection of the interaction strength within the region enclosed by the isosurface. Therefore, if the distribution of interaction strength within the isosurface is uneven, corresponding non-uniform color distributions will appear on the isosurface, with the corresponding color bar provided below the figure. Empirically, green represents van der Waals interactions, blue represents stronger attractive interactions, such as hydrogen bonds, and red represents stronger repulsive interactions.

The most representative configurations—2-leg, 0-leg, and $\theta = 60^{\circ}$—are analyzed and presented in the first three lines of Figure~\ref{fig:SI_IGMH}. The isosurfaces are uniformly green, indicating that the interaction strength between water and graphene is consistently even across the board, demonstrating the characteristics of typical van der Waals interactions, and no areas of particularly strong interactions are present. Notably, across different angular configurations, the morphology of the interaction isosurfaces remains similarly bowl-shaped, and there are no significant changes in the intensity or range of interactions. 

For comparison, the system of CH$_4$ adsorbed on graphene is also analyzed. CH$_4$, having no polarity due to the complete overlap of its positive and negative polarity centers, exhibits almost identical interacting energies between different adsorption configurations when interacting with the graphene surface~\cite{thierfelder2011methane}. The most representative configurations for CH$_4$@Graphene are 1d and 4d. In the 1d configuration, the carbon atom of CH$_4$ is located above the hole in graphene, with hydrogen atoms directed towards the adjacent carbon atoms on the graphene, whereas in the 4d configuration, CH$_4$'s carbon atom is directly above a carbon atom of graphene, and the hydrogen atoms point toward the holes adjacent in the graphene. In the calculations, graphene is modeled using PAH, and the calculations are done under OBC. For both 1d and 4d configurations, structural optimization are carried out on PAH(4) using $\omega$B97M-V, with all atoms on PAH(4) being fixed. The HF calculations are done by replacing PAH(4) to PAH(6) in the optimized structures. The results, as shown in the last two lines of the Figure~\ref{fig:SI_IGMH}, clearly show that, the morphology and intensity of interaction region between two CH$_4$@PAH(6) configurations and three H$_2$O@PAH(6) configurations remain highly consistent.

This comparison demonstrates that graphene is indeed capable of canceling out the anisotropy in interactions with water, allowing it to exhibit similar interaction strengths for the configurations across different water rotation configurations. However, this is not commonly observed in studies related to water monomer adsorption. Water is a highly polar molecule and prone to forming hydrogen bond, therefore it usually shows preference towards specific adsorption sites and orientations when interacting with surfaces.


\section{Carbonaceous molecules on various surfaces}
\label{sec:SI_benchmark}

\subsection{CO@MgO(001)}

\begin{figure}[ht]
    \centering
    \includegraphics[width=0.5\linewidth]{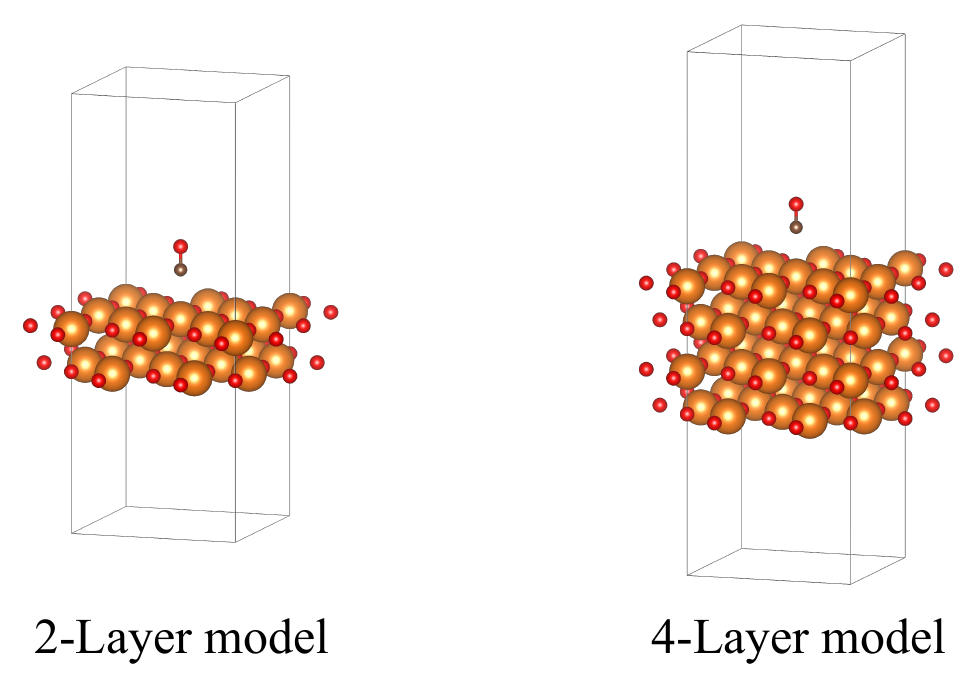}
    \caption{\label{fig:SI_CO@MgO_structure}
     2-Layer model and 4-Layer model for CO@MgO(001).
    }
\end{figure}

The CO@MgO(001) adsorption energy calculation primarily follows the approach detailed in a referenced study~\cite{jacs_mgo_co}. The adsorption energy is composed of two parts: interacting energy and geometry relaxation. We directly employ the geometry relaxation from the original study, which is approximately 8~meV, while the interacting energy part is calculated using SIE+CCSD(T). Note that two structures were used to obtain the adsorption energy in the original article. One is 4-Layer model, which contains 4 layers of MgO. The other one is the MgO 2-Layer model, derived from the MgO 4-Layer model. The structures for those two models are shown in Figure~\ref{fig:SI_CO@MgO_structure}. The 2-Layer model was employed for p-CCSD(T) in the original paper. The difference between 4-Layer model and 2-Layer model was estimated using MP2 and was regarded as a correction added back to the 2-Layer model p-CCSD (T) results. However, thanks to engineering efforts and the relatively lower complexity of the SIE algorithm, SIE+CCSD(T) does not need to do such things and directly employs the 4-Layer model of MgO(001) for calculations. 

During the calculation, CO adsorbed on the substrate is treated as a complete fragment, with the BNO threshold set to 10$^{-7.5}$. The remaining fragments in the substrate MgO are obtained by treating each Mg and O atom as a fragment, and using the BNO threshold of 10$^{-8.0}$. The BNO thresholds used in the calculations have essentially reached the upper limit we can afford due to the OOM issue. The basis set employed is the recommended aug-cc-pV$n$Z basis set~\cite{kendall1992electron} as the original cluster CCSD(T) used~\cite{jacs_mgo_co}. In this case, we employed a full electron basis set without using any pseudopotentials or frozen core approximations.

Hartree-Fock energy at CBS is obtained through aug-cc-pV(T,Q)Z two-point extrapolation, and correlation energy at CBS is calculated using SIE+CCSD(T) with aug-cc-pV(D,T)Z two-point extrapolation. 
Finally, the interacting energy and the corresponding adsorption energy are estimated as $-213$~meV and $-205$~meV ($-4.7$~kcal/mol), respectively. 
Note that the reference adsorption energy calculated by p-CCSD(T) and the experimental data are $-193$~meV ($-4.5$~kcal/mol),  $-198 \pm 19$~meV ($-4.6 \pm 0.4$~kcal/mol), respectively. In this sense, our method agrees well with these methods within 0.2~kcal/mol.

\subsection{Organic Molecules@Coronene}
\label{sec:SI_OM@Coronene}
Since the optimized structures for the Organic Small Molecules@Coronene composite were not provided in the original article~\cite{jacs_graphene_small_organic_molecules}, we follow the recent paper~\cite{ajala2019assessment} and use the double hybrid meta-GGA functional $\omega$B97M-V~\cite{mardirossian2016omegab97m} for the geometry optimization. We first constructed an initial structure roughly based on the structural diagrams presented in the original article's supporting information. Then DFT geometry optimization is conducted using the cc-pVTZ basis set. Initially, optimization was performed using PBE until convergence, and then it was followed by further optimization with $\omega$B97M-V. Note here, we used a model of coronene~(PAH(2)) under OBC to represent the substrate graphene as was done in the original paper. SIE+CCSD(T) calculates the interacting energy on the optimized geometries.

However, the experimental data provided is adsorption enthalpy, which cannot directly compare to our adsorption energy. To derive the experimental adsorption energy, the difference offered in the original paper~\cite{jacs_graphene_small_organic_molecules} between the optB88-vdW calculated adsorption enthalpy and the interacting energy is used. This difference is subtracted from the experimental adsorption enthalpy to obtain the experimental interacting energy, which then serves as the experimental reference.

In this SIE+CCSD(T) calculation, the cc-pV(D,T)Z basis sets are utilized for basis set extrapolation to CBS. Note that in this case, we treat each atom as an individual fragment for calculations, with a BNO threshold of 10$^{-8.0}$. All data have been listed in Table~\ref{tab:SI_SOM@Graphene}, and the energy has been taken as the root-mean-square value and represented in the main text Fig.~\ref{fig:benchmark}.

\begin{table}[!ht]
    \centering
    \caption{Interacting energy~(in kcal/mol) comes from SIE+CCSD(T) calculation and experiment.}
    \label{tab:SI_SOM@Graphene}
    \begin{tabular}{lcc}
    \toprule\midrule
    Adsorbate & $\Delta E_{\text{int}}^{\text{SIE+CCSD(T)}}$ & $\Delta E_{\text{int}}^{\text{exp}}$ \\
    \midrule
    Acetone & -8.7 & -8.8 \\
    Acetonitrile & -7.0 & -7.4  \\
    Dichloromethane & -8.3 & -6.9 \\
    Ethanol & -6.5 & -8.7 \\
    Ethyl Acetate & -10.7 & -10.8\\
    Toluene & -14.4 & -12.9  \\
    \midrule
    RMS & -9.6 & -9.5 \\
    \midrule\bottomrule
    \end{tabular}
\end{table}

In this study, we also investigate whether SIE+CCSD(T) preserves the size consistency. It is well known that MP2 suffers from lacking size consistency due to its omission of higher-order excitations, manifesting that its accuracy cannot be consistently preserved in the systems with different sizes. On the other hand, CCSD and CCSD(T), which employ exponential excitation operators that naturally consider higher-order excitations, are recognized for preserving size consistency. This attribute is a key reason why CCSD and CCSD(T) are regarded as effective for handling large systems. However, since SIE+CCSD(T) is a kind of embedding method, it is worth investigating whether its size consistency can still be maintained.

\begin{figure}[!ht]
    \centering
    \includegraphics[width=1.0\linewidth]{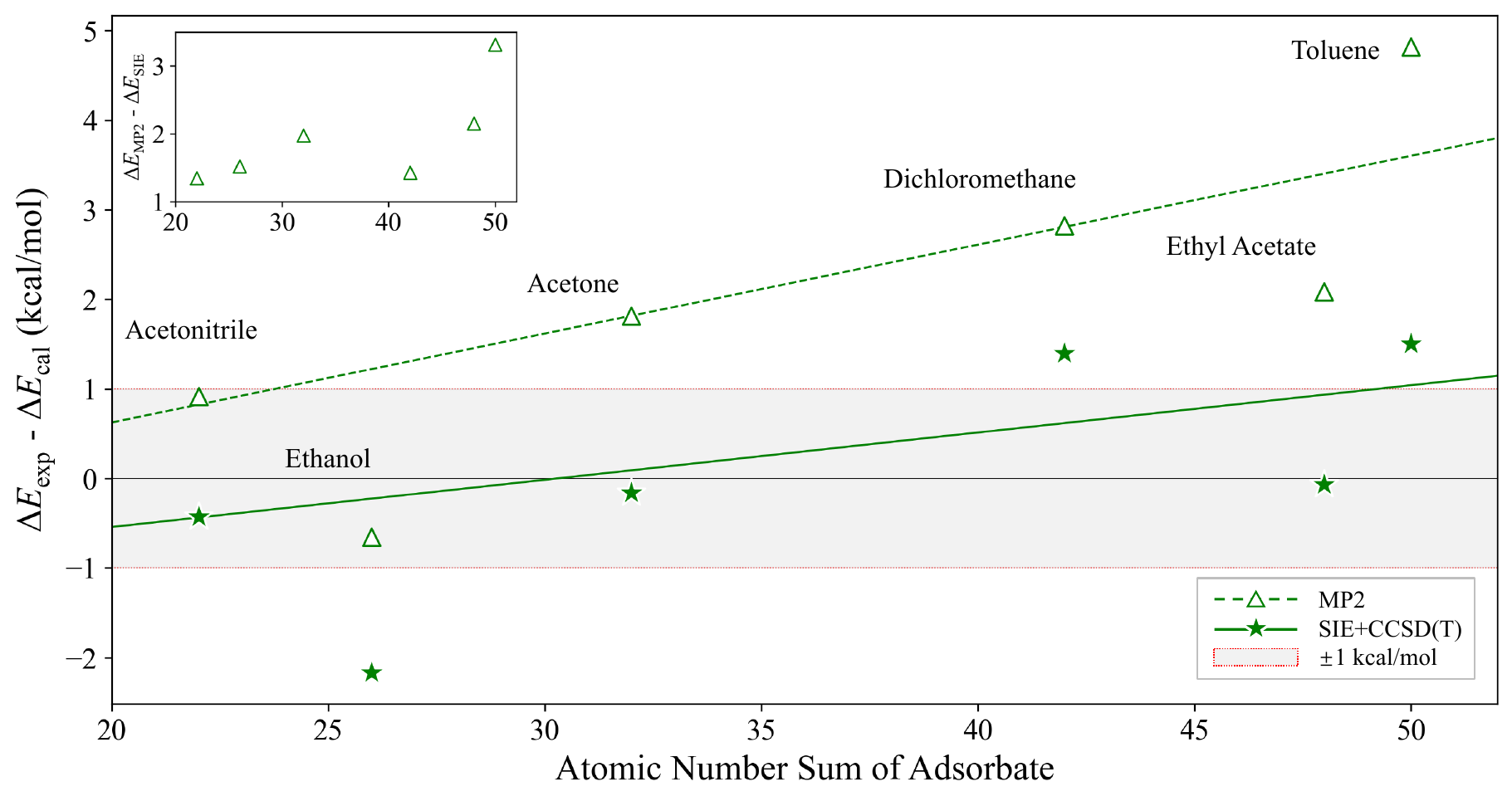}
    \caption{\label{fig:SI_size_consistency}
    The interacting energies difference between SIE+CCSD(T) or MP2 and experiment for Organic molecules@Coronene. The inset figure shows the interacting energies difference between MP2 and SIE+CCSD(T).
    }
\end{figure}

The group of organic small molecules adsorbed on graphene has gradually increased size, making those adsorption systems suitable for size consistency tests. Specifically, the size of all organic small molecule adsorption systems is labeled by the sum of the atomic numbers of all atoms within the organic small molecules. The interacting energies difference between MP2/SIE+CCSD(T) and experiments are plotted in Figure~\ref{fig:SI_size_consistency} related to the size of the system. The relationship between the difference and system size is fitted linearly. Notably, the results for Ethanol are considered as outliers because the SIE+CCSD(T) result deviates further from the experimental value compared to MP2, indicating that there are additional significant errors not considered in the Ethanol@coronene SIE+CCSD(T)/MP2 calculation compared to the real Ethanol@graphene system, such as finite size error. Therefore, the results for Ethanol are not included in the fitting analysis in both MP2 and SIE+CCSD(T).

It can be observed that the slope of the fitting line for SIE+CCSD(T) is significantly lower than that of MP2, suggesting that SIE+CCSD(T) is more size consistent compared to MP2. The same conclusion can be drawn from the difference in interacting energies between MP2 and SIE+CCSD(T), as shown in the inset picture in Figure~\ref{fig:SI_size_consistency}, where the correction by SIE+CCSD(T) to MP2 increases with the system size, indicating that SIE+CCSD(T) strives to maintain size consistency. However, it is worth mentioning that since SIE+CCSD(T) employs MP2 to correct bath truncation errors, the potential for size inconsistency errors may still exist. Considering alternative methods to correct bath truncation errors might avoid this issue, which will be further explored in the future.

\subsection{CO/CO$_2$@CPO-27-Mg Metal-Organic Framework}

The structures are chosen following ref.~\cite{jacs_MOF_CO}.
Specifically, CO is adsorbed within a CPO-27-Mg PBC structure that features two large pores, for which the PBE+D2 method is applied for calculation. Then the area close to CO is directly cut out, referred to as the 6\textbf{B} cluster, and calculated by MP2 under OBC. Further, a smaller structure, named the 2\textbf{B} cluster, is extracted from the 6\textbf{B} cluster and subjected to CCSD(T) calculations. The difference between the results from CCSD(T) and MP2 on the 2\textbf{B} cluster is employed as a correction, contributing to the total adsorption energy. In our benchmarking case, we choose the most stable adsorption configuration that is the configuration, with the C atom in CO pointing towards the Mg atom in CPO-27-Mg.

Finally, the adsorption energy $\Delta E_{\rm{ads}}$ can be expressed as
\begin{equation}
    \Delta E_{\rm{ads}} = \Delta E^{\rm{MP2}}(\rm{C_{\text{6\textbf{B}}}}) + \Delta_{\rm{LR}}(\rm{S,C_{\text{6\textbf{B}}}}) + \Delta_{\rm{CCSD(T)}}(\rm{C}_{\text{2\textbf{B}}}) + \textit{E}_{\rm{pair}},
\end{equation}
where the term $\Delta E^{\rm{MP2}}(\rm{C_{\rm{6\textbf{B}}}})$ represents the part of adsorption energy calculated using canonical MP2 for the 6\textbf{B} cluster under OBC, $E_{\rm{pair}}$ is the pair correction caused by monomer interaction in different adsorption site in PBC structure. $\Delta_{\rm{LR}}(\rm{S,C_{\rm{6\textbf{B}}}})$ symbolizes the difference between the PBC structure and the OBC 6\textbf{B} cluster, utilized to correct a portion of the cluster approach in the 6\textbf{B} cluster. $\Delta_{\rm{LR}}(\rm{S,C_{\rm{6\textbf{B}}}})$ is estimated by PBE+D2. Therefore, $\Delta_{\rm{LR}}(\rm{S,C_{\rm{6\textbf{B}}}})$ can be expressed as
\begin{equation}
    \Delta_{\rm{LR}}(\rm{S,C_{\rm{6\textbf{B}}}}) = \Delta \textit{E}^{\rm{PBE+D2}}(\rm{PBC}) - \Delta \textit{E}^{\rm{PBE+D2}}(\rm{OBC}).
\end{equation}
$\Delta_{\rm{CCSD(T)}}(\text{C}_{\rm{2\textbf{B}}})$ is the CCSD(T) correction perform on 2\textbf{B} cluster which could be described as
\begin{equation}
    \Delta_{\rm{CCSD(T)}}(\rm{C_{2\textbf{B}}}) = \Delta \textit{E}^{\rm{CCSD(T)}}(\rm{C_{\rm{2\textbf{B}}}}) - \Delta \textit{E}^{\rm{MP2}}(\rm{C_{\rm{2\textbf{B}}}}).
\end{equation}

In the case of SIE+CCSD(T), we can handle $\Delta E^{\rm{SIE+CCSD(T)}}(\rm{C_{\rm{6\textbf{B}}}})$ component to replace the $\Delta E^{\rm{MP2}}(\rm{C_{\rm{6\textbf{B}}}}) + \Delta_{\rm{CCSD(T)}}(C_{\rm{2\textbf{B}}})$ part of $\Delta E_{\rm{ads}}$. However, Our calculated $\Delta E^{\rm{MP2}}(\rm{C_{\rm{6\textbf{B}}}})$, $-49.4$ kJ/mol, is 4.0 kJ/mol lower than the  $-45.4$ kJ/mol value reported in ref.~\cite{jacs_MOF_CO}. 
Such small difference may be attributed to only valence shell electrons (2\textit{p} electrons of Mg have been treated as a valence
shell) have been used for MP2 and CCSD(T) calculation in original work~\cite{jacs_MOF_CO}, but full electrons are used for MP2 and SIE+CCSD(T) calculation in this work.
Actually, even for $\Delta E^{\rm{SIE+CCSD(T)}}(\rm{C_{\rm{6\textbf{B}}}})$, the gap is really small compared to the value of $\Delta E^{\rm{MP2}}(\rm{C_{\rm{6\textbf{B}}}}) + \Delta_{\rm{CCSD(T)}}(C_{\rm{2\textbf{B}}})$ in original paper. The $\Delta E^{\rm{SIE+CCSD(T)}}(\rm{C_{\rm{6\textbf{B}}}})$ is $-46.2$ kJ/mol, in comparison the $\Delta E^{\rm{MP2}}(\rm{C_{\rm{6\textbf{B}}}}) + \Delta_{\rm{CCSD(T)}}(C_{\rm{2\textbf{B}}})$ is $-43.1$ kJ/mol, where the absolute difference is 3.1 kJ/mol which is already below than chemical accuracy. To maintain consistency with reference data, we reuse the data provided in the original paper as much as possible. Hence, we chose to calculate $\Delta_{\rm{SIE+CCSD(T)}}(\rm{C_{6\textbf{B}}})$ on the 6\textbf{B} cluster to replace $\Delta_{\rm{CCSD(T)}}(\rm{C_{2\textbf{B}}})$ where $\Delta_{\rm{SIE+CCSD(T)}}(\rm{C_{6\textbf{B}}})$ is denoted as
\begin{equation}
    \Delta_{\rm{SIE+CCSD(T)}}(\rm{C_{6\textbf{B}}}) = \Delta \textit{E}^{\rm{SIE+CCSD(T)}}(\rm{C_{\rm{6\textbf{B}}}}) - \Delta \textit{E}^{\rm{MP2}}(\rm{C_{\rm{6\textbf{B}}}}).
\end{equation}
The $\Delta_{\rm{SIE+CCSD(T)}}(\rm{C_{6\textbf{B}}})$ has the value of 3.2 kJ/mol, combining it with the original $\Delta E^{\rm{MP2}}(\rm{C_{\rm{6\textbf{B}}}})$ ($-45.4$ kJ/mol), $\Delta_{\rm{LR}}(\rm{S,C_{\rm{6\textbf{B}}}})$ ($-0.1$ kJ/mol) and $\textit{E}_{\rm{pair}}$ ($-0.3$ kJ/mol) in the paper, the adsorption energy with SIE+CCSD(T) correction on 6\textbf{B} cluster is $-42.4$ kJ/mol in agreement with both the original paper, $-43.3$ kJ/mol and the experiment reference~\cite{sauer2019ab} value $-43.8 \pm 1.0$ kJ/mol.

The adsorption energy of CO$_2$@CPO-27-Mg was estimated using the same process. The reference calculated adsorption energy using the isolated model~\cite{jpcc_MOF_CO2_2023} is $-50.4$ kJ/mol; After a series of corrections, the final adsorption enthalpy was determined to be $-41.4$ kJ/mol. The difference $9.0$ kJ/mol between the calculated adsorption energy and adsorption enthalpy is the zero-point energy and thermal correction and $-RT$ correction. The experimentally measured adsorption enthalpy is $-43 \pm 4$ kJ/mol~\cite{MOF_CO2_exp1, MOF_CO2_exp2, MOF_CO2_exp3, MOF_CO2_exp4, MOF_CO2_exp5}. Thus, the estimated experimental reference adsorption energy is  $-52 \pm 4$ kJ/mol after subtracting the zero-point energy, thermal correction and $-RT$ correction.

The $\Delta_{\rm{SIE+CCSD(T)}}(\rm{C_{6\textbf{B}}})$ correction obtained using SIE+CCSD(T) on the 6\textbf{B} cluster is $-1.3$ kJ/mol, which is $-3.5$ kJ/mol less than the previous corrected CCSD(T) value used in~\cite{jpcc_MOF_CO2_2023}, Therefore, when using $\Delta_{\rm{SIE+CCSD(T)}}(\rm{C_{6\textbf{B}}})$ correction, the adsorption energy value is estimated as $-53.9$ kJ/mol. In SIE+CCSD(T) calculation, the recommended aug-cc-pV(D,T)Z basis sets are used for basis set extrapolation to CBS. Here, we also treat each atom as an individual fragment for calculations, with a BNO threshold of 10$^{-8.0}$.

\clearpage
\section{Computational details for SIE calculations}
\label{sec:SI_setting_summary}

\renewcommand{\arraystretch}{1.1}
\begin{table}[!h]
    \centering
    \caption{
     H$_2$O@Graphene.
    }
    \begin{tabular}{c|c}
        \toprule\midrule
        System & H$_2$O@Graphene\\
        \midrule
        Boundary condition &  OBC and PBC \\
        Basis & ccECP-cc-pV(D,T)Z \\
        BNO threshold & $10^{-6.5}$ for H$_2$O $10^{-8.0}$ for graphene \\
        Partition & See section~\ref{sec:SI_Partition} \\
        \midrule\bottomrule
    \end{tabular}
\end{table}
\renewcommand{\arraystretch}{1.1}
\begin{table}[!h]
    \centering
    \caption{
     CO@MgO(001)
    }
    \begin{tabular}{c|c}
        \toprule\midrule
        System & CO@MgO(001)\\
        \midrule
        Boundary condition & PBC \\
        Basis & aug-cc-pV(T,Q)Z for HF and aug-cc-pV(D,T)Z for SIE \\
        BNO threshold & $10^{-7.5}$ for CO $10^{-8.0}$ for MgO \\
        Partition & Take CO as a fragment, take every single atom in MgO as a fragment.  \\
        \midrule\bottomrule
    \end{tabular}
\end{table}
\renewcommand{\arraystretch}{1.1}
\begin{table}[!h]
    \centering
    \caption{
     Organic molecule @graphene
    }
    \begin{tabular}{c|c}
        \toprule\midrule
        System & Organic small molcule@graphene\\
        \midrule
        Boundary condition & OBC \\
        Basis & cc-pV(D,T)Z \\
        BNO threshold & $10^{-8.0}$ \\
        Partition & Take every single atom as a fragment.  \\
        \midrule\bottomrule
    \end{tabular}
\end{table}
\renewcommand{\arraystretch}{1.1}
\begin{table}[!h]
    \centering
    \caption{
     CO/CO$_2$@CPO-27-Mg
    }
    \begin{tabular}{c|c}
        \toprule\midrule
        System & CO/CO$_2$@CPO-27-Mg\\
        \midrule
        Boundary condition & OBC \\
        Basis & aug-cc-pV(D,T)Z \\
        BNO threshold & $10^{-8.0}$ \\
        Partition & Take every single atom as a fragment. \\
        \midrule\bottomrule
    \end{tabular}
\end{table}

\clearpage
\printbibliography